\documentclass[useAMS,usenatbib,a4paper]{mn2e}

\newcommand{\lt}{<}

\newcommand{\lya}{Ly$\alpha~$}
\newcommand{\lyat}{Ly$\alpha$} 
 
\newcommand{\msun}{\ifmmode M_{\odot} \else M$_{\odot}$\fi}
\newcommand{\msunyr}{\ifmmode M_{\odot} {\rm yr}^{-1} \else
M$_{\odot}$ yr$^{-1}$\fi}
\newcommand{\kms}{\ifmmode {\rm km s}^{-1} \else km s$^{-1}$\fi}

\newcommand{\hi}{\ifmmode {\textrm{H\textsc{i}}} \else H\textsc{i} \fi}
\newcommand{\hii}{\ifmmode {\textrm{H\textsc{ii}}} \else H\textsc{ii} \fi}
\newcommand{\affil}[1]{$^{\rm #1}$}

\defcitealias{garel2012a}{Paper I}	

\usepackage{times}
\usepackage[pdftex]{graphicx}
\usepackage{epsfig}
\usepackage{amssymb}
\usepackage{natbib}
\usepackage[latin1]{inputenc}
\usepackage{amsmath}
\usepackage{stfloats}
\usepackage{fixltx2e}
\usepackage{array}
\usepackage{wrapfig}
\usepackage{relsize}
\usepackage{color}
\usepackage{subfig}
\usepackage[section] {placeins}
\usepackage[cyr]{aeguill}     
\usepackage[margin=2cm]{geometry} 
\usepackage{hyperref}
\usepackage{arydshln}

\definecolor{r}{rgb}{1,0,0}

\newcommand{\modi}{\textcolor{black}}
\newcommand{\modibis}{\textcolor{black}}

\title{The UV, Lyman-$\alpha$, and dark matter halo properties of high redshift galaxies}
\author[Garel et al.]{{\Large T. Garel\affil{1,\dagger}, J. Blaizot\affil{2}, B. Guiderdoni\affil{2}, L. Michel-Dansac\affil{2}, M. Hayes\affil{3} and A. Verhamme\affil{4}}\\
\vspace{-0.3cm}
{\affil{1}\,Centre for Astrophysics and Supercomputing, Swinburne University of Technology, Hawthorn, Victoria 3122, Australia}\\
\vspace{-0.3cm}
{\affil{2}\,Universit\'e de Lyon, Lyon, F-69003; Universit\'e Lyon 1, Observatoire de Lyon, 9 avenue Charles Andr\'e, Saint-Genis Laval, F-69230;}\\
\vspace{-0.3cm}
{\affil{}\,CNRS, UMR 5574, Centre de Recherche Astrophysique de Lyon; Ecole Normale Sup\'erieure de Lyon, Lyon, F-69007, France}\\
\vspace{-0.3cm}
{\affil{3}\,Department of Astronomy, Oskar Klein Centre, Stockholm University, AlbaNova University Centre, SE-106 91 Stockholm, Sweden}\\
\vspace{-0.3cm}
{\affil{4}\,Observatoire de Gen\`eve, Universit\'e de Gen\`eve, 51, Ch. des Maillettes, CH-1290 Versoix, Switzerland}}

\begin{document}

\pdfminorversion=5
\pdfobjcompresslevel=2

\pagerange{\pageref{firstpage}--\pageref{lastpage}} \pubyear{2014}

\maketitle
\label{firstpage}

\begin{abstract}
We explore the properties of high-redshift Lyman-alpha emitters (LAE), and their link with the Lyman-Break galaxy population (LBG), using a semi-analytic model of galaxy formation that takes into account resonant scattering of \lya photons in gas outflows. We can reasonably reproduce the abundances of LAEs and LBGs from z $\approx$ 3 to 7, as well as most UV LFs of LAEs. The stronger dust attenuation for (resonant) \lya photons compared to UV continuum photons in bright LBGs provides a natural interpretation to the increase of the LAE fraction in LBG samples, $X_{\rm LAE}$, towards fainter magnitudes. The redshift evolution of $X_{\rm LAE}$ seems however very sensitive to UV magnitudes limits and EW cuts. In spite of the apparent good match between the statistical properties predicted by the model and the observations, we find that the tail of the \lya equivalent width distribution (EW $\gtrsim$ 100 \AA) cannot be explained by our model, and we need to invoke additional mechanisms. We find that LAEs and LBGs span a very similar dynamical range, but bright LAEs are $\sim$ 4 times rarer than LBGs in massive halos. Moreover, massive halos mainly contain weak LAEs in our model, which might introduce a bias towards low-mass halos in surveys which select sources with high EW cuts. Overall, our results are consistent with the idea that LAEs and LBGs make a very similar galaxy population. Their apparent differences seem mainly due to EW selections, UV detection limits, and a decreasing \lya to UV escape fraction ratio in high SFR galaxies.\\
\end{abstract}


\section{Introduction}
{\let\thefootnote\relax\footnotetext{$\dagger$ Australian Research Council Super Science Fellow.}}
{\let\thefootnote\relax\footnotetext{Email: \href{mailto:tgarel@astro.swin.edu.au}{tgarel@astro.swin.edu.au}}}
Star-forming galaxies are commonly detected nowadays via two main channels at high redshift (z $\gtrsim$ 3). On the one hand, the Lyman-Break technique is quite efficient at selecting objects with strong stellar UV continuum ($\lambda \approx 1500$ \AA), using a set of broad band colour-colour criteria \citep[e.g.][]{steidel1996a,steidel99,steidel2003a,gabasch,bouwens,bouwens2011a}. On the other hand, narrow-band imaging and spectroscopic surveys have also been able to probe large samples of high redshift sources via their (nebular) \lya emission line \citep[$\lambda=1215.67$ \AA; e.g.][]{hu98, rhoads00,shima06,ouch08, rauch08, hu2010a, hayes10,cassata2011a, blanc2011a}. 

The joint study of these two populations is essential to improve our understanding of the formation and evolution of galaxies in the early Universe, and their impact on the intergalactic gas notably during the epoch of reionisation.
Large observational datasets have enabled to constrain many statistical properties of Lyman-Break Galaxies (LBG) and Lyman-alpha Emitters (LAEs). The evolution of the UV luminosity function (LF) of dropout galaxies appears to be much stronger than the \lya LF of LAEs at z $\approx$ 3-7. In terms of physical properties, typical LBGs correspond to massive objects, with stellar mass correlated to UV luminosity \citep{shapley2001a,bouwens10,gonzalez2011}. LAEs are commonly thought to be less massive galaxies and to form a highly inhomogeneous population in terms of age, mass or dust content \citep{gawiser06,finkel,pirzkal2007a,finkelstein2009a,ono2010a}. Although part of the difference between the LBG and LAE populations is inherent to their respective, both biased, methods of selection, their link is not quite well established yet, and it is still unclear to what extent they are representative of the underlying galaxy population.

In the modern picture of hierarchical structure formation, galaxies form through gas accretion within virialised halos of dark matter located at the density peaks of the background matter field \citep{birnboim2003,keres2005,dekel_nature}. The internal properties of galaxies are tightly linked to the characteristics of the host halos. The host halo masses of typical LBGs are inferred to be one order of magnitude larger than those of currently observed LAEs from angular auto-correlation function measurements \citep{hamana2004a,gawiser2007a,hildebrandt2009a,mclure2009,ouchi2010a} and halo occupation distribution models \citep[][]{lee2006a,jose2013}. 

Accurate estimates on the halo masses of LBGs and LAEs at high-redshift are crucial to constrain formation history of local galaxies and identify the progenitors of the Milky-Way in the hierarchical context.
The observed LBGs and LAEs luminosities span several orders of magnitude and their host halo properties are thus expected to cover a wide range. Although it is essential to investigate the connection between UV/\lyat-selected galaxies and their dark matter halos within cosmological simulations, special care has first to be taken to describe the complex transfer of \lya photons in the interstellar medium.

The line emission originates in \hii regions in the interstellar medium as ionising radiation produced by massive, short-lived, stars is reprocessed into \lya photons.
The \textit{intrinsic} \lya emission is then a direct tracer of recent star formation activity in galaxies. However, the \textit{observed} flux can be strongly reduced by dust extinction in the interstellar medium (ISM) and intervening hydrogen absorption in the intergalactic medium (IGM). Due to a large absorption cross-section combined with substantial \hi column densities, the medium becomes optically thick to \lya photons which undergo a resonant scattering process. \lya photons diffuse both in frequency and physical space which can strongly alter the shape of the observed line. The travelling path in neutral gas, as well as the probability of encounter with dust grains, may be dramatically increased, and interactions with atoms in the tail of the velocity distribution can scatter photons off to the wing.  \lya radiation transfer is thus highly sensitive to the geometry, the ionisation state, and the kinematics of the ISM. 

Only very idealised cases can be investigated analytically \citep[e.g.][]{neufeld1990a,dijk06} and more realistic configurations require numerical schemes, mostly based on the Monte-Carlo technique \citep{ahn01,zheng02,verh06,dijk06,hansen06,laursen2007a,laursen2013a}.  Although a plethora of codes have been developed, only a few authors have intended to address the \lya RT issue on galactic scales within hydrodynamical simulations \citep{tasitsiomi,laursen09,barnes2011a,yajima2012a,verhamme2012}. These studies are extremely useful because they can follow the propagation of \lya photons through more realistic, non-linear, ISM density/velocity fields. Even though some trends are identified (anti-correlation between halo mass and \lya escape fraction, orientation effects, etc), the results still appear to depend on the resolution and the physics implemented in the underlying hydrodynamical simulation \citep[e.g.][]{verhamme2012}. Moreover, only a handful of high-resolution galaxies can be studied at once due to computing time issues. 

Semi-analytic models (SAM) of galaxy formation provide a unique alternative to investigate statistical samples of mock galaxies and their evolution with redshift. Recently, several models accounting for the \lya RT in galaxies based on numerical libraries have been presented. Rather simplistic physical pictures are usually adopted, such as plane-parallel slab geometries \citep{forero-romero2011} or thin expanding shell models \citep{orsi2012a,garel2012a}. Nonetheless, these approaches provide a more accurate treatment than previous models which used simple parametrisations to describe the \lya transfer effects and the \textit{observed} \lya properties of galaxies, assuming a constant \lya escape fraction \citep[e.g.][]{le-delliou2005a,nag}, simple dust models neglecting \lya resonant scattering \citep{mao07}, or using phenomenological prescriptions to account for its effect \citep[][]{koba07,koba10,dayal2011}.

In the present paper, we use the semi-analytic model presented in \citet[][]{garel2012a} to investigate the UV, \lyat, and dark matter halos properties of high redshift galaxies. In \citet[][hereafter Paper I]{garel2012a}, we described the coupling of GALICS \citep[\textit{GALaxies In Cosmological Simulations;}][]{hatton}, a hybrid model of galaxy formation based on a N-body cosmological simulation, with the library of \citet{schaerer2011a} which computes numerically the \lya RT through expanding gas shells. We present an overview of our model in Section \ref{sec:model}. In Section \ref{sec:uvlyalf}, we show that this model is able to reproduce the observed UV and \lya luminosity functions from z $\approx$ 3 to z $\approx$ 7. In Section \ref{sec:cross_prop1} and \ref{sec:cross_prop2}, we investigate the observed cross properties of LAEs and LBGs to study the connection between these two populations, focussing on the \lya equivalent width distributions. In Section \ref{sec:halos}, we present our model predictions for the host halo properties of LAEs and LBGs in terms of halo mass and occupation number. Finally, Section \ref{sec:summary} gives a discussion and a summary of our results.  

All quantities used throughout this paper assume the following cosmological parameter values: $h=H_0/(100  {\rm km \: s}^{-1} {\rm \: Mpc}^{-1}) = 0.70$, $\Omega_\Lambda = 0.72$, $\Omega_{\rm m} = 0.28$, $\Omega_{\rm b} = 0.046$, and $\sigma_8 = 0.82$ \modibis{\citep[WMAP-5;][]{komatsu09}}. All magnitudes are expressed in the AB system.

\section{Model overview}
\label{sec:model}

Our modelling of the high-redshift galaxies and their \lya emission properties has been fully described in \citetalias{garel2012a}. Here, we briefly summarise the main aspects of our approach.

We use an updated version of the GALICS semi-analytic galaxy formation model to predict the abundances and physical properties of galaxies, based on the hierarchical evolution of dark matter halos. The formation and evolution of dark matter structures is described by a cosmological N-body simulation post-processed with a Friends-of-Friends algorithm \citep[][]{davis85} to identify bound regions (i.e. dark matter halos). The merging histories of dark matter halos are computed according to the procedure of \citet{tweed09}. Our N-body simulation has been run with GADGET \citep[][]{gadget2} using 1024$^3$ particles in a periodic, comoving box of $100h^{-1}$ Mpc on a side. The mass resolution (M$_{\rm halo} \geq 2 \times 10^9$ \msun, corresponding to bound groups of at least 20 particles) and the box size have been chosen to allow us to investigate the statistical properties of LAEs and LBGs currently detectable at high-redshift. While in \citetalias{garel2012a} we assumed parameter values from the WMAP-3 release \citep{spergel}, here we adopt a more recent set of cosmological parameters consistent with the WMAP-5 data, which are described above.

Our modelling of galaxies and \lya emission is very similar to Paper
I, except for a few changes that we detail below. In \citetalias{garel2012a}, we
presented the modifications made to the original GALICS model
described in \citet{hatton} \citep[see also][]{cattaneo06}. The main updates
consisted of a new implementation of gas accretion onto galaxies (cold
flows), dust extinction of the UV continuum (spherical geometry so as
to be consistent with the \lya radiative transfer within outflowing
shells of gas - see below), and star formation. We compute star
formation rates directly from the cold gas surface density using the
Kennicutt-Schmidt law (i.e. $\Sigma_{\rm SFR} \propto \Sigma_{\rm gas}^{1.4}$), and we had to increase the star formation
efficiency by a factor of 25 in order to match the observed high redshift UV luminosity functions. This departure from the z$=$0 normalization of \citet{kennicutt98} was certainly due to the underlying (WMAP-3) cosmology that we adopted for the dark matter simulation, and especially the low value of $\sigma_8$ \citep[0.76;][]{spergel} which delays the growth of structure at early times. To this extent, boosting star formation was necessary to compensate the low abundance of star forming galaxies at high-redshift. With our new N-body simulation, $\sigma_8$ is slightly higher which eases the formation of larger structures at earlier epochs. We find that an increasing factor for a star formation efficiency equal to 5 in our model is sufficient to obtain a reasonable match to the high redshift data (see section \ref{sec:uvlyalf}). 

We use the stellar libraries of \citet[][STARDUST]{devriendt} to compute the intrinsic UV and \lya emission of galaxies for a Kennicutt IMF \citep[0.1-120 \msun;][]{kenn83}. Under the case B approximation \citep{osterbrock2006}, two third of the ionising photons emitted by stars are reprocessed into \lya photons when HII regions recombine, and we assume that no ionising photon can escape the medium (optically thick limit).The intrinsic \lya luminosity of galaxies is therefore given by $L_{Ly\alpha}^{\rm intr}= 0.67 Q(H)\frac{h_Pc}{\lambda_\alpha}$, where $Q(H)$ is the production rate of hydrogen-ionising photons, $c$ the speed of light, $h_P$ the Planck constant, and $\lambda_\alpha$ is the Ly$\alpha$ wavelength at line center. For the shape of the intrinsic line, we assume a Gaussian profile centered on $\lambda_\alpha$ with a characteristic width given by the circular velocity of the galaxy disc \citep{santos04}.

We take into account the transfer of UV continuum and \lya radiation through outflowing gas and dust. This picture is motivated by the apparent ubiquity of galactic winds seen at high redshift (powered by supernovae and/or stellar winds), as well as their impact on the observed \lya spectral signature in Lyman Break galaxies and Lyman Alpha Emitters \citep{shapley03,steidel2010a,mclinden,finkelstein2011c,kulas2012a,berry2012a}. Following the work of \citet{verh06}, we describe galactic outflows as homogenous, spherical, expanding shells of cold gas mixed with dust. \citet{verh08} have shown that high redshift \lya line profiles can be well reproduced by 3D Monte-Carlo radiation transfer (RT) models in expanding shells, when adjusting the values of various relevant shell parameters: the bulk motion expansion velocity $V_{\rm exp}$, the gas velocity dispersion $b$, the dust opacity $\tau_{\rm dust}$, and the neutral hydrogen column density of the shell N$_{\rm \hi}$. We compute shell parameters for GALICS galaxies in a similar fashion to \citetalias{garel2012a}. In our model, the shell speed $V_{\rm exp}$ has a (weak) dependence on the star formation rate, the value of $b$ is fixed to $20$ km s$^{-1}$, and the dust opacity is estimated from the \citet{mathis83} extinction curve at a given wavelength, the gas metallicity and column density, and a redshift-dependent dust-to-gas ratio \citep[see][for a detailed description of our model]{garel2012a}. The column density is proportional to ${\rm M}^{\rm gas}_{\rm shell} / (4\pi {\rm R_{shell}}^2)$, where ${\rm M}^{\rm gas}_{\rm shell}$ and ${\rm R_{shell}}$ are the shell mass and size respectively. We use the cold gas in galaxies as a proxy for the shell mass, and the disk scalelength ${\rm R_{disk}}$ to compute ${\rm R_{shell}}$ such that ${\rm R_{shell}}=\xi {\rm R_{disk}}$ where $\xi$ is a free parameter of the order of unity. While $\xi$ was $1.0$ in \citetalias{garel2012a}, we now set the value to 1.2 to improve the agreement of our model with the luminosity functions data over our redshift range of interest. 

The escape fraction of UV continuum photons is $e^{-\tau_{\rm dust}}$, as for a screen model, consistent with the thin shell approximation we make here. On the other hand, we predict the \lya properties of galaxies using the grid of models of \citet{schaerer2011a} which contains the results of a large number of \lya RT simulations in expanding shells, as described above \citep[see][for more details]{verh06}. The grid provides us with the \lya escape fraction $f_{\rm esc}^{\rm grid}$ and the emergent line profile $\Phi^{\rm grid}(Ly\alpha)$ for more than 6000 quadruplets of parameter values ($V_{\rm exp}$, $b$, $\tau_{\rm dust}$, N$_{\rm \hi}$). We interpolate the shell parameters predicted by GALICS onto the grid so as to compute $f_{\rm esc}$ and $\Phi(Ly\alpha)$ for each individual galaxy. The \textit{observed} \lya luminosities are thus given by $L_{\rm Ly\alpha} = L_{\rm Ly\alpha}^{\rm intr} \times f_{\rm esc}$. 

A main result of this modelling is that $f_{\rm esc}$ is of the order of unity for galaxies with a low star formation rate (SFR $\lesssim 1 \msunyr$) because they are predicted to have low gas column densities and dust opacities \citepalias[see section 3.2.3 in][]{garel2012a}. This suggests that the observed \lya luminosity could be used on average as a tracer of the SFR for faint LAEs. On the other hand, $f_{\rm esc}$ is greatly dispersed between 0 and 1 for high-SFR galaxies because star forming galaxies can have a wide range of N$_{\rm \hi}$ and $\tau_{\rm dust}$ values in our model. For these galaxies, \lya is no longer a reliable tracer of the star formation rate.

Our model only assumes the internal attenuation of the \lya line by dust in the shell, and the effect of the intergalactic medium is neglected. We will discuss this choice in more details in Section \ref{subsec:lyall}.

\section{Abundances of LBGs and LAEs}
\label{sec:uvlyalf}

In this section, we present the UV and \lya luminosity functions at z $=$ 3-7, a redshift range that corresponds to the post-reionisation epoch where most dropout galaxy and Lyman-alpha emitter surveys have been conducted so far. 
\begin{figure*}
\vskip-14ex 
\hspace{-1cm}
\begin{minipage}[]{0.3\textwidth}
\centering
\includegraphics[width=6.1cm,height=8.5cm]{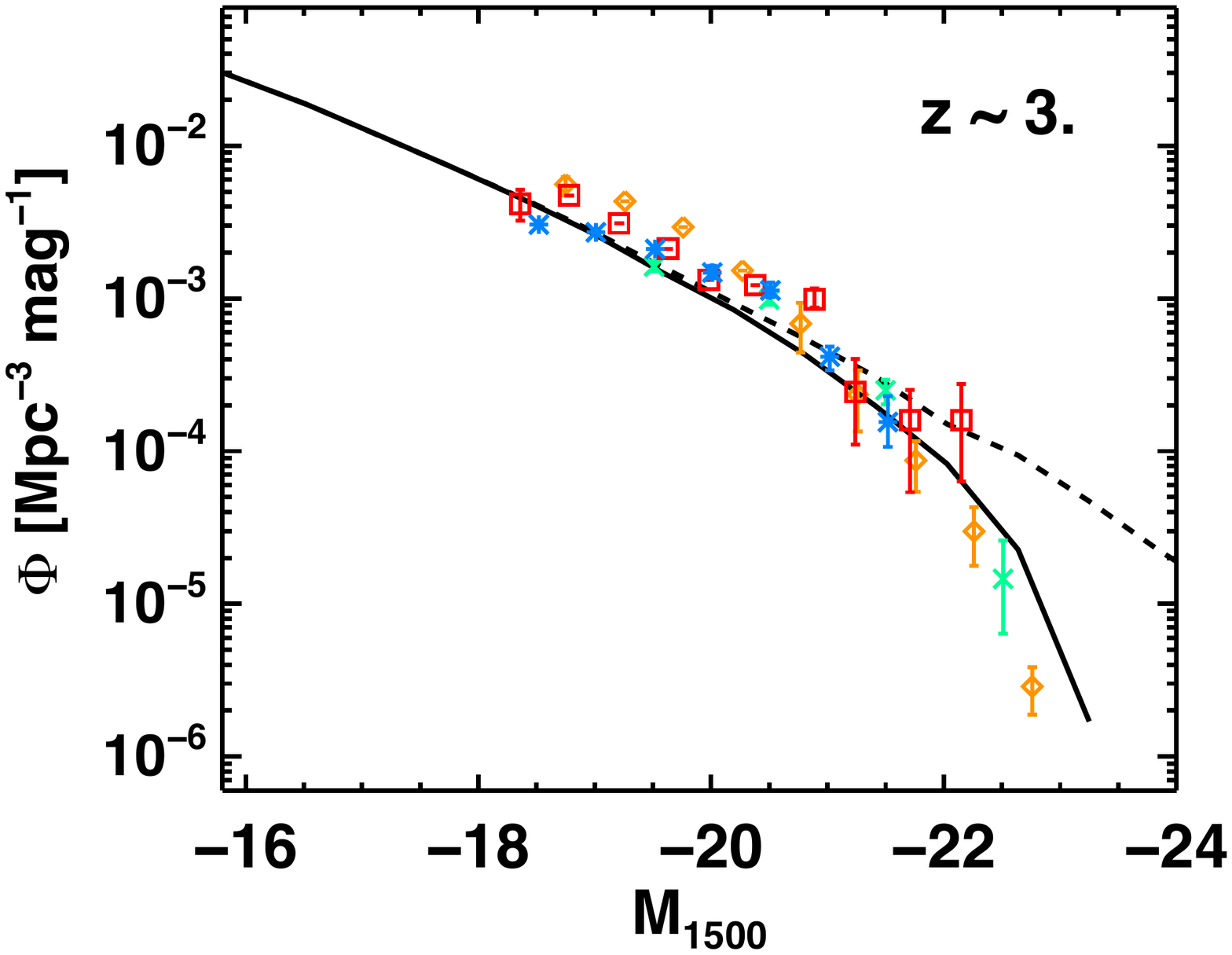}
\end{minipage}
\hspace{0.4cm}
\begin{minipage}[]{0.3\textwidth}
\centering
\includegraphics[width=6.1cm,height=8.5cm]{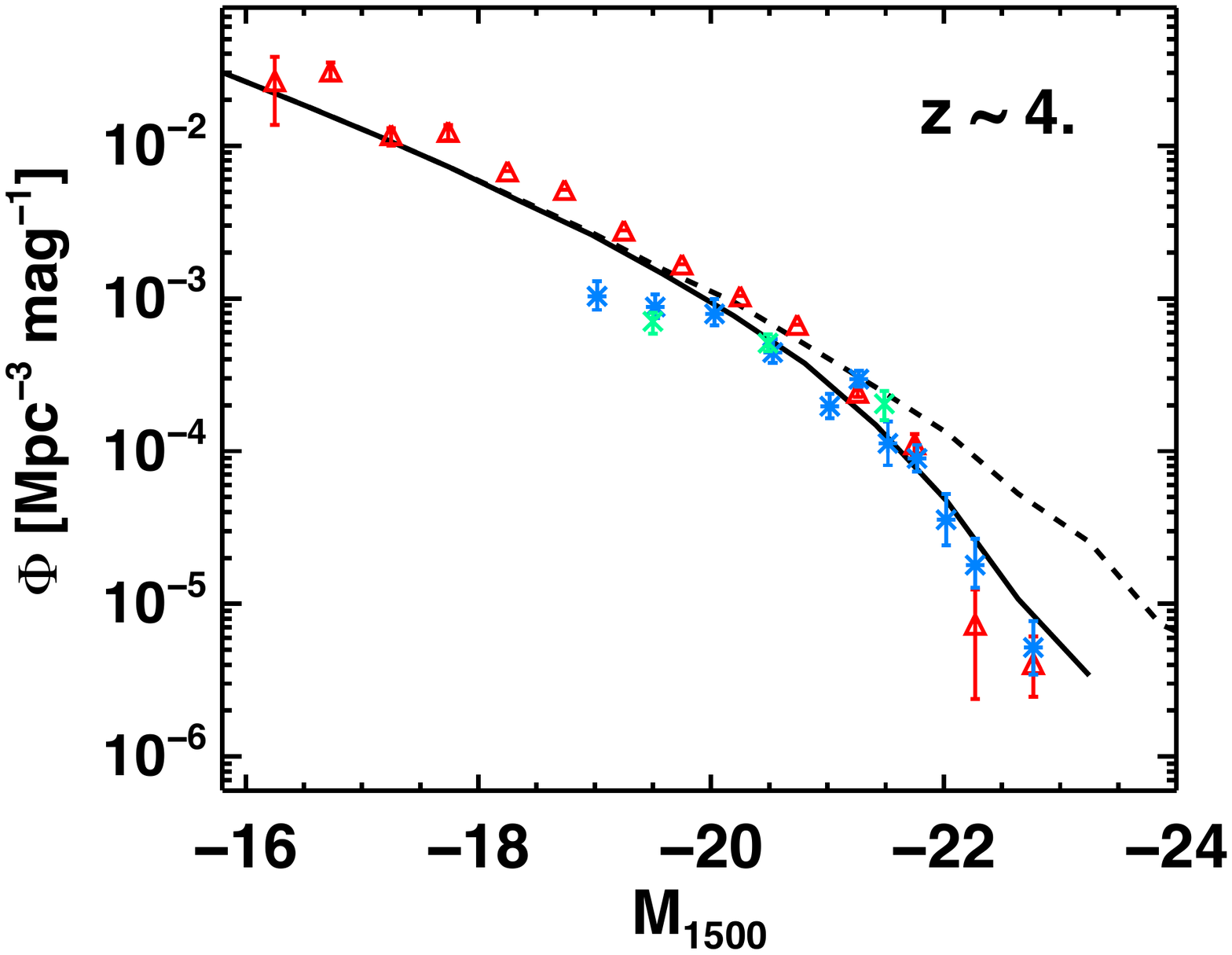}
\end{minipage}
\hspace{0.4cm}
\begin{minipage}[]{0.3\textwidth}
\centering
\includegraphics[width=6.1cm,height=8.5cm]{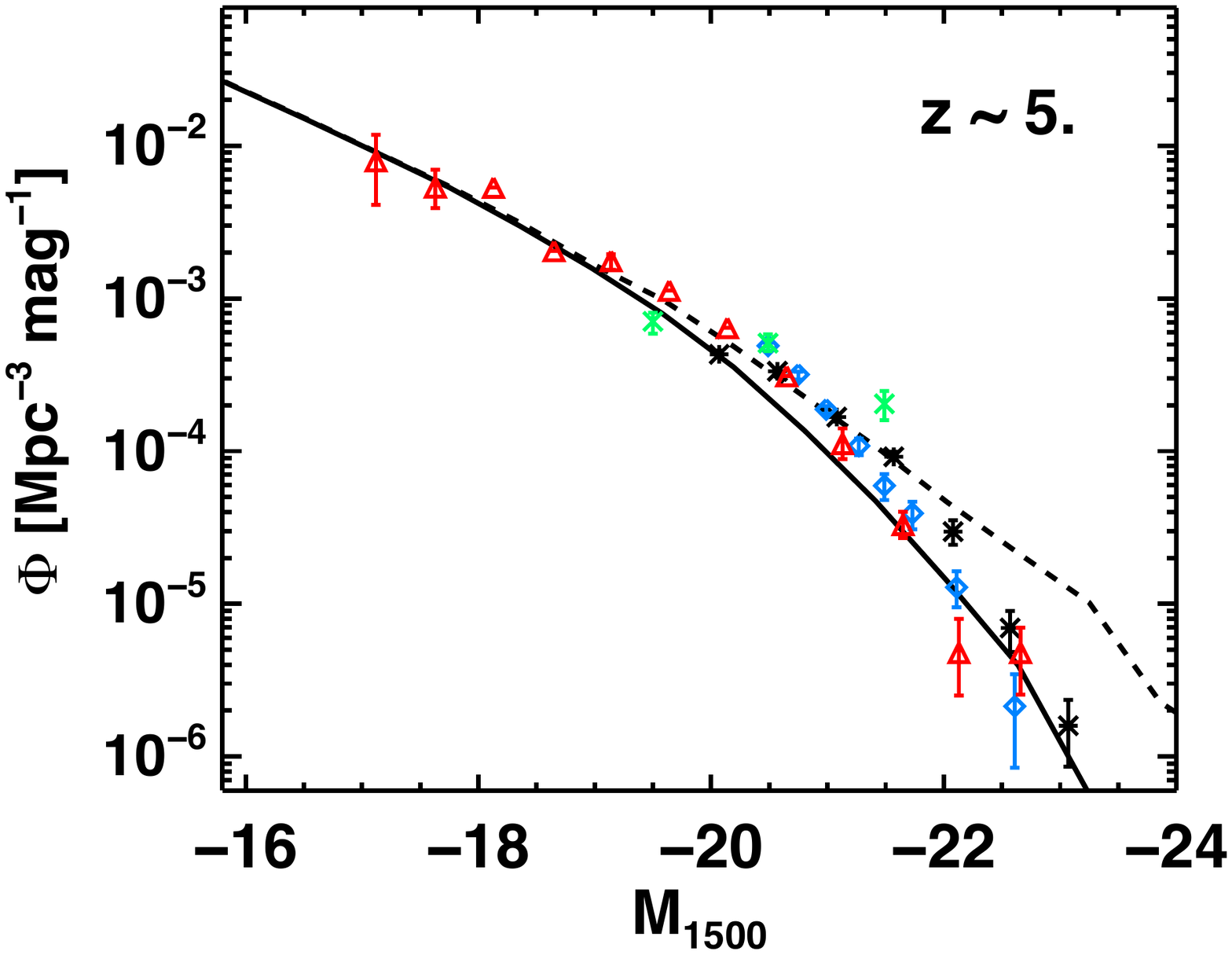}
\end{minipage}
\vskip-30ex 
\hspace{-1cm}
\begin{minipage}[]{0.3\textwidth}
\centering
\includegraphics[width=6.1cm,height=8.5cm]{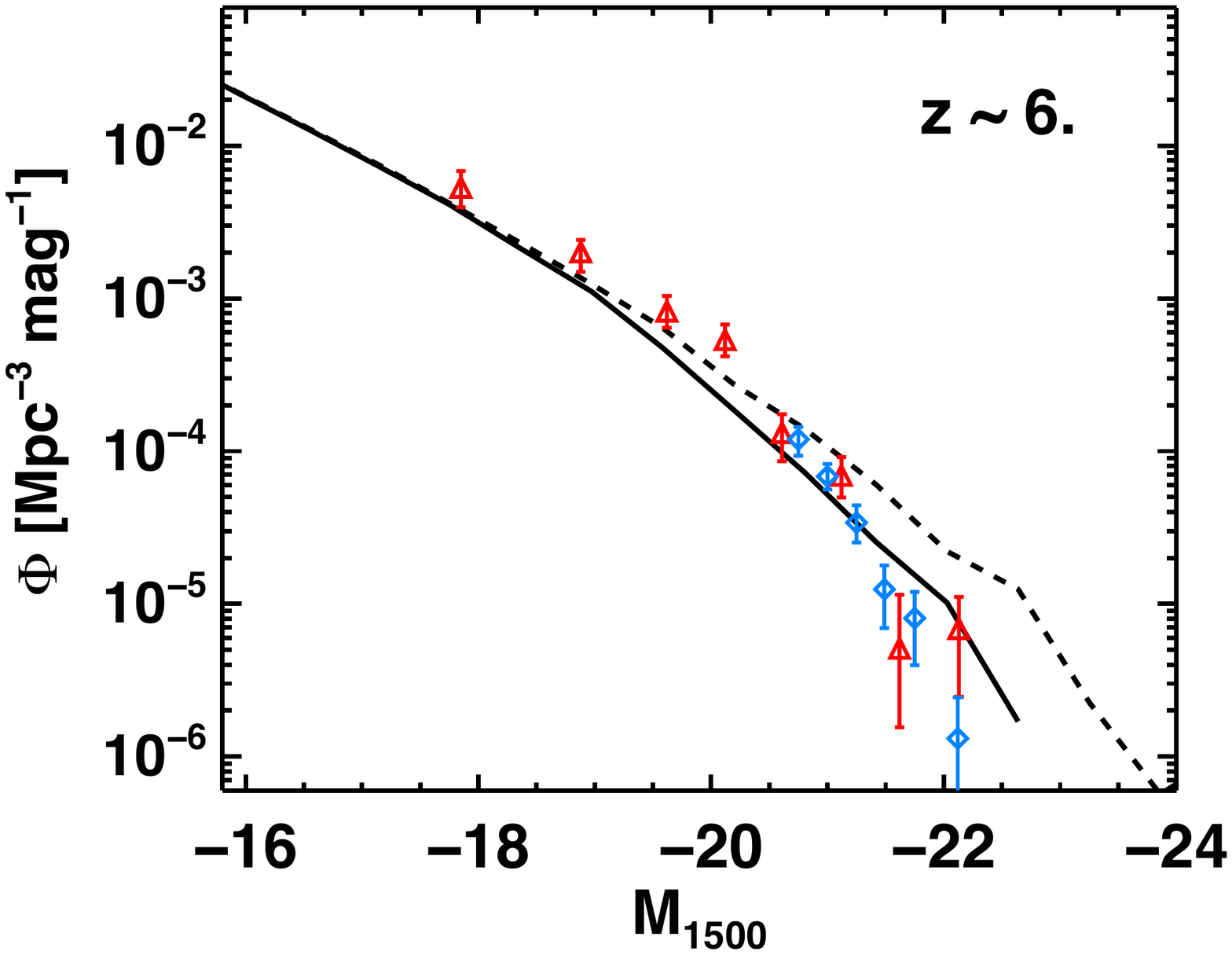}
\end{minipage}
\hspace{0.4cm}
\begin{minipage}[]{0.3\textwidth}
\centering
\includegraphics[width=6.1cm,height=8.5cm]{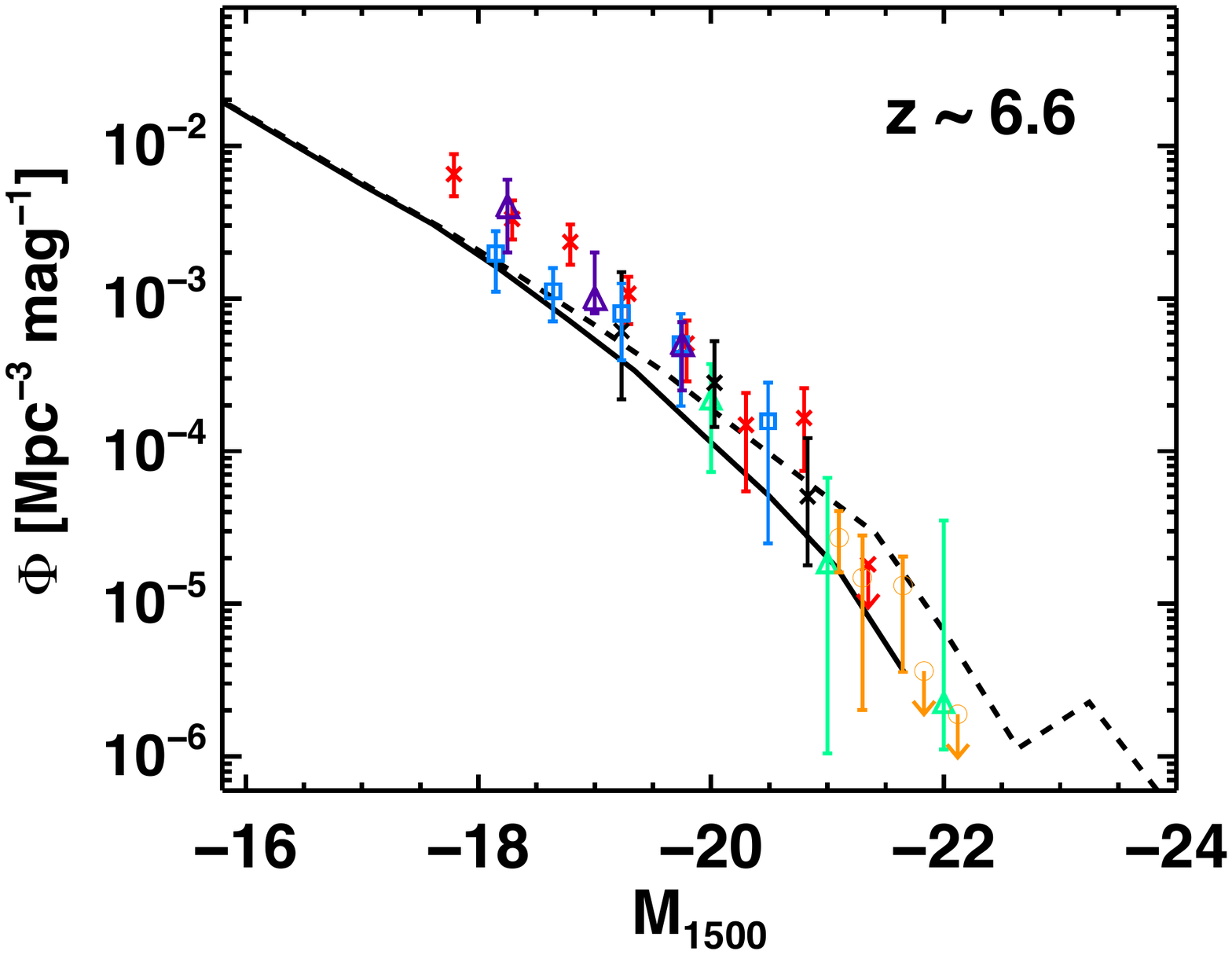}
\end{minipage}
\hspace{0.4cm}
\begin{minipage}[]{0.3\textwidth}
\centering
\includegraphics[width=6.1cm,height=8.5cm]{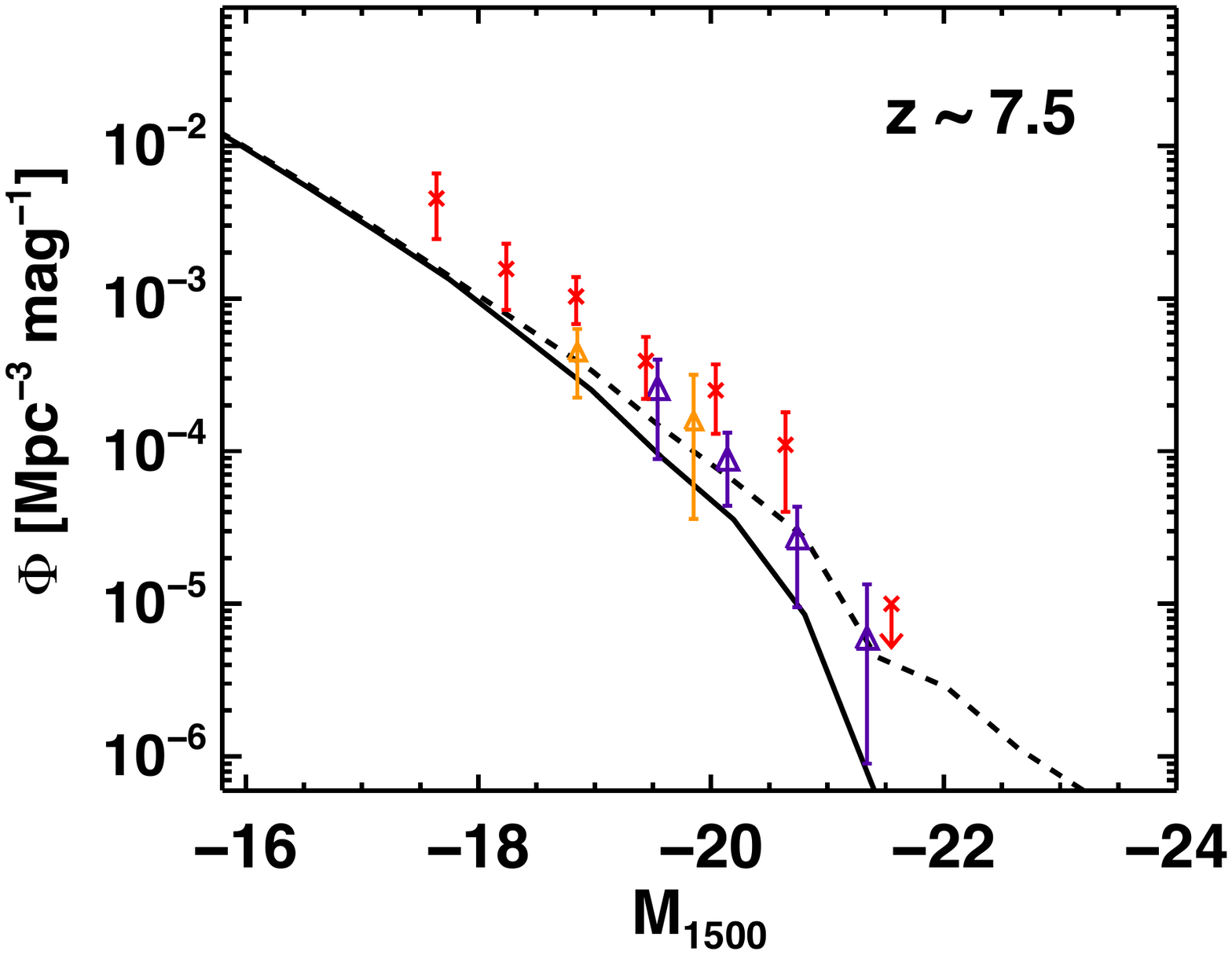}
\end{minipage}
\vskip-12ex 
\caption{UV luminosity functions at z $\approx$ 3, 4, 5, 6, 7 and 7.5. The solid (dashed) line is the model with (without, i.e. intrinsic) dust extinction. Symbols are observational data: \citet{reddy08} (orange diamonds), \citet{arnouts} (red squares), \citet{sawicki} (blue asterisks), \citet{gabasch} (green crosses), \citet{bouwens} (red triangles), \citet{iwata} (black asterisks), \citet{mclure2009} (blue diamonds), \citet{castellano} (light green triangles), \citet{bouwens2011a} (red crosses), \citet{bouwens08} (black crosses), \citet{mclure10} (blue squares), \citet{ouchi2009a} (orange circles), \citet{oesch2010,oesch2012} (purple triangles).}
\label{fig:uvlfs}
\end{figure*}

\subsection{UV luminosity functions}

Observationally, rest-frame UV luminosity functions (LF) are now rather well constrained up to z $\approx$ 6. At higher redshift, significant scatter in the data remains, mainly due to smaller statistics, larger cosmic variance effects, and more numerous contamination by foreground sources.
In Figure \ref{fig:uvlfs}, we show the UV LFs at z $\approx$ 3, 4, 5, 6, 7 and 7.5. In each panel, the solid black line corresponds to the LF with dust attenuation, and the dashed curve gives the intrinsic LF, i.e. without dust attenuation. The observations, represented by the symbols with error bars, come from various surveys of LBGs.

Overall, our model provides a good fit to the data over the redshift range of interest. While we note some tension at the faint end of the z $=$ 3 LF, the agreement between the model and the observations is excellent at these magnitudes for z $=$ 4 and 5. We somehow underpredict the luminosities and/or number densities of dropout galaxies at z $=$ 7.5 (bottom right panel in Figure \ref{fig:uvlfs}). Given that the intrinsic LF agrees with the data, our modelling of dust attenuation is perhaps too strong at z $=$ 7.5. We do not think of this small mismatch between the model and the observations as a real issue since the observed UV LF is still uncertain at this redshift (as the fraction of interlopers could be quite high in z $=$ 7-8 samples of dropout galaxies). 

The LBG selection probes bright UV-continuum sources (at $\sim 1300-1700$ \AA{} in rest-frame) and applies colour-colour criteria to ensure the detections at a given redshift, minimising the contamination by interlopers. The selection and filters vary from one survey to another, so, in our model, we choose to directly measure the restframe absolute UV magnitude of galaxies in an effective rectangular filter at 1500 $\pm$ 100 \AA{} to compute the UV LF \citep[as quoted by][]{gabasch}, without any colour-colour selection. These different selections only introduce a little variation on the UV LFs but they could still be responsible for part of the slight differences seen in the LFs between the various surveys, and between the data and the model.\\

\subsection{\lya luminosity functions}   
\label{subsec:lyall}
\begin{figure*}
\vskip-14ex 
\hspace{-1cm}
\begin{minipage}[]{0.3\textwidth}
\centering
\includegraphics[width=6.1cm,height=8.5cm]{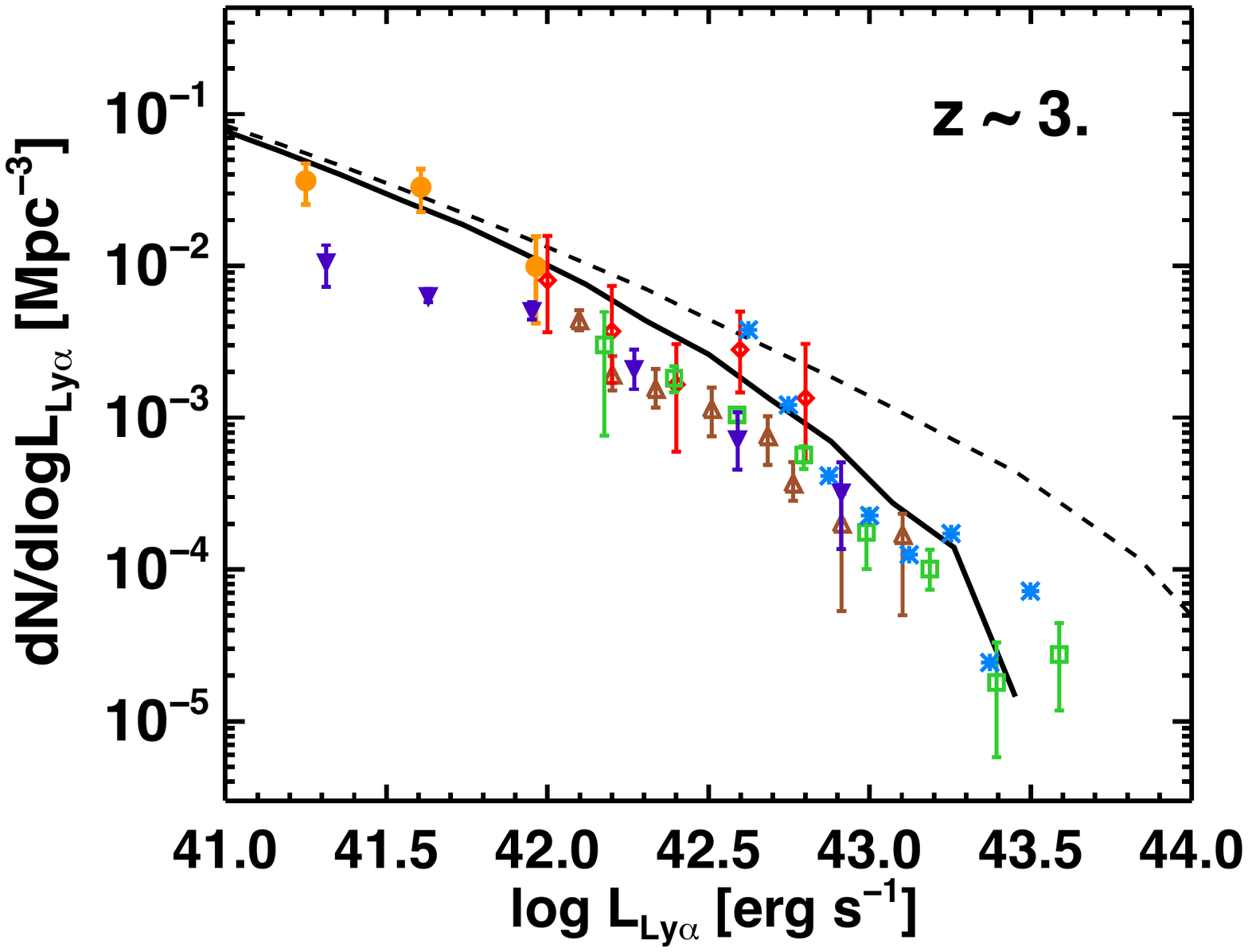}
\end{minipage}
\hspace{0.4cm}
\begin{minipage}[]{0.3\textwidth}
\centering
\includegraphics[width=6.1cm,height=8.5cm]{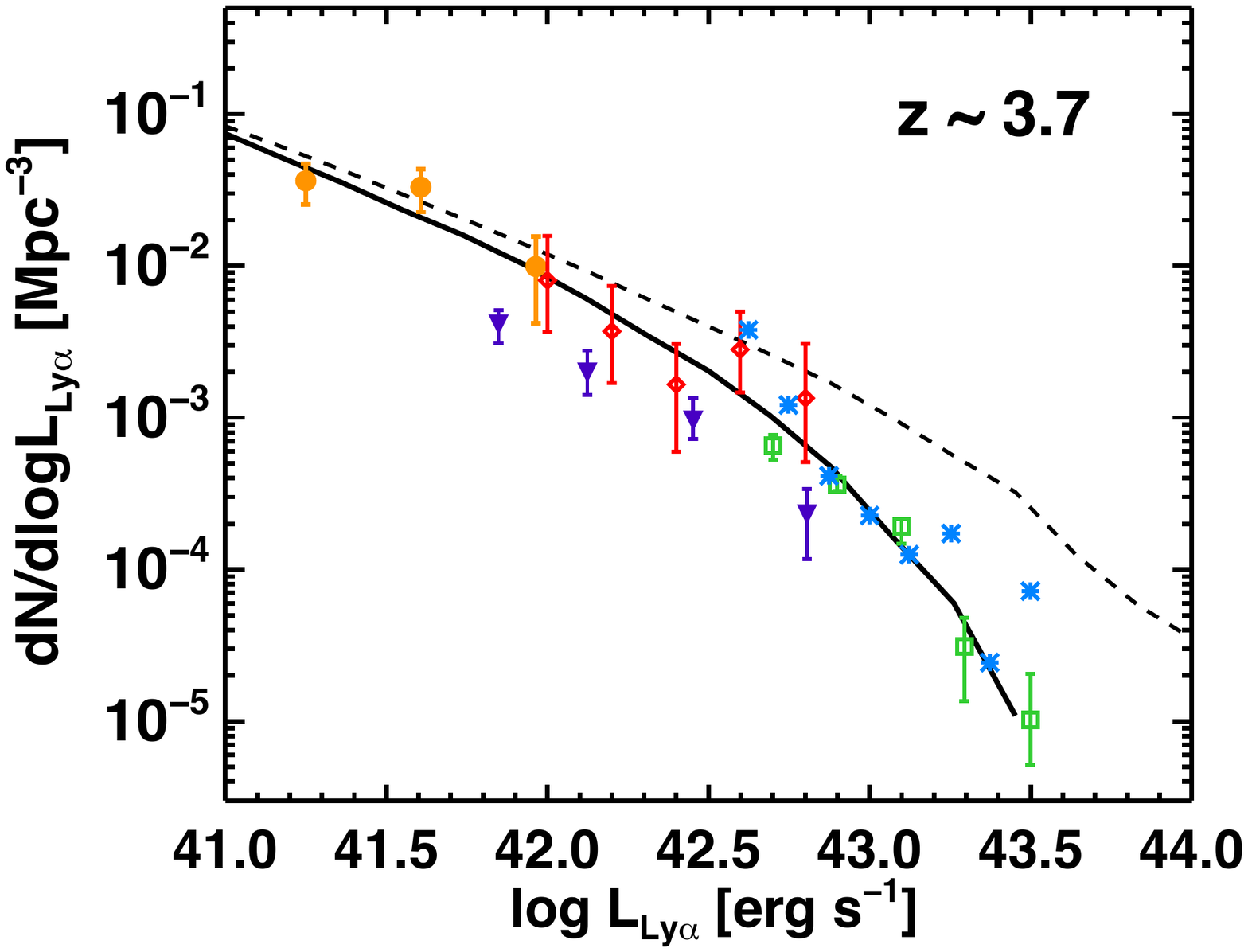}
\end{minipage}
\hspace{0.4cm}
\begin{minipage}[]{0.3\textwidth}
\centering
\includegraphics[width=6.1cm,height=8.5cm]{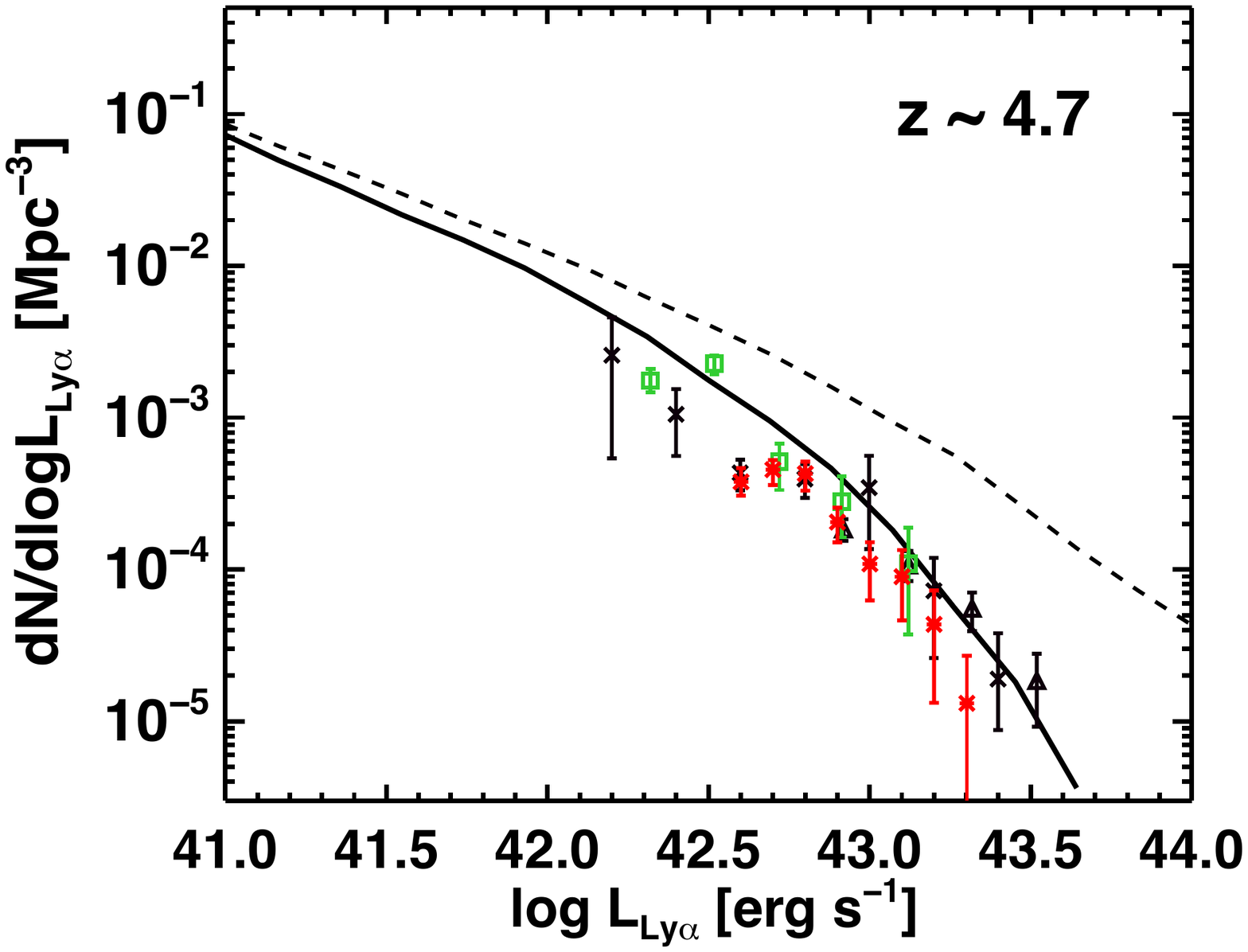}
\end{minipage}
\vskip-30ex 
\hspace{-1cm}
\begin{minipage}[]{0.3\textwidth}
\centering
\includegraphics[width=6.1cm,height=8.5cm]{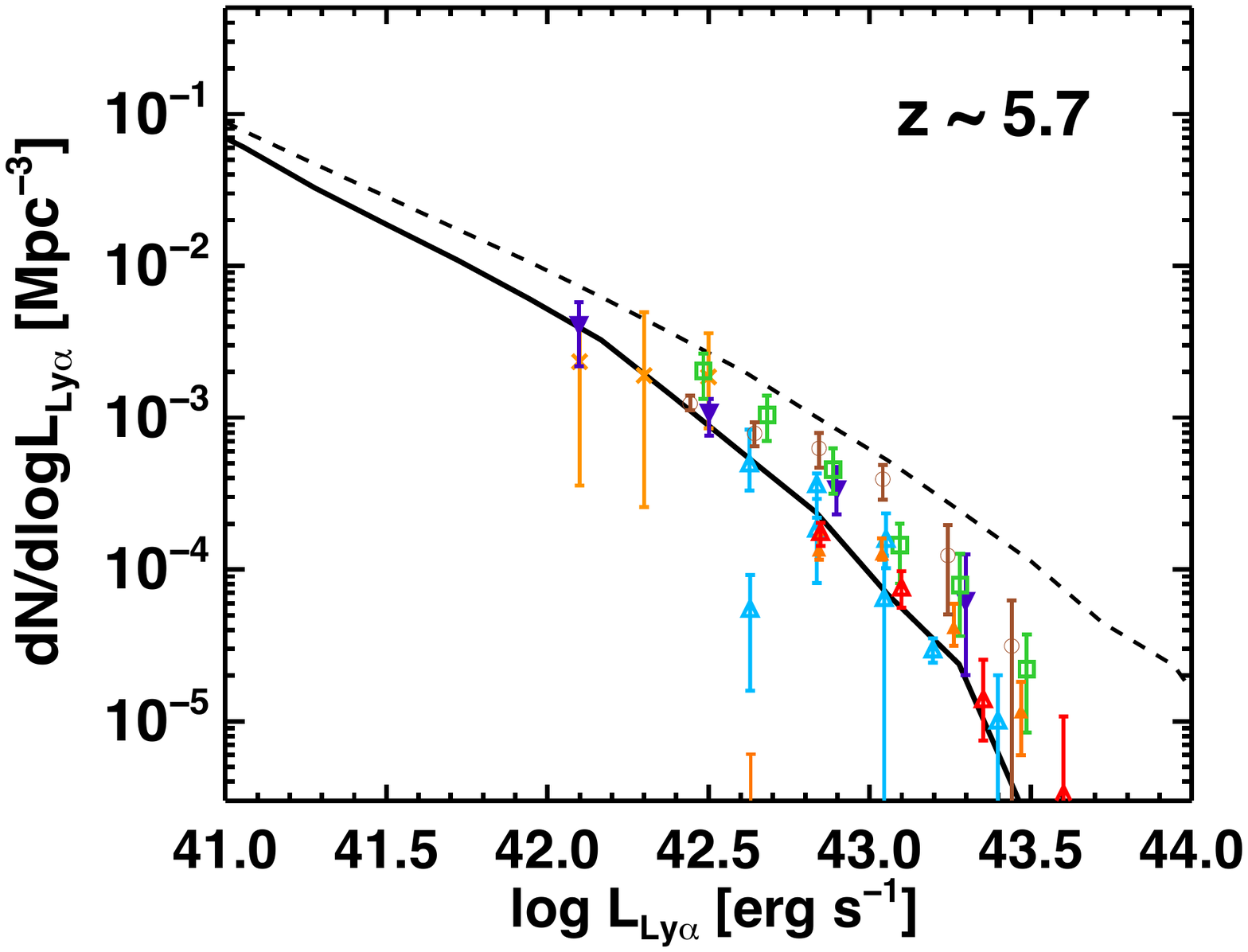}
\end{minipage}
\hspace{0.4cm}
\begin{minipage}[]{0.3\textwidth}
\centering
\includegraphics[width=6.1cm,height=8.5cm]{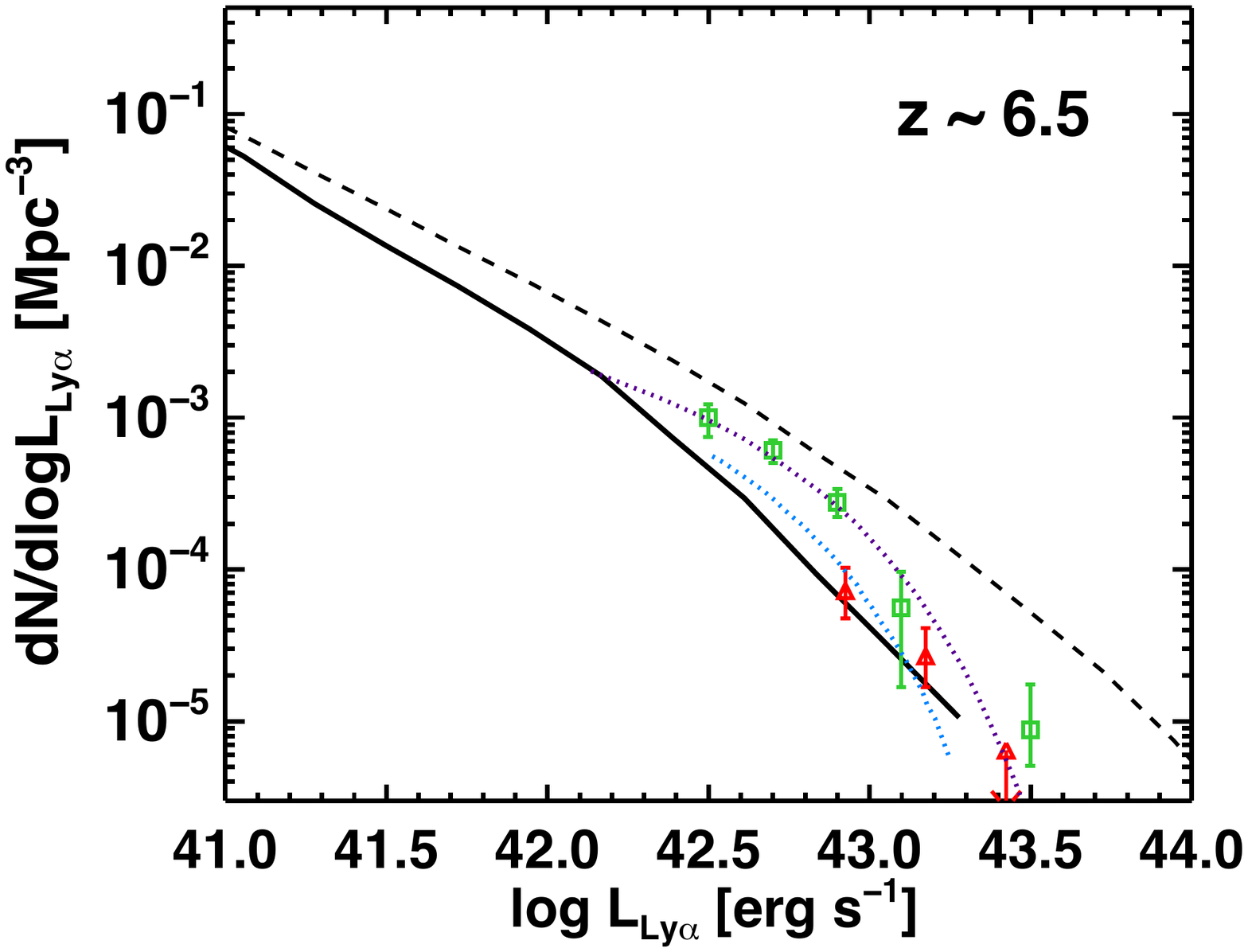}
\end{minipage}
\hspace{0.4cm}
\begin{minipage}[]{0.3\textwidth}
\centering
\includegraphics[width=6.1cm,height=8.5cm]{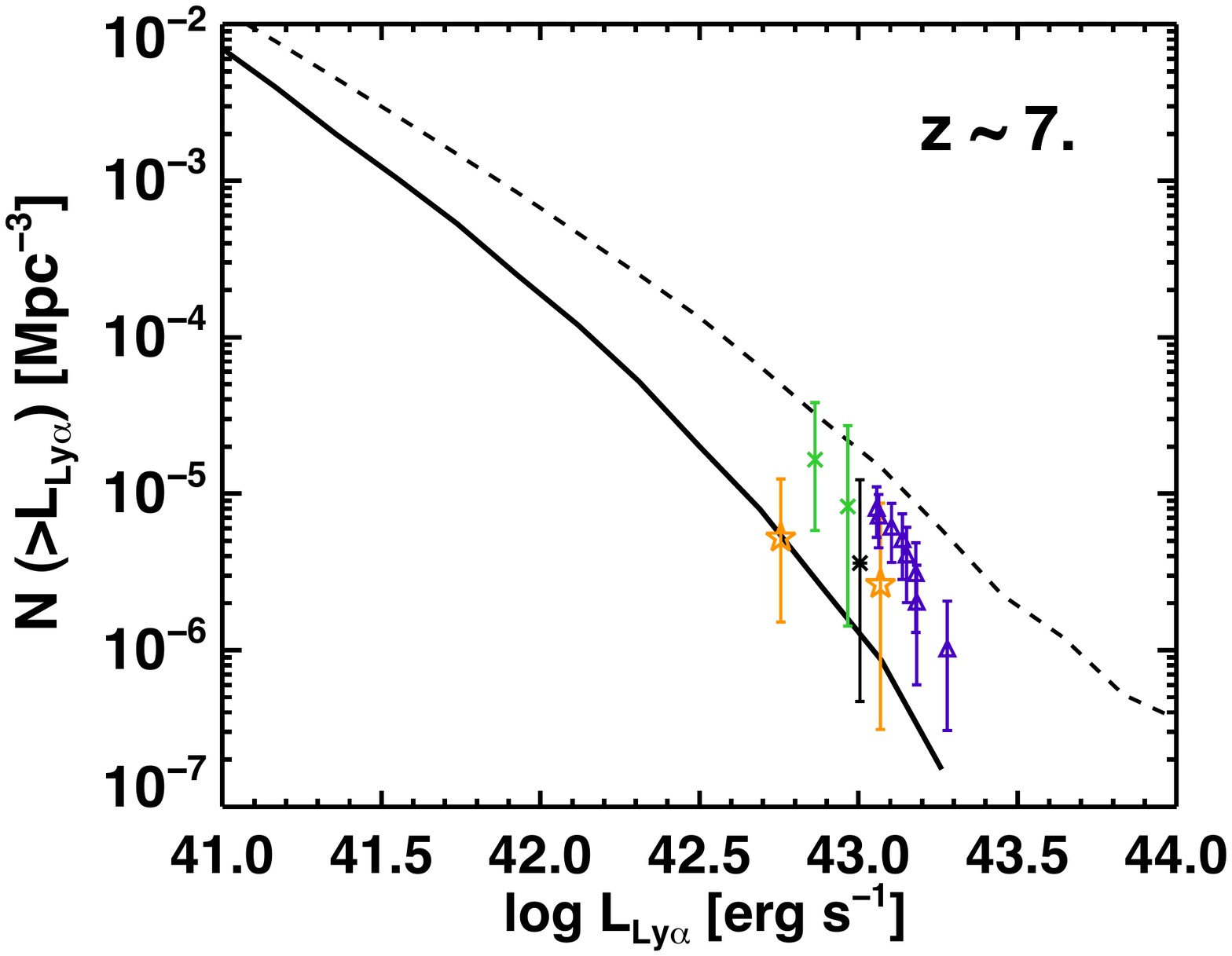}
\end{minipage}
\vskip-12ex 
\caption{\lya luminosity functions at z $\approx$ 3, 3.7, 4.7, 5.7, 6.5 and 7. The solid (dashed) line is the model with (without, i.e. intrinsic) dust. Symbols are observational data  from \citet{vanb05} (red diamonds, $2.3 < z < 4.6$), \citet{gronwall07} (brown triangles, z $= 3.1$), \citet{ouch03,ouch08,ouchi2010a} (green squares, z $= 3.1$, $3.7$, $4.9$, $5.7$ and $6.6$), \citet{blanc2011a} (blue asterisks, $2.8 < z < 3.8$), \citet{rauch08} (orange circles, $2.67 < z < 3.75$), \citet{cassata2011a} (purple downward triangles, z $\approx 3$, $4$, $6$), \citet{dawson07} (black crosses, z $= 4.5$), \citet{wang09} (red asterisks, z $= 4.5$), \citet{shioya} (black triangles, z $= 4.9$), \citet{henry2012} (orange crosses, z $= 5.7$), \citet{shima06} (brown circles, z $= 5.7$), \citet{hu2010a} (red triangles, z $= 5.7$ and $6.6$), \citet{aji03,aji04,aji06} (light blue triangles, z $= 5.7$), \citet{murayama07} (orange triangles, z $= 5.7$), \citet{malhotra04} (blue dotted line, z $= 6.5$), \citet{kashikawa2011a} (purple dotted line, z $= 6.5$), \citet{shibuya2012a} (orange stars, z $\approx 7$), \citet{hibon2012a} (purple triangles, z $\approx 7$), \citet{iye06} (black asterisk, z $\approx 7$) and \citet{vanzella2011} (green crosses, z $\approx 7$).}
\label{fig:lyalfs}
\end{figure*}	    
\hspace{-15cm}

In comparison to UV LFs, the \lya LFs are computed observationally over smaller samples so cosmic variance effects remain significant. Moreover, most LAEs are detected using the narrow-band (NB) technique which introduces a selection in \lya equivalent width (EW)\footnote{All equivalent width values discussed in this paper are restframe equivalent widths.}. The \lya LFs are therefore computed over EW-limited samples that are not complete in terms of \lya luminosity, so caution must be taken when comparing LFs from various surveys which select LAEs differently (i.e. with different EW cuts). 

In Figure \ref{fig:lyalfs}, we compare the \lya LF from our model with available observational data at six redshifts (z $=$ 3, 3.7, 4.7, 5.7, 6.5 and 7). To build the LFs, we only consider the \lya luminosity of galaxies, and, unlike observations (symbols with error bars), we do not apply any EW thresholds. Hence, we expect our LFs to lie above data points which come from LAE samples selected with high EW thresholds. For instance, \citet{ouch08,ouchi2010a} obtained four NB-selected samples of LAEs at z $\approx$ 3.1, 3.7, 5.7, and 6.6, which are often taken as a reference to study the statistical properties of LAEs, and notably the evolution of \lya LF, given the rather large number of objects contained in their samples (a few hundreds at each redshift). In these surveys, the selection of LAEs implied EW thresholds which decline with increasing redshift (i.e. 64 \AA, 44 \AA, 27 \AA, and 14 \AA{} at z $=$ 3.1, 3.7, 5.7, and 6.6 respectively), and it is unclear how this selection effect affects our interpretation of the redshift evolution of the LAE population. Especially, the EW threshold of 64 \AA{} at z $=$ 3.1 coincides with the typical EW value predicted for galaxies with constant star formation \citep{charlot93}. Then, a significant fraction of LAEs can possibly be missed with such observational threshold.

From Figure \ref{fig:lyalfs}, we first note that the overall agreement is quite acceptable given the complexity of the mechanisms involved in the \lya transfer, and the simplicity of the physical picture we assume in this paper. Especially, our LFs compare very well to the observed ones at z $\approx$ 3.7 and 4.7 (top center and right panels in Figure \ref{fig:lyalfs}). At z $\approx$ 3 (top left panel), although our model agrees well with the spectroscopic data of \citet{blanc2011a} and \citet{vanb05} at $L_{\rm Ly\alpha} \gtrsim 10^{42}$ erg s$^{-1}$ (blue asterisks and red diamonds respectively), it overpredicts by a factor of $\sim$ 1.5-2 the number density reported by \citet[][green squares]{ouch08}. As already discussed in \citetalias{garel2012a}, our LF can be reconciled with their data if we impose high EW cuts (EW$_{\rm Ly\alpha} \gtrsim 50$ \AA{} provides the best match). 

At the faint end of the LF, our predictions seem to favour the number density reported by \citet{rauch08}. These observations are however in slight disagreement with the findings of \citet{cassata2011a} who measure a density of LAEs a few times lower for a similar detection limit ($L_{\rm Ly\alpha} \gtrsim 10^{41}$ erg s$^{-1}$ at z $\approx$ 3). Cosmic variance effects, due to the rather small and elongated volumes that are probed, as well as incompleteness issues or slit losses may explain the discrepancy between both measurements. Larger homogeneous datasets are therefore still needed to better constrain the \lya LF at the faint end. This will be one of the key objectives of forthcoming instruments like the Multi Unit Spectrograph Explorer \citep[MUSE;][]{bacon06} which recently started to operate at VLT. We will address these issues in more details in a next paper (Garel et al. in prep).

At z $\approx$ 5.7 and 6.5, the EW cuts employed in NB surveys are small ($\approx 15-25$ \AA) so the impact on the selection of LAEs should be minor. As can be seen from Figure \ref{fig:lyalfs}, the shape and the normalisation of the observed \lya LF at z $\approx$ 5.7 and 6.5 varies significantly from one survey to another, and our model agrees better with the lower end of the envelope of data points.
For instance, at z $\approx$ 6.5, while our model reproduces nicely the observed LFs of \citet[][red diamonds in Figure \ref{fig:lyalfs}]{hu2010a}, it underpredicts by a factor of 2-4 the number densities of  \citet[][green squares]{ouch08,ouchi2010a} and \citet[][purple dotted line]{kashikawa2011a}. The LF of \citet{hu2010a} was computed from their spectroscopic sample and they find that only $\approx$ 50 \% of the LAE candidates could be confirmed by spectroscopy. To explain the difference between the LFs, \citet{hu2010a} argue that the photometric samples of \citet{ouchi2010a} might contain high fractions of interlopers.
However, \citet{kashikawa2011a} report that the rate of contamination in the photometric sample of \citet{taniguchi2005a} at z $\approx$ 6.5 is less than 20\%, and their LF is in much better agreement with the results of \citet{ouchi2010a} than those of \citet{hu2010a}. \citet{kashikawa2011a} claim that the difference may come from the lack of completeness at the faint end in the sample of \citet{hu2010a}, due to the shallow depth of their spectroscopic follow-up survey.

Therefore, we conclude that our model is in agreement with observations at z $\gtrsim$ 6 only when comparing with the data that report the lowest densities of sources at the bright end of the LF. A weaker of effect of dust, or higher intrinsic \lya luminosities would be required to match the data of \citet{ouchi2010a} or \citet{kashikawa2011a}.\\


We note that we do not take the effect of the IGM into account in our model whereas it is well-known that neutral hydrogen atoms can scatter photons on the blue side of the \lya resonance off the line of sight. This could strongly reduce the transmitted \lya flux, especially at z $\gtrsim$ 6. However, as already shown by e.g. \citet{santos04} and \citet{dijkstra2010a}, \lya radiative transfer through gas outflows can Doppler-shift \lya photons towards longer wavelengths and then considerably reduce the impact of IGM. In \citetalias{garel2012a} (Section 4.4), we used the prescription of \citet{madau} to compute the mean contribution of the \lya forest as a function of redshift. We showed that the IGM had a negligible impact on the \lya luminosities of galaxies in our model up to z $\approx$ 5 because of the peak of the \lya lines being redshifted sufficiently away from line centre by the scattering in the expanding shells.
Here, we did a similar test, and we found that the \lya LFs up to redshift 7 remain unaffected by the IGM transmission, for the same reasons as explained above. We also tested the recent model of \citet{Inoue2014} which predicts a slightly higher (lower) \lya transmission at z $\lesssim$ 4.7 (z $\gtrsim$ 4.7) than the recipe of \citet{madau}, and gives a better fit to the observational data. Again, as the \lya lines of most our galaxies are Doppler-shifted, this IGM attenuation model has also almost no effect on the \lya fluxes at all redshifts. Nevertheless, we note that \citet{laursen2011} have shown that the IGM can non-negligibly reduce the flux on the red side of the \lya line at z $\gtrsim$ 6. Moreover, additional contributions to the IGM opacity might be important at z $\gtrsim$ 6, such as (i) the potential high number density of Lyman-Limit systems \citep{bolton2012a}, or (ii) the increasing neutral fraction of the diffuse component of the IGM, $x_{\hi}$, before reionisation is fully complete \citep[see ][for more details]{dijkstra2014}. Thus, a more refined IGM model with full RT treatment would be needed to assess the exact effect of the IGM on the \lya lines in our model.

The end of the epoch of reionisation is still a subject of debate, and the value of $x_{\hi}$ at z $=$ 6-7 is not fully constrained yet \citep[e.g.][]{dijkstra2014}. In addition, \lya observations are often used as a probe of $x_{\hi}$, e.g. by measuring the evolution of the \lya LF \citep{ouchi2010a}. In this context, it is interesting to note that our model intrinsically predicts a decrease of 50 \% of the number density of bright LAEs at $L_{\rm Ly\alpha} = 10^{43}$ erg s$^{-1}$ between z $=$ 5.7 and 6.5 without accounting for the effect of IGM (Figure \ref{fig:lyalfs}).

We now turn our interest to the cross UV/\lya properties of the LBG and LAE populations to study their link with one another. Owing to different methods of selection, the dropout and narrow band techniques do not necessarily probe the same galaxies, but both populations are a subset of a common parent population, and it is worth asking how they overlap. 

\section{UV/\lya properties of LAEs}
\label{sec:cross_prop1}	    
\subsection{The UV LFs of LAEs}
\begin{figure*}
\vskip-14ex 
\hspace{-1cm}
\begin{minipage}[]{0.3\textwidth}
\centering
\includegraphics[width=6.1cm,height=8.5cm]{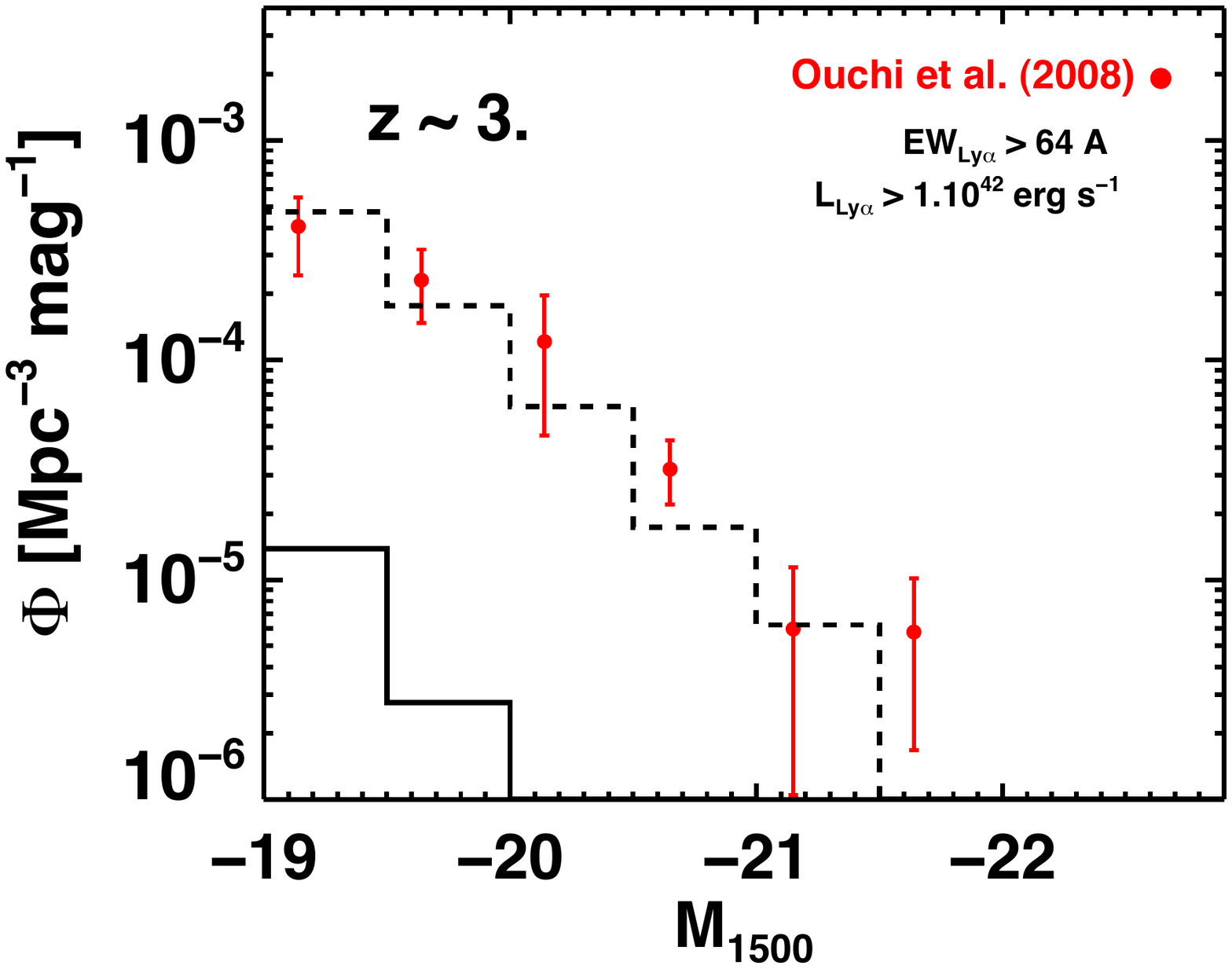}
\end{minipage}
\hspace{0.4cm}
\begin{minipage}[]{0.3\textwidth}
\centering
\includegraphics[width=6.1cm,height=8.5cm]{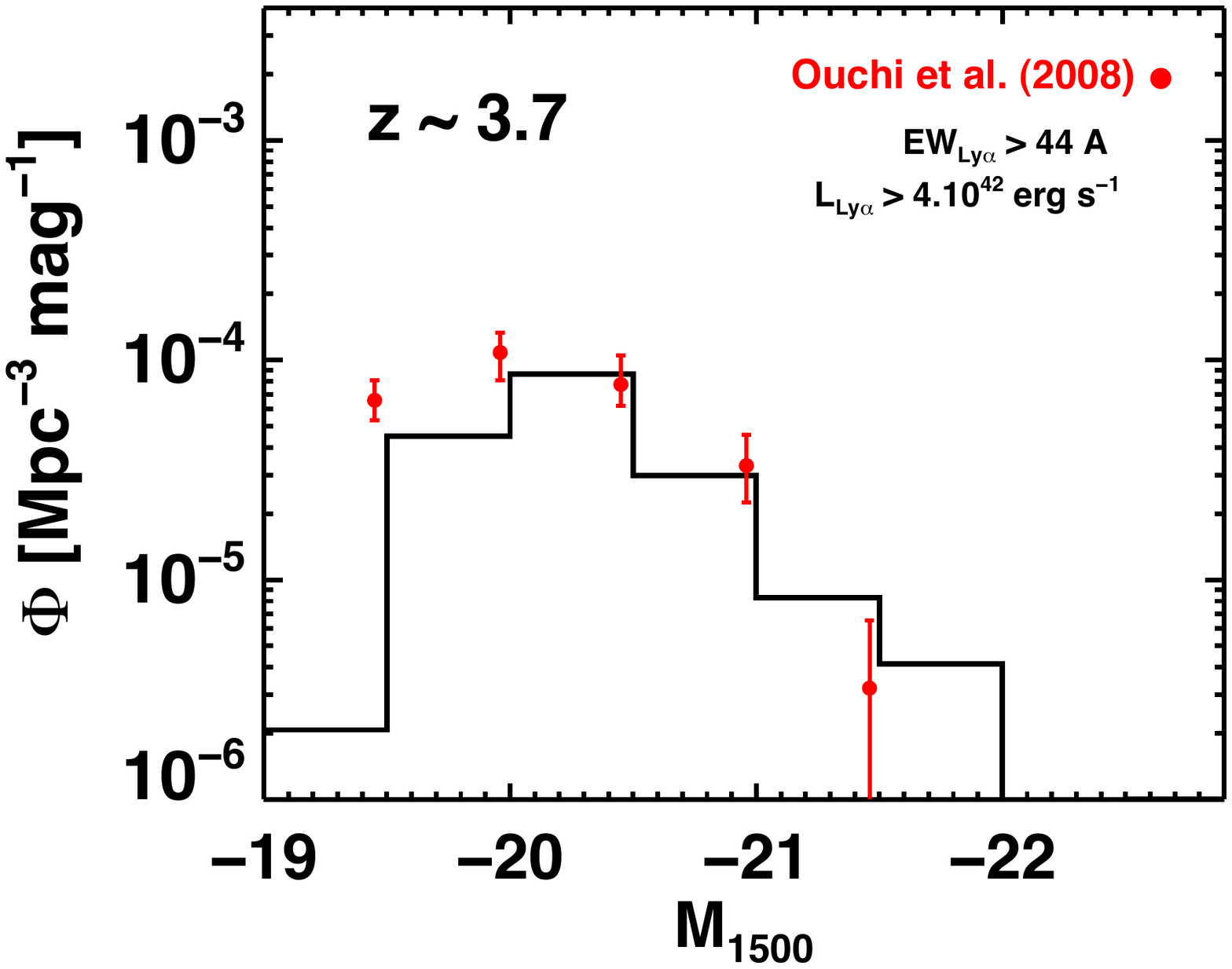}
\end{minipage}
\hspace{0.4cm}
\begin{minipage}[]{0.3\textwidth}
\centering
\includegraphics[width=6.1cm,height=8.5cm]{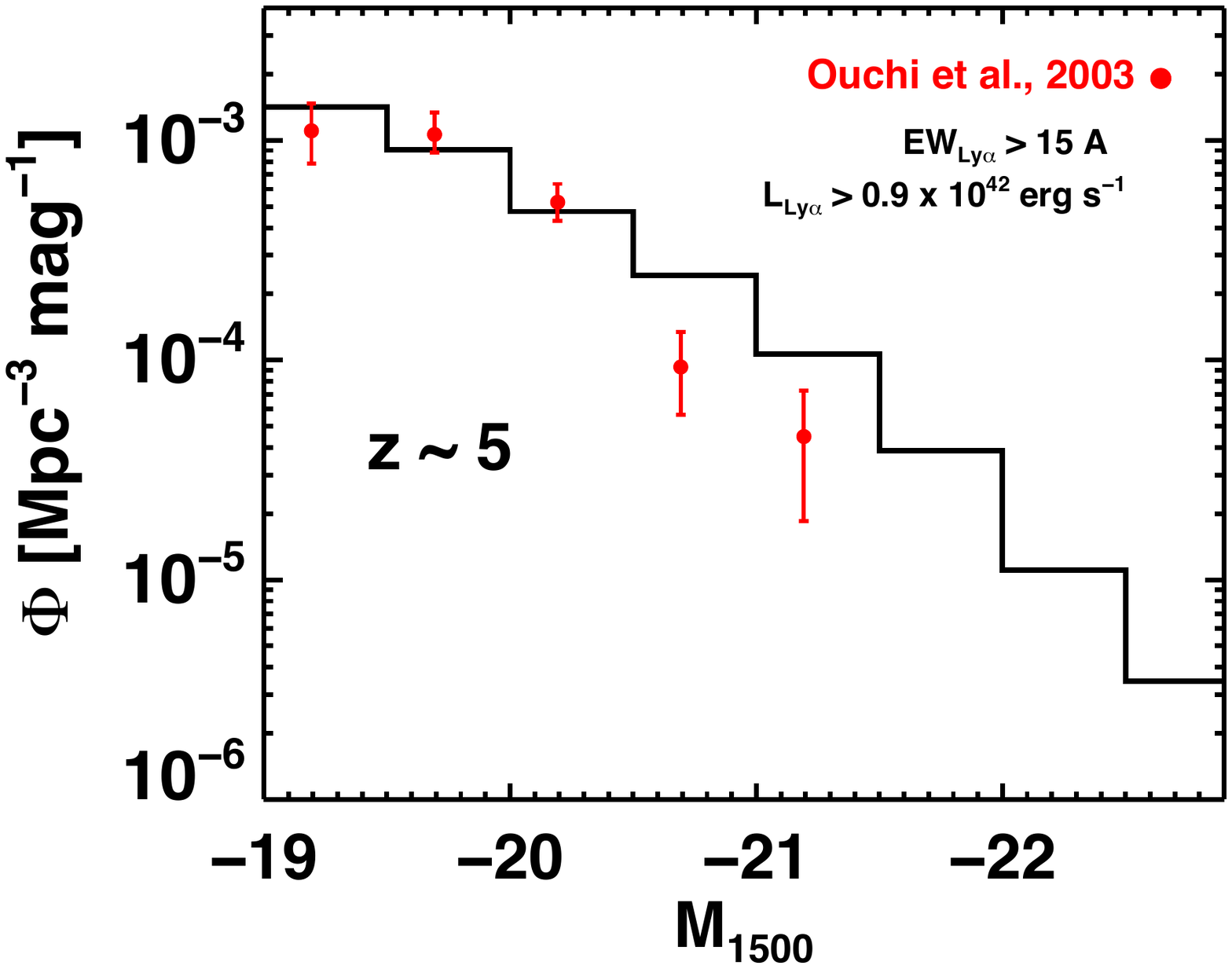}
\end{minipage}
\vskip-30ex 
\hspace{-1cm}
\begin{minipage}[]{0.3\textwidth}
\centering
\includegraphics[width=6.1cm,height=8.5cm]{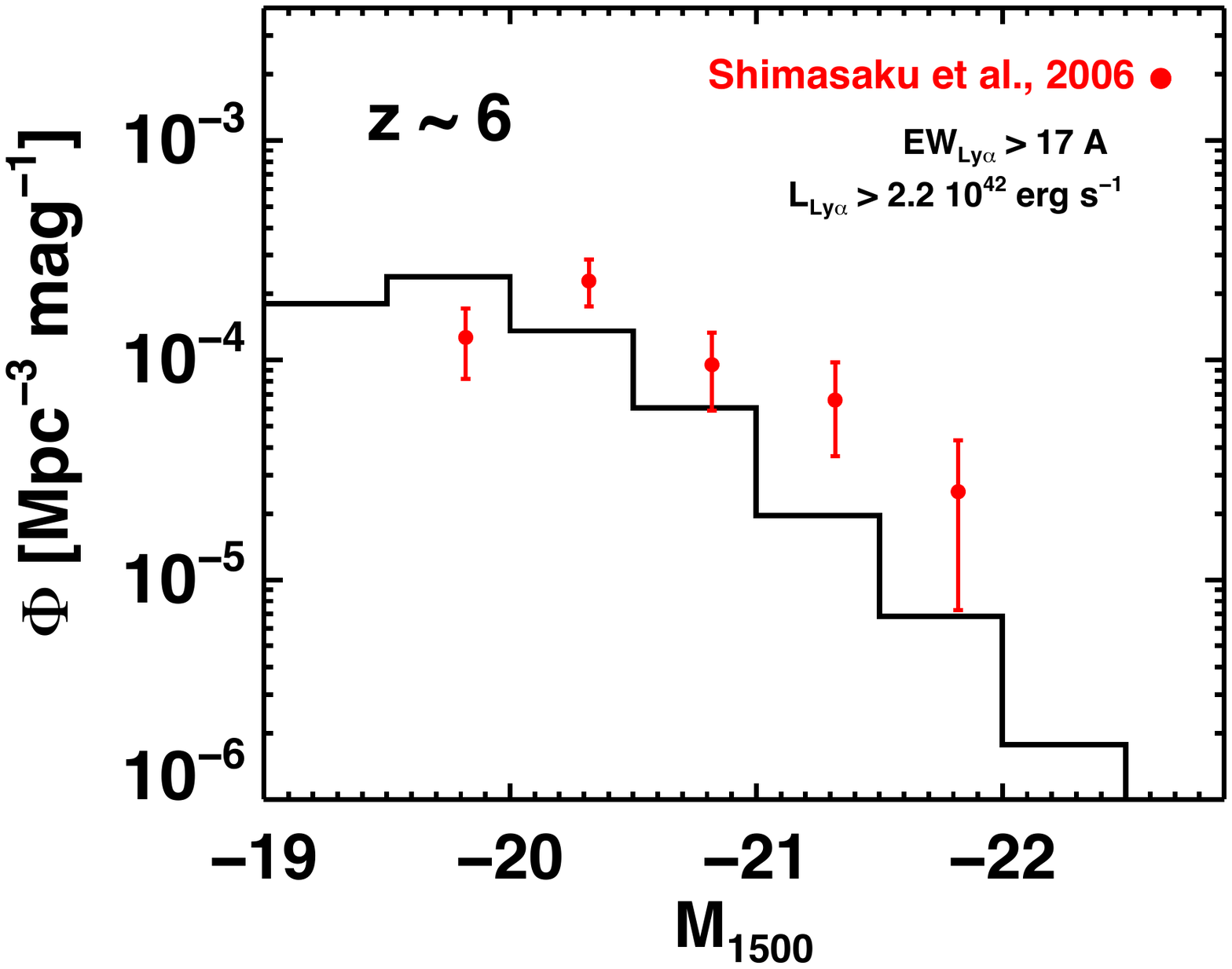}
 \end{minipage}
 \hspace{0.4cm}
\begin{minipage}[]{0.3\textwidth}
\centering
\includegraphics[width=6.1cm,height=8.5cm]{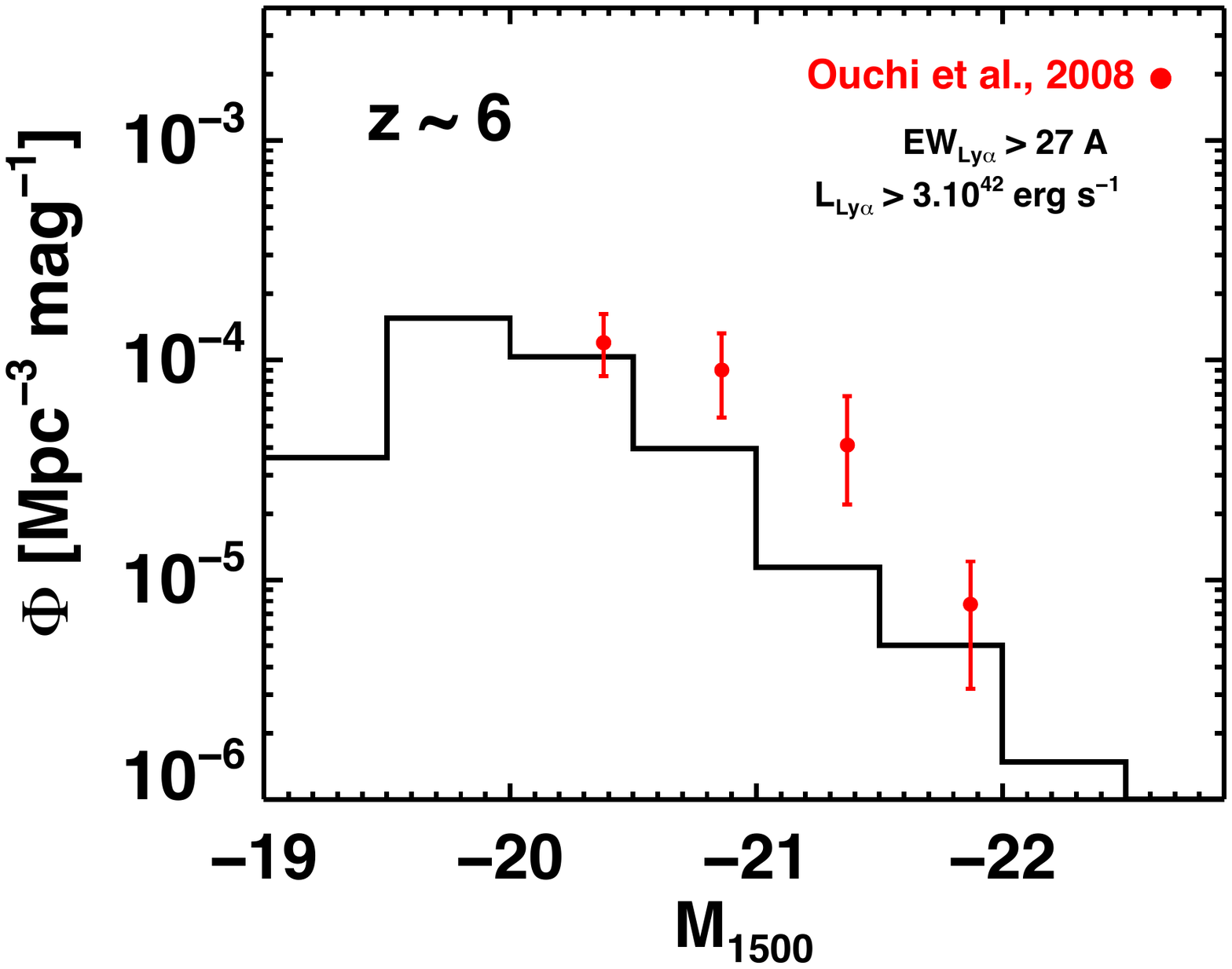}
\end{minipage}
\hspace{0.4cm}
\begin{minipage}[]{0.3\textwidth}
\centering
\includegraphics[width=6.1cm,height=8.5cm]{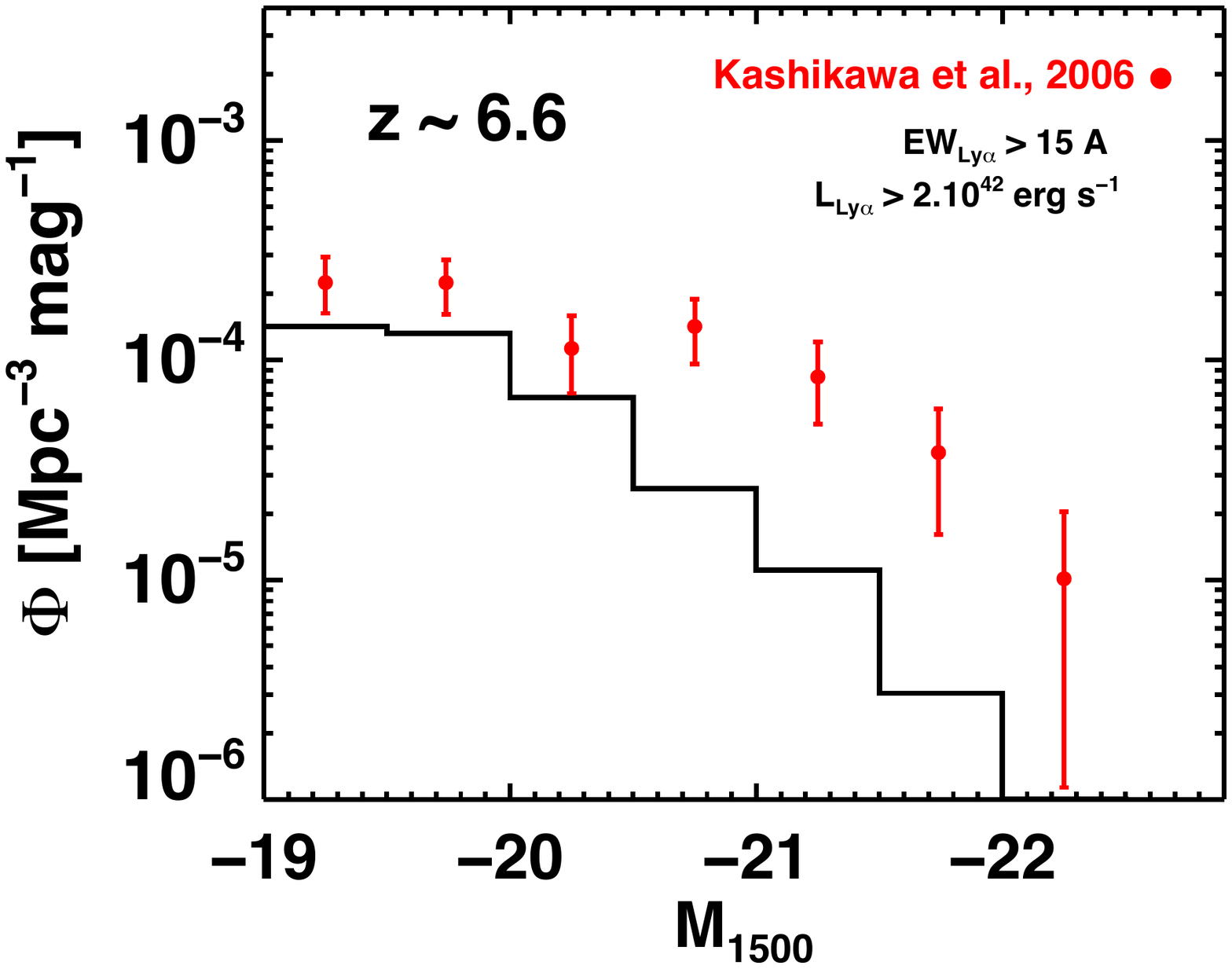}
\end{minipage}
\vskip-12ex 
\caption{UV luminosity functions of \lyat-selected galaxies at z $\approx$ 3, 3.7, 5, 6, and 6.6. The solid lines correspond to the model with dust attenuation included and the red filled circles are observational data \citep{ouch08,ouch03,shima06,kashi06}. We compared our model to each dataset individually, applying selections in terms of \lya luminosities and equivalent width so as to mimic real observations (see legends). The dashed line in the top left panel corresponds to a selection where the EW threshold of \citet[][]{ouch08} has been decreased from 64 to 50 \AA. With such a small change ($\approx 20\%$), our model agrees quite well with the data.}
\label{fig:uvlfs_lae}
\end{figure*}

Here, we compute the UV luminosity functions of LAEs selected using \lya luminosity and equivalent width thresholds similar to various narrow band surveys (Figure \ref{fig:uvlfs_lae}). In practice, we allow the observational cuts to vary by $\lesssim$ 20 \% in order to improve the agreement with the data. This is justified by the fact that the quoted \lya luminosity and equivalent width thresholds do not exactly mimic the actual colour-magnitude selections \citep{ouch08,dijkstra2012a}. 

Our model reproduces the UV LF of LAEs from \citet{ouch08} and \citet{ouch03} at z $\approx$ 3.7 and 4.9 respectively using selections similar to those reported by these authors (top centre and top right panels in Figure \ref{fig:uvlfs_lae}). At z $\gtrsim$ 6, our model reproduces correctly the shape of the LFs observed by \citet{ouch08}, \citet{shima06} and \citet{kashi06} but the overall normalisation is somehow too low. Varying the \lya luminosity and equivalent width threshold has little effects because the cuts are low enough such that few galaxies with $M_{1500} \lt -18$ are removed by the \lya selection. As for the comparison of our \lya LFs with the data from the same surveys, it is plausible that the number density measured from photometric samples in these observations is overestimated due to non-negligible fractions of contaminants.

At z $\approx$ 3.1 however, the observed LF is strongly underestimated if we use the equivalent width  threshold of \citet{ouch08}, as shown by the solid histogram in the top left panel. The value of 64 \AA{} assumed by these authors is close to the median \lya equivalent width in our model, hence a large fraction of galaxies are below this selection limit.  We need to decrease the EW threshold from 64 \AA{} to 50 \AA{} to bring the model into agreement with the data (dotted line in the top left panel of Figure \ref{fig:uvlfs_lae}). In other terms, at a given UV magnitude, the \lya luminosities predicted by our model are not high enough compared to the LAEs of \citet{ouch08}. In the next section, we investigate the correlation between UV and \lya emission in more details, and we quantitatively discuss plausible causes for the missing high EWs in the model.

\subsection{Expected relation between intrinsic \lya and UV emission}
\label{subsec:expected_rel}

Assuming that star formation is at the origin of all the \lya and UV emission of galaxies, the relation between SFR and the intrinsic \lya luminosity writes $L_{{\rm Ly}\alpha} = 1.2 \times 10^{42} ({\rm SFR}/ \msunyr) \;{\rm erg \: s}^{-1}$ \citep[][]{kennicutt98,osterbrock2006}. Following \citet{madau1998a}, the relation between SFR and UV luminosity density evaluated at 1500 \AA{} is $L_{\lambda {\rm ,UV}} = 1.4 \times 10^{40} ({\rm SFR}/ \msunyr) \: {\rm erg \: s}^{-1}$ \AA$^{-1}$.

\begin{figure}
\vskip-27ex
\includegraphics[width=8.9cm,height=12.7cm]{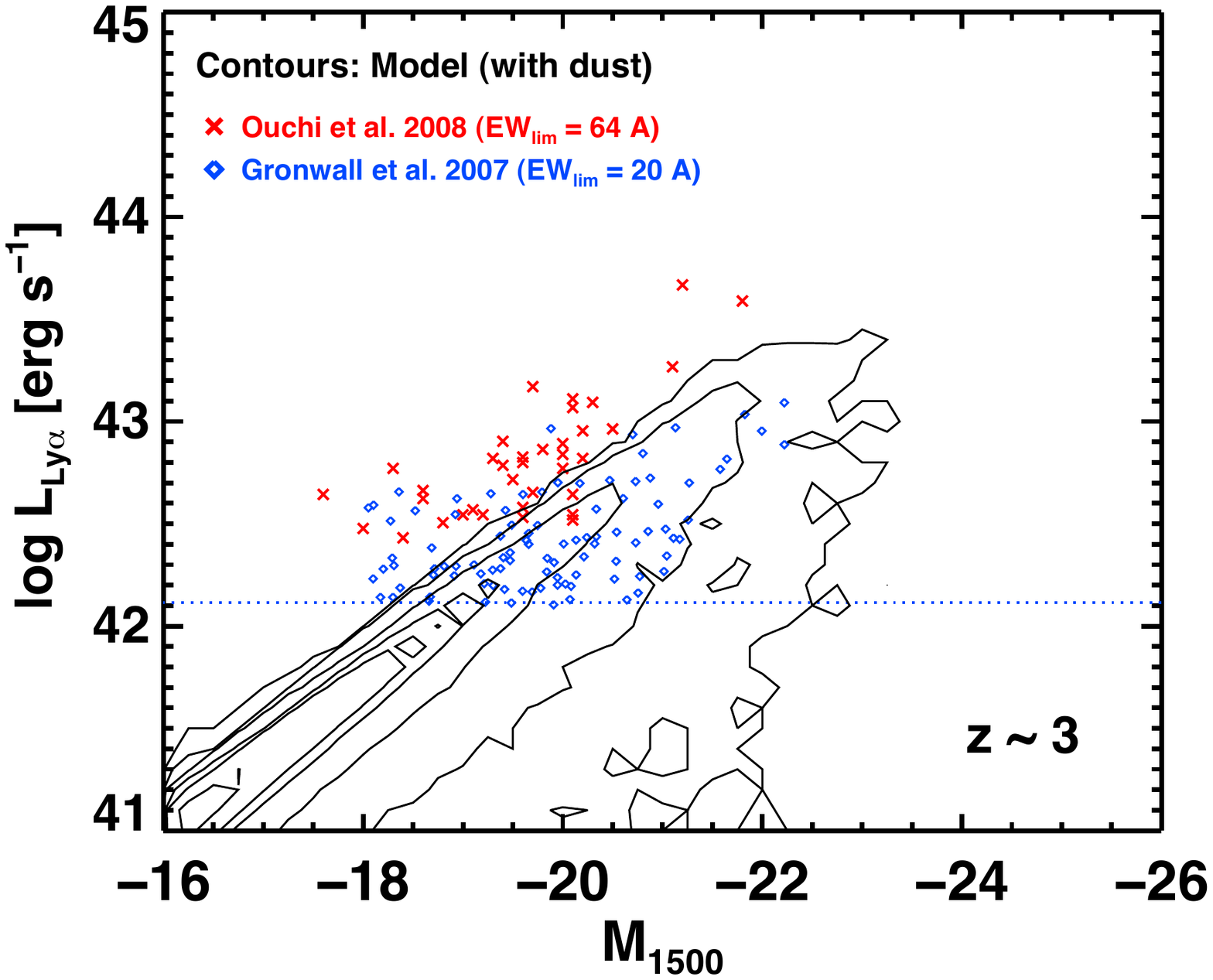}
\vskip-50ex
\includegraphics[width=8.9cm,height=12.7cm]{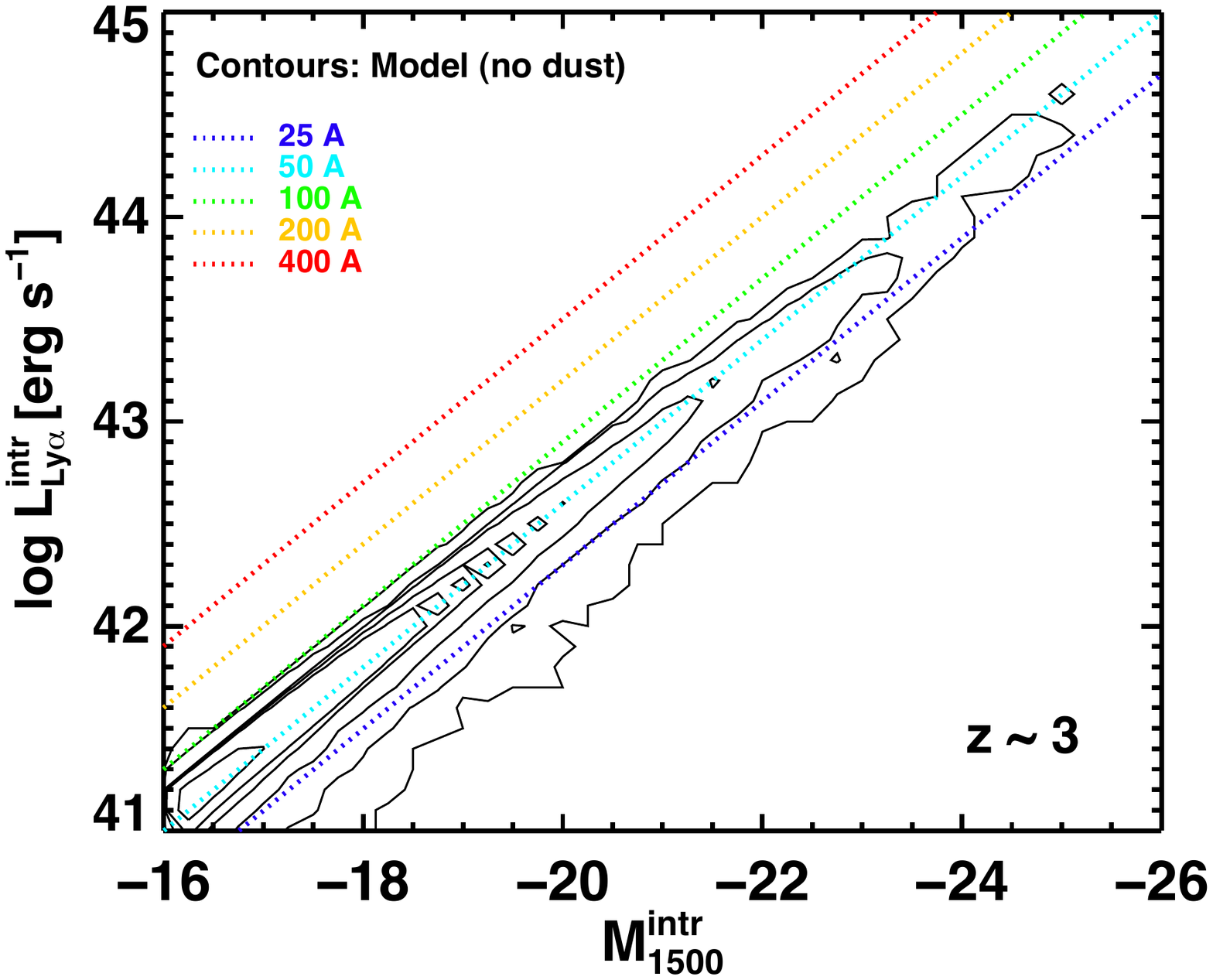}
\vskip-50ex
\includegraphics[width=8.9cm,height=12.7cm]{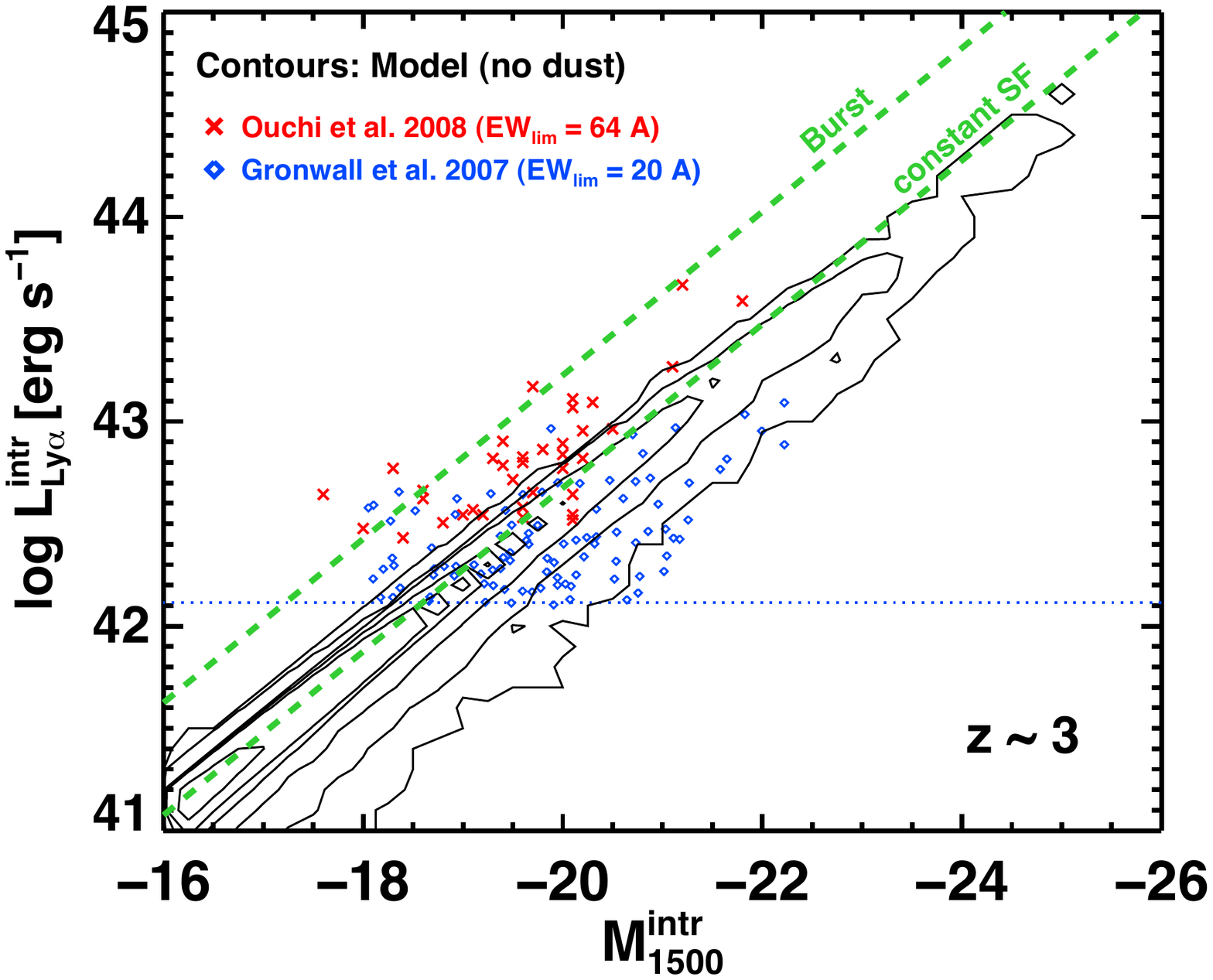}
\vskip-24ex
\caption{Distribution of LAEs in the \lya luminosity versus UV magnitude plane at z $\approx$ 3. The black contours show the number distribution of objects in the model (without any \lya luminosity or EW cut). 
\textit{Top:} \lya luminosity versus UV magnitude (with dust attenuation included). The data of \citet{gronwall07} (blue diamonds) and \citet{ouch08} (red crosses) are also shown. The blue dotted line show the \lya detection limit of  \citet{gronwall07}.
\textit{Middle:} Intrinsic \lya luminosity versus intrinsic UV magnitude (without dust attenuation included). Overlaid are the expected relations for various \lya equivalent widths, according to Eq. \ref{eq:llya_ew_exp}.
\textit{Bottom:} Intrinsic \lya luminosity versus intrinsic UV magnitude (without dust attenuation included). The green dashed lines show the expected relation in the case of constant star formation (Eq. \ref{eq:llya_muv_kennimf}), and in the case of an instantaneous starburst (see text). Same data as above (i.e. not corrected for dust).}
\label{fig:muv_llya_z3}
\end{figure}

Combining these formulae, $L_{\rm Ly\alpha}$ and $M_{1500}$ scale with one another as follows\footnote{These scaling relations assume a Kennicutt IMF, solar metallicity, the production of \lya photons through Case B recombination at T$=10^4$K, and are valid for constant star formation.}:

\begin{equation}
M_{{\rm 1500}} = -2.5 \: {\rm log}_{10} \left(\frac{\raisebox{0.55ex}{\textit L}_{{\rm Ly}\alpha} }{10^{42} \: {\rm erg \: s}^{-1}} \right) - 18.3
\label{eq:llya_muv_kennimf}
\end{equation}
In this framework, galaxies with intrinsic luminosities of $L_{\rm Ly\alpha} = 10^{41}, 10^{42}$ and $10^{43}$ erg s$^{-1}$ should display $M_{1500} = -15.8, -18.3$ and $-20.8$, corresponding to star formation rates of $\sim 0.08, 0.8$ and $8 \msunyr$ respectively. 

In Figure \ref{fig:muv_llya_z3}, we investigate the relation between \lya luminosity and UV magnitude at z $\approx$ 3. In the upper panel, the distribution of galaxies from the model, represented by the black contours, shows the relation between \lya luminosity and UV magnitude after the effect of dust. Overlaid are measurements from two sets of narrow-band selected LAEs. The cloud of data points from \citet{gronwall07} (blue diamonds) covers a similar region to the model above their detection limit ($\approx 1.2 \times 10^{42}$ erg s$^{-1}$; blue dotted line). We note however that some data points lie above the model, i.e. their \lya luminosity is larger than predicted by the model for a given UV magnitude. Now comparing with data of \citet{ouch08} (red crosses), we see this time that many LAEs lie above the model predictions. These objects belong to the tail of the \lya equivalent width distribution, that is EW $\gtrsim$ 100 \AA, which our model does not reproduce. Moreover, we note that most LAEs from the survey of \citet{ouch08} are above the ones of \citet{gronwall07} in the $L_{\rm Ly\alpha}$-$M_{{\rm 1500}}$ plane, which is due to very different EW thresholds used to select LAEs (20 \AA{} vs 64 \AA).

In the middle panel of Figure \ref{fig:muv_llya_z3}, we now show the model distribution of galaxies before applying the effect of dust. To show how the \lya and UV intrinsic luminosities relation are expected to scale with one another as a function \lya equivalent width, we overplot the expected intrinsic $L_{\rm Ly\alpha}^{\rm intr}$-$M_{{\rm 1500}}^{\rm intr}$ relation for various EWs (as labelled on the plot), according to:
\begin{equation}
{\rm EW}_{\rm exp} = \frac{L_{{\rm Ly}\alpha}}{L_{\lambda{\rm ,UV}}} \left( \frac{\lambda_{\rm UV}}{\lambda_{{\rm Ly}\alpha}}\right)^{\beta}
\label{eq:llya_ew_exp}
\end{equation}
where ${\beta}$ is the UV slope used to renormalise the continuum luminosity density around the \lya line. Here, we have assumed ${\beta} = -1.5$, which is a typical value for observed LAEs \citep[e.g.][]{venemans2005a,blanc2011a}.

From the bottom panel of Figure \ref{fig:muv_llya_z3}, it is obvious that many sources from the samples of \citet{gronwall07} and \citet{ouch08} have observed EW larger than in the model, even \modi{before taking the effect of dust into account (contours)}. At the same time, we note that the relation between intrinsic \lya luminosity and UV magnitude agrees well with the formula for constant star formation (SF) from Eq. \ref{eq:llya_muv_kennimf} (dashed green line). 
\modi{The production rate of \lya photons, dominated by short-lived hot stars, traces the very recent star formation, so the intrinsic \lya luminosity, $L_{\rm Ly\alpha}^{\rm intr}$, is proportional to the star formation rate averaged over the last $\approx$ 10 Myr, SFR$(t\approx10$ Myr$)$ \citep[][]{charlot93}. The UV continuum intensity is however a direct function of the mean SFR over longer timescales, of the order of a few hundreds Myr \citep[$L_{\lambda {\rm ,UV}}$ $\propto$ SFR($t\approx100$ Myr);][]{madau1998a}. In the constant SF scenario, i.e. SFR($t\approx10$ Myr) $=$ SFR($t\approx100$ Myr), the \lya equivalent width reaches an equilibrium value of approximatively 60 \AA{} after $\approx$ 10 Myr, which is consistent with the median intrinsic EW value we find in our model.}
 
In GALICS, star formation is iteratively computed at each timestep using the Kennicutt law, namely $\Sigma_{\rm SFR} \propto \Sigma_{\rm gas}^{1.4}$, and a gas surface density threshold: $\Sigma_{\rm gas}^{\rm thresh} = 10^{21}$ cm$^{-2}$. This threshold, inferred from late-type galaxies observations, is in broad agreement with the Toomre model for gravitational instability-triggered star formation \citep{kennicutt1989}.  In GALICS, this criterion is met most of the time in LAEs and LBGs we consider in this paper, hence the star formation appears to be close to a constant process, although there is scatter in the $L_{\rm Ly\alpha}^{\rm intr}$-$M_{{\rm 1500}}^{\rm intr}$ relation due to small variations of the recent star formation history or stellar metallicity.
We also consider here the $L_{\rm Ly\alpha}^{\rm intr}$-$M_{{\rm 1500}}^{\rm intr}$ relation \citep[see Fig. 15 of][]{verh08} that is expected a few Myr after a burst of star formation (dashed green line). We see from the \modi{bottom} panel of Figure \ref{fig:muv_llya_z3} that the high-EW sources of \citet{ouch08} can be interpreted by this scenario, which produces higher \lya luminosity for a given UV emission than ongoing constant SF. Thus, the high-EW LAE population probed by \citet{ouch08} could correspond to starburst galaxies, rather than constantly star forming objects. 

Although our model can recover the standard \lya equivalent widths (EW $\sim$ 0-70 \AA) which seem to make the bulk of the LAE population \citep{cassata2011a}, it certainly requires extra ingredients to explain the high-EW galaxies\footnote{Here, we have focussed on z $=$ 3 LAEs, but the issue is similar at higher redshifts, as large EW values are also measured in LAEs up to z $\approx$ 6 \citep[e.g.][]{kashikawa2011a}.}.

\subsection{High \lya equivalent widths}
\label{subsec:high_ews}

As discussed in the previous section, the commonly observed of high-EW LAEs (EW $\gtrsim$ 100 \AA) are missing in our model, even when considering intrinsic values. Here, we discuss various potential causes responsible for this mismatch. 

First, it is often claimed that \lya radiation transfer effects could enhance the intrinsic EW, as predicted by the analytic model of \citet{neufeld} for a multiphase medium. \citet{laursen2013a} and \citet{duval2014} have shown that the (angle-averaged) boost of \lya EWs in a clumpy multiphase medium would require very special conditions unlikely to exist in most high-redshift galaxies. However, the recent model of \citet{gronke2014} suggests that significant \lya enhancements can be found along particular sight lines where UV continuum photons are absorbed by dust clouds, whereas \lya photons escape isotropically. Alternatively, \citet{verhamme2012} have reported a strong inclination effect using Monte-Carlo \lya RT in hydrodynamical high-resolution simulations of galactic disks, in which the \lya equivalent width measured perpendicularly to the disk is larger than in the edge-on direction. While both UV continuum and \lya are emitted isotropically in the disk, a fraction of the \lya photons emitted edge-on may be scattered off by hydrogen in the face-on direction, therefore boosting the observed EW along the face-on line-of-sight. 

\begin{figure}
\vskip-24ex 
\hskip-4ex
\includegraphics[width=8.7cm,height=11.4cm]{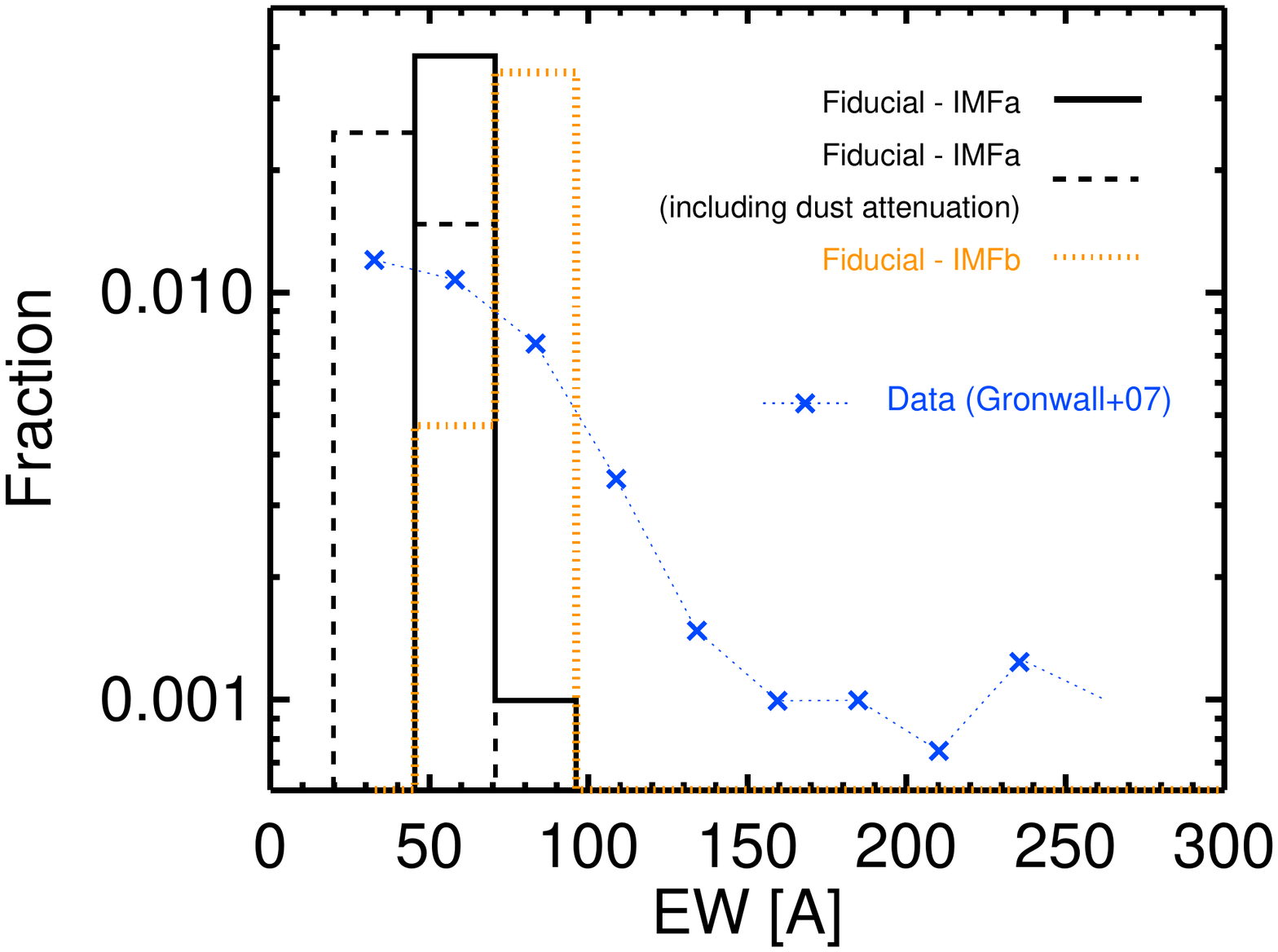}
\vskip-40ex 
\hskip-4ex
\includegraphics[width=8.7cm,height=11.4cm]{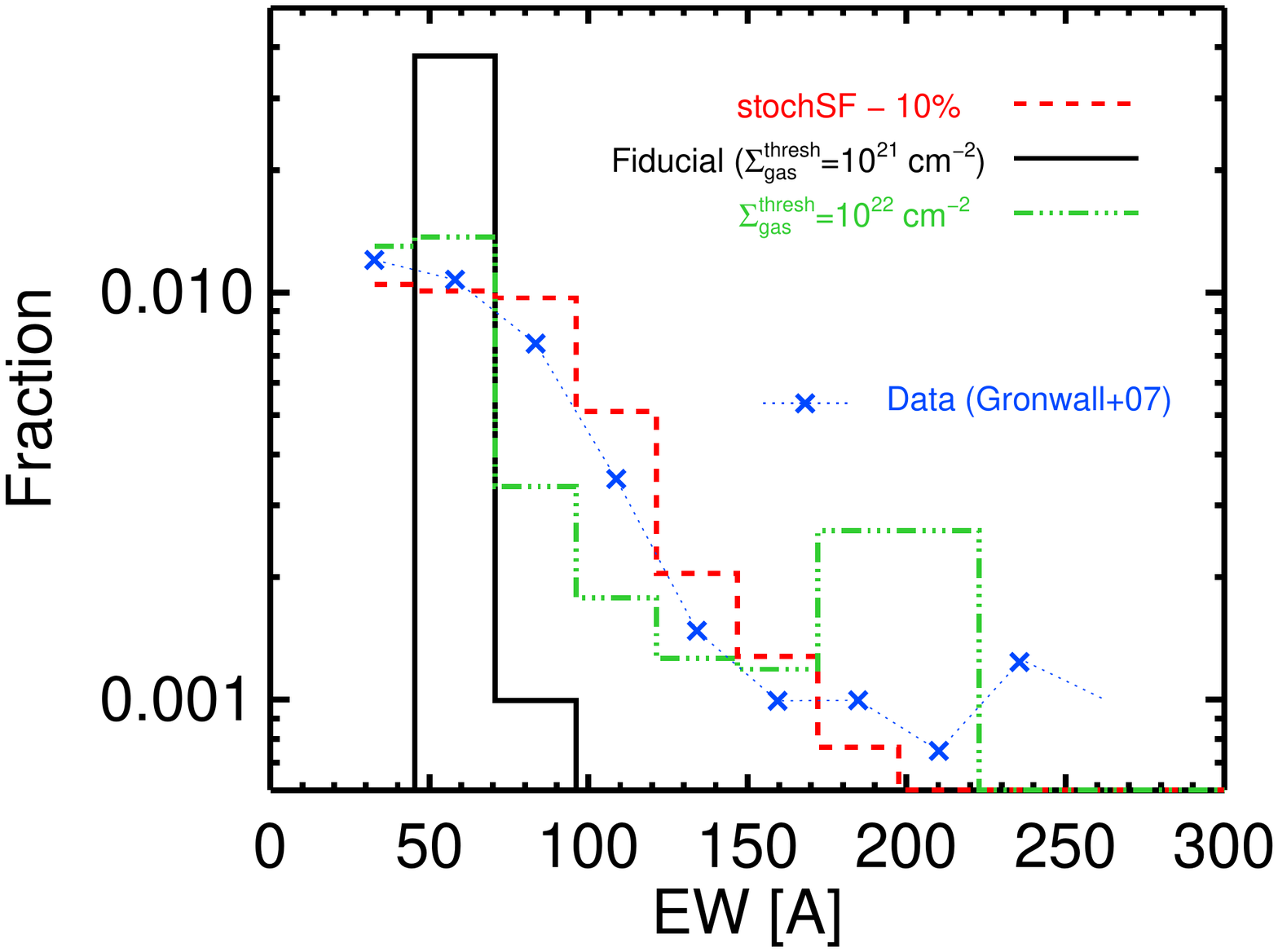}
\vskip-21ex 
\caption{\lya equivalent widths distribution of LAEs at z $=$ 3.
\textit{Top:} The black solid and dashed histograms show the EW distributions of LAEs predicted by our model before and after the effect of dust \modi{included, respectively}. Our fiducial model assumes a Kennicutt IMF, with a low-mass cut-off of 0.1 \msun (IMFa). For comparison, we also show the EW distribution (without dust) using a more top-heavy Kennicutt IMF, with a low-mass cut-off of 4 \msun (IMFb; orange dotted curve). LAEs have been selected using $L_{{\rm Ly}\alpha} \geq {1.2 \times 10^{42} \: {\rm erg \: s}^{-1}}$ and EW $\geq$ 20 \AA{}, as in the survey of \citet{gronwall07} (blue crosses). \textit{Bottom:} To highlight the effect of starbursts on the \lya equivalent widths, we compare the intrinsic EW distribution from our fiducial model (\modibis{solid black line}), with similar models in which (i) star formation occurs stochastically with a duty cycle of 10\% (\modibis{dashed red line}),  and (ii)  star formation only occurs when the gas surface density is larger than $10^{22}$ cm$^{-2}$ in galaxies (\modibis{dot-dashed green line}). }
\label{fig:high_ews}
\end{figure}

Second, high EWs can be intrinsically produced in galaxies. For instance, \citet{schaerer03} show that the EW strongly depend on the metallicity of stars and the IMF. Our SED libraries \citep{devriendt} only take into account metallicities larger than $Z=2.10^{-3}$ so we cannot investigate the impact of metal-poor stars in our model, but we can compare the EW variation between the results from our fiducial Kennicutt IMF with a low-mass cut-off of 0.1 \msun (labelled IMFa) and from a more top-heavy Kennicutt IMF that assumes a low-mass cut-off of 4 \msun (IMFb). In Figure \ref{fig:high_ews} (top panel), we show the z $=$ 3 model distributions of intrinsic EWs (\modi{before dust attenuation included}; solid histogram), and observed EW (\modi{after dust attenuation included}; dashed histogram) with IMFa. This model is in reasonable agreement with the data of \citet{gronwall07} (blue crosses) in the range 20-70 \AA, which corresponds to the peak of the EW distribution. However, the high-EW tail extending to $\sim$ 250 \AA{} is poorly matched, as expected from the above discussion\footnote{It is worth pointing out here that our simulation box is few times larger than current LAE surveys, so it is unlikely that these very strong emitters are missing in our model because of our finite volume.}. IMFb tends to produce higher EWs due to a larger ratio of high-mass to low-mass stars (orange dotted curve), but the distribution remains as narrow as for IMFa, and it does not reproduce the high-EW tail either. Alternative IMFs could have a stronger impact on the EW distribution, as shown by \citet{orsi2012a} who use an extreme top-heavy, flat IMF. They find a much broader EW distribution than ours, but it turns out to be even too broad compared to observations, and it peaks to higher EW than observed (see their Figure 11).  We also note that, rather than invoking very top-heavy IMFs, \citet{forero-romero2013} demonstrate that the stochastic sampling of the IMF can lead to strong variations in terms of EW, and significantly broaden the EW distribution.

As mentioned in Sec. \ref{subsec:expected_rel}, a starburst should be able to produce higher EWs than ongoing star formation at a constant rate, as it is the case in GALICS. We test this scenario in the bottom panel of Figure \ref{fig:high_ews}, where we compare the intrinsic EW distribution from our fiducial model (black \modibis{solid} histogram) with a modified version of GALICS in which star formation has a stochastic duty cycle of 10\% ('stoch10'; red \modibis{dashed} histogram). In GALICS, star formation (and all other galaxy formation processes) are integrated over timesteps of 1 Myr. Here, we randomly switch on star formation once every ten timesteps to see the effect of bursty star formation onto the \lya equivalent widths. With this ad-hoc model, we clearly see that the EW distribution is broader than the fiducial model's distribution, and agrees much better with the data of \citet{gronwall07}. 

An obvious method to mimic a SF duty-cycle is to set a high gas surface density threshold. While the criterion $\Sigma_{\rm gas}^{\rm thresh}=10^{21}$ cm$^{-2}$ is nearly always met in our fiducial model for galaxies considered here, we tried to increase this value to $10^{22}$ cm$^{-2}$. In this case, accreted gas needs to accumulate for longer periods in the galaxy in order to reach the surface density threshold, at least in some objects. The corresponding intrinsic EW distribution plotted as a green \modibis{dot-dashed} histogram on the bottom panel of Figure \ref{fig:high_ews} is similar to what we find for our ad-hoc 'stoch10' model, with an extended tail towards high large EW values. 
\modi{Although star formation occurs more rarely in these two models than in our fiducial one, we note that the intrinsic \lya and UV LFs (and the stellar mass functions) are not changed by a large amount. Indeed, more gas accumulates in the galaxy for the 'stoch10' and high-$\Sigma_{\rm gas}^{\rm thresh}$ models, so SF events turn more gas into stars at once using the Kennicutt law. These models even produce slightly larger SFRs, hence larger \lya and UV luminosities, due to the power-law index of the Kennicutt law being greater than 1. However, the higher gas content would increase the effect of dust attenuation, as both quantities are correlated \citepalias[Section 2.2 in][]{garel2012a}, so these models will no longer necessarily give a good match to the observed UV and \lya LFs. Overall, stochastic star formation scenarios constitute a viable mechanism to produce high \lya EWs in high redshift galaxies, and we will investigate these models in more details in future studies.}

\section{UV/\lya properties of LBGs}
\label{sec:cross_prop2}

\begin{figure}
\vskip-4ex
\hskip1ex
\includegraphics[width=8.cm,height=6.7cm]{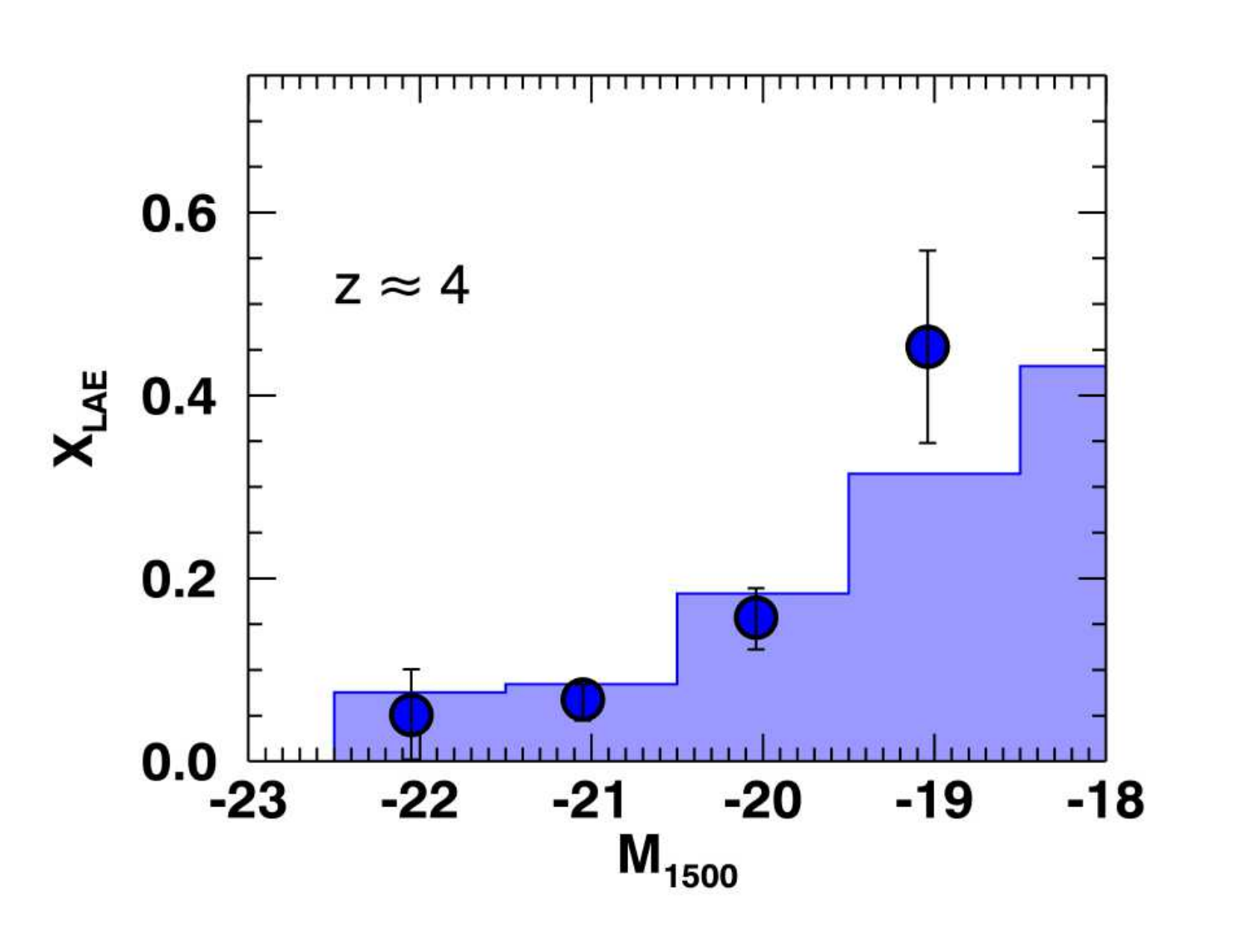}
\vskip-2ex
\includegraphics[width=8.4cm,height=6.3cm]{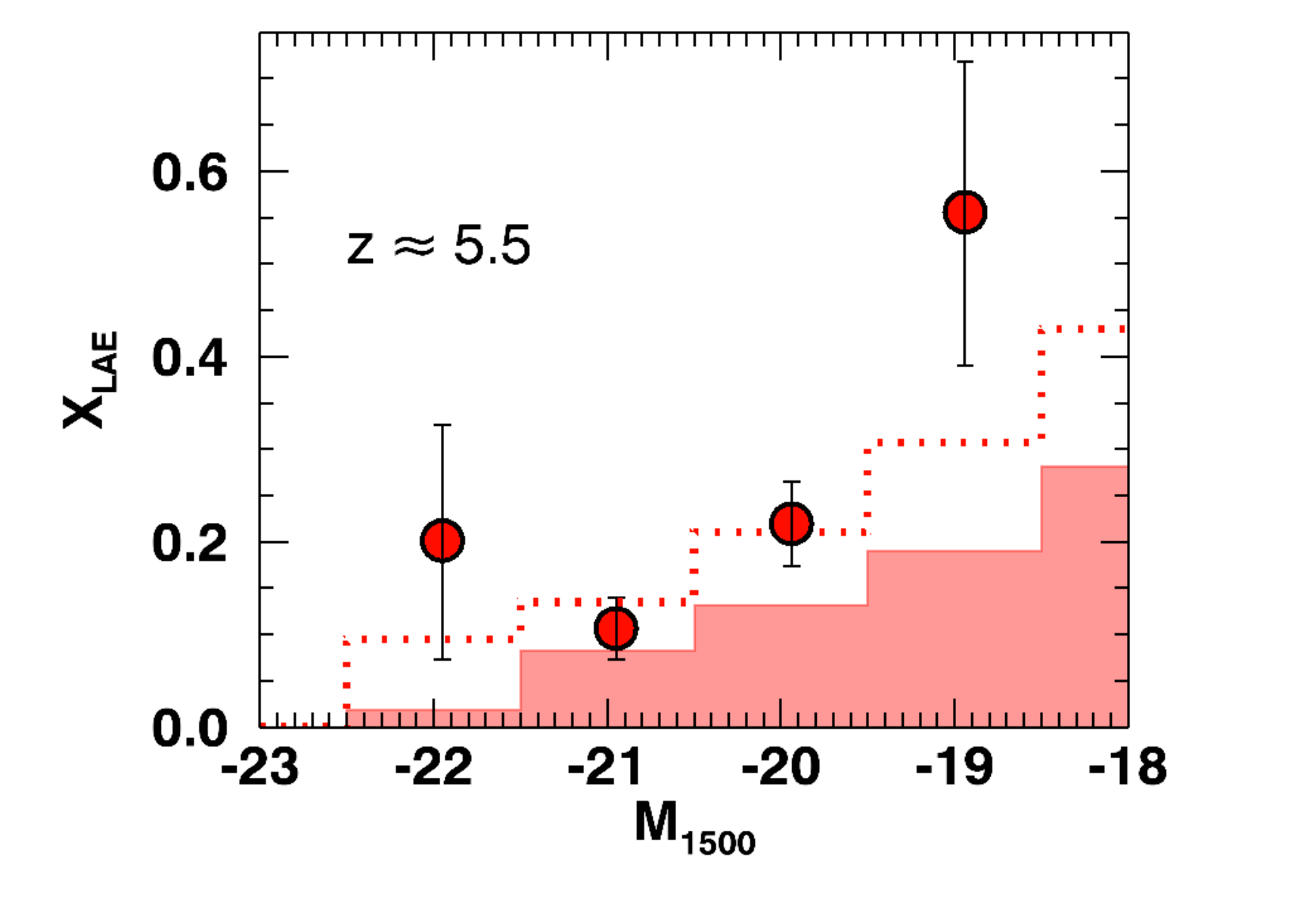}
\vskip-2ex
\caption{Fraction of strong \lya emitters among samples of LBGs at z $\approx$ 4 (top) and 5.5 (bottom). In each panel, the histogram shows the predicted fraction of LAEs $X_{\rm LAE}$ in five bins of rest-frame absolute UV magnitude $M_{1500}$. The model is compared to the data of \citet{stark10} (represented by circles with error bars) who measured the fraction of LBGs with \lya equivalent widths larger than 50 \AA. As shown by the dotted line in the bottom panel, we find a slightly better agreement with the observations at z $\approx$ 5 with an EW cut of 45 \AA{} instead of 50 \AA{} (solid line).}
\label{fig:frac_lae}
\end{figure}

\subsection{The fraction of emitters within LBGs}
Spectroscopic follow-ups of LBG samples show that \lya emission is more often detected in UV-fainter galaxies. In Figure \ref{fig:frac_lae}, we test our (fiducial) model against the observations of \citet{stark10} who found that the fraction of strong \lya emitters \modi{(EW $>$ 50 \AA)}, $X_{\rm LAE}$, increases from $\approx$ 7 \% (20 \%) at $M_{1500} = -22$ to 45 \% (55 \%) at $M_{1500} = -19$ for z $\approx$ 4 (z $\approx$ 5). We find a very similar trend at both z $\approx$ 4 and 5.5, as shown by the histograms in the top and bottom panels. The agreement with the data of \citet{stark10} (represented as circles) is excellent at $M_{1500} < -19.5$ for z $\approx$ 4. At z $\approx$ 5.5, the fraction of LAEs with EW $>$ 50 \AA{} in the model is slightly lower than the observations. However, as shown by the red dotted histogram, we find a good match to the data if we decrease the EW threshold by only 10 \% (EW $>$ 45 \AA). The fractions predicted by our model in the faintest bin are a bit lower than their observational measurement, but the error bars are significant at this magnitude. 

\begin{figure}
\hskip-1ex
\includegraphics[width=8.6cm,height=6.6cm]{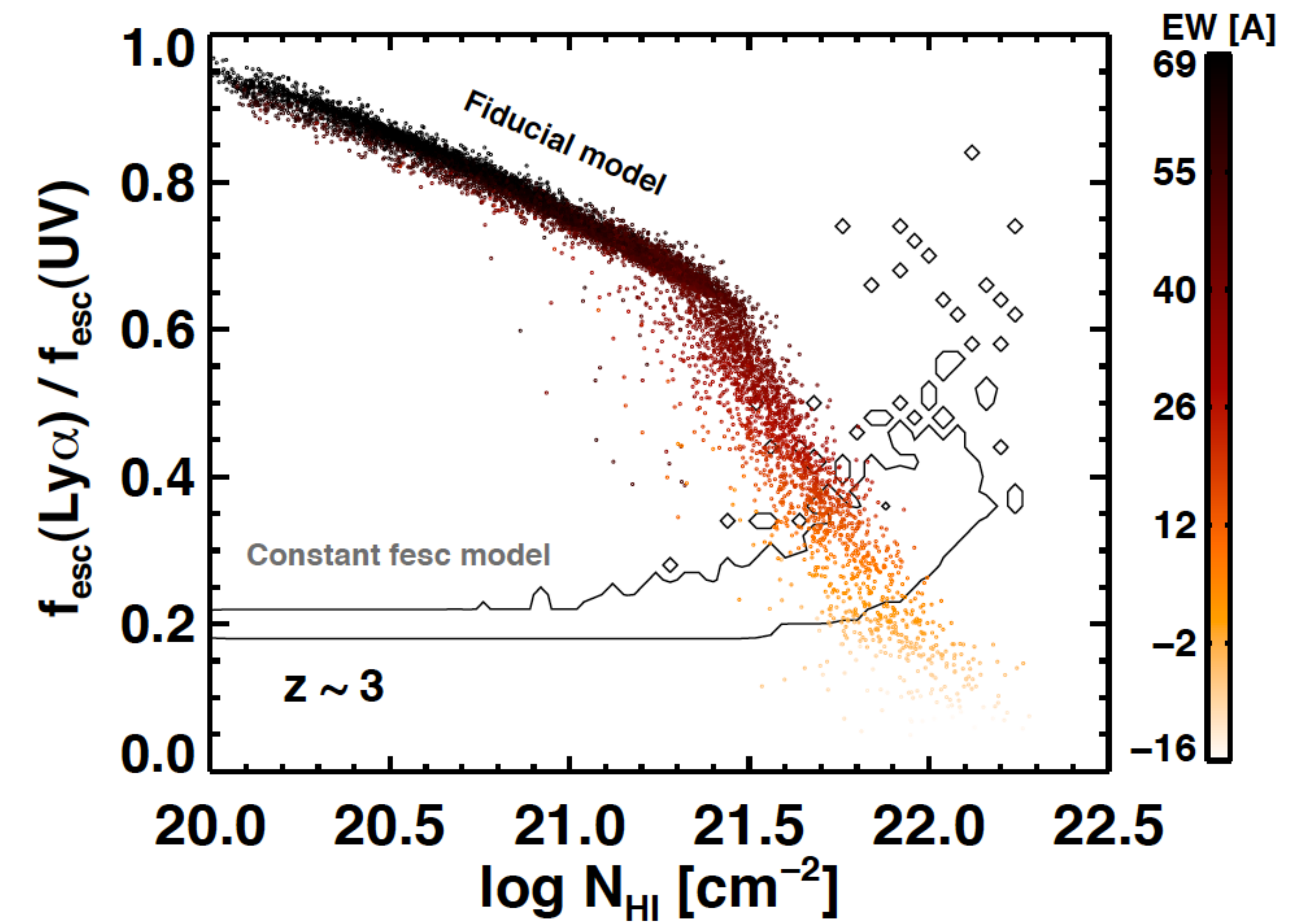}
\hskip-8ex
\caption{\lyat/UV escape fraction ratio in z $=$ 3 LAEs as a function of \hi shell column density. The dots represent individual galaxies in our model, colour-coded relatively to their \lya EW (\modi{after applying dust attenuation}). For comparison, we show the \lyat/UV escape fraction ratio expected from constant \lya escape fraction models \citep{le-delliou2005a,nag}, assuming f$_{\rm esc}$(\lyat) $=$ 20\% (\modibis{contours}).}
\label{fig:fesc_ratio}
\vskip-6ex
\end{figure}

The $X_{\rm LAE}$ evolution as a function of $M_{1500}$ is related to the anti-correlation between \lya equivalent width and UV magnitude which is commonly observed in LBGs and LAEs \citep{ando,stanway07,ouch08}. As discussed in \citetalias{garel2012a}, we find that UV faint galaxies are slightly more likely to display high intrinsic EW than bright LBGs. However, the main driver of the observed $X_{\rm LAE}$-$M_{1500}$ relation in our model remains the differential dust attenuation experienced by \lya and UV photons, caused by \lya resonant scattering in \hi gas. This is illustrated on Figure \ref{fig:fesc_ratio}, where we plot the ratio of UV to \lya escape fractions as a function of the shell \hi column density, N$_{\rm \hi}$, for galaxies at z $=$ 3 which have $L_{\rm Ly\alpha} > 10^{42}$ erg s$^{-1}$. Each dot represents a galaxy, colour-coded as a function of its \lya equivalent width. The f$_{\rm esc}$(\lyat)$/$f$_{\rm esc}$(UV) ratio is approximatively one at low column density, but it sharply decreases towards larger values, which highlights the stronger effect of dust on \lya due to resonant scattering in \hi optically thick media. High-SFR galaxies (i.e. UV-bright galaxies) having a larger \hi column density on average in GALICS,  the N$_{\rm \hi}$-dependent f$_{\rm esc}$(\lyat)$/$f$_{\rm esc}$(UV) ratio unavoidably leads to a lower \lya fraction in bright LBGs compared to faint LBGs (Fig. \ref{fig:frac_lae}). 

For comparison, we have overlaid on Figure \ref{fig:fesc_ratio} the predictions for the constant \lya escape fraction model (\modibis{contours}), which is often used in the literature to fit the observed \lya luminosity functions \citep[e.g.][]{le-delliou2005a,nag}. Fixing f$_{\rm esc}$(\lyat)$=$20\%, which is the value needed to roughly match the bright-end of the z $=$ 3 LF in our model, the f$_{\rm esc}$(\lyat)$/$f$_{\rm esc}$(UV) ratio increases towards large N$_{\rm \hi}$ values. A consequence of this trend, opposite to what is found with our fiducial model, would be to predict a higher LAE fraction in brighter LBGs, at odds with the observations. 

The scaling of the ratio of UV continuum escape fraction to resonant \lya escape fraction with N$_{\rm \hi}$ and  SFR, is then a key factor to explain the $X_{\rm LAE}$-$M_{1500}$ trend, as also reported by \citet{forero-romero2012a}, who coupled hydrodynamical simulations with \lya RT in galaxies approximated as dusty slabs. 

\subsection{The evolution of the \lya fraction in LBGs with redshift}

According to recent studies, the fraction of strong \lya emitters among LBGs, $X_{\rm LAE}$, seems to increase from z $\approx$ 3 to 6 \citep{stark2011a,cassata2014a}. This trend appears to invert at z $=$ 7, as reported by \citet{ono2012a} and \citet{schenker2012a} who find that $X_{\rm LAE}$ dramatically drops between z $\approx$ 6 and 7 \citep[see also][]{pentericci2011a}.  The redshift evolution of $X_{\rm LAE}$ could be due to a rise of the mean internal \lya escape fraction from galaxies at z $\approx$ 3-6, and by a sharp increase of neutral IGM opacity at z $\gtrsim$ 6. Nonetheless, it remains plausible that the apparent trend of the fraction of emitters in LBGs with redshift is also (i) affected, or even driven, by the intrinsic evolution of the \lyat, UV, ionising properties of galaxies \citep{dayal2012,dijkstra2014b}, (ii) due to a reduction of the \lya transmission caused by the increase of the number of Lyman Limit systems at the end of the epoch of reionisation \citep{bolton2012a}, and/or (iii) altered by small number statistics \citep[see][for a more detailed discussion]{dijkstra2014,mesinger2014}.

\begin{figure}
\hskip-5ex
\includegraphics[totalheight=0.32\textheight,width=0.61\textwidth]{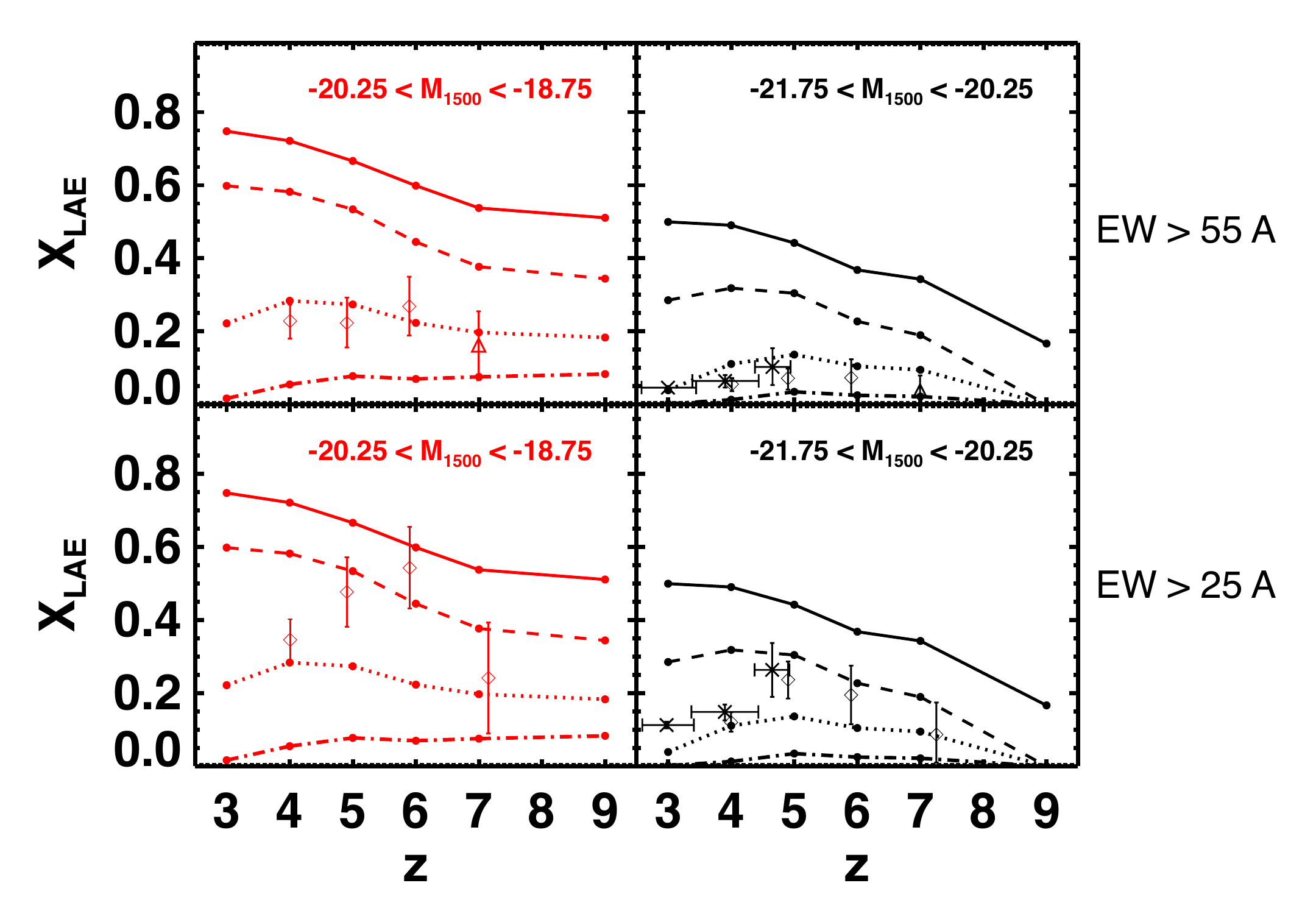}
\vskip-2ex
\caption{\modi{Fraction of \lya emitters, $X_{\rm LAE}$, as a function of redshift among samples of faint LBGs ($-20.25 < M_{1500} < -18.75$; left panels, in red) and bright LBGs ($-21.75 < M_{1500} < -20.25$; right panels, in black).}
\modi{We show the observed fraction of strong \lya emitters (EW $\geq$ 55 \AA) and weaker emitters (EW $\geq$ 25 \AA) in the top and bottom panels respectively. \textit{Data: } Crosses \citep{cassata2014a}, diamonds \citep{stark2011a,schenker2012a}, and triangles \citep{ono2012a}. In all panels, we compare the observations to the LAE fraction predicted by our model for various EW cuts (solid line: EW $\geq$ 25 \AA; dashed line: EW $\geq$ 35 \AA; dotted line: EW $\geq$ 45 \AA; dot-dashed line: EW $\geq$ 55 \AA).}}
\label{fig:frac_lae_z}
\vskip-4ex
\end{figure}

Even though the study of reionisation is out of the scope of this paper, it is interesting to compare our predictions to the observed redshift evolution of $X_{\rm LAE}$ without taking the effect of IGM into account. Figure \ref{fig:frac_lae_z} shows the evolution of $X_{\rm LAE}$ of faint (in red) and bright LBGs (in black). The symbols are observational estimates of $X_{\rm LAE}$ where a LAE is defined as having EW $>$ 55 \AA{} in the top panels, and EW $>$ 25 \AA{} in the bottom panels. The solid, dashed, dotted and dot-dash curves are predictions from our model for several EW cuts: EW $\geq$ 25 \AA, EW $\geq$ 35 \AA, EW $\geq$ 45 \AA, and EW $\geq$ 55 \AA{} respectively. In all panels, we clearly see that the LAE fraction is reduced at all redshifts when higher EW cuts are applied. There is a decreasing trend with redshift for EW $\geq$ 25 \AA, but it flattens for higher EW cuts.

From the upper panels (EW $>$ 55\AA), we see that we need to lower the EW cut from 55 \AA{} to 45 \AA{} to match the data (dotted lines). In this case, the $X_{\rm LAE}$ computed from the model is based on small statistics because there are few LBGs in the corresponding ranges of magnitude with EW $\geq$ 55 \AA, so it is hard to draw robust conclusions from our model. For the EW $\geq$ 25 \AA{} measurements however (\modi{bottom} panels), a better agreement is obtained if we increase the threshold to select only emitters stronger than EW $=$ 35 \AA.

\begin{figure*}
\vskip-4ex
\centering
\includegraphics[width=16.cm,height=6.8cm]{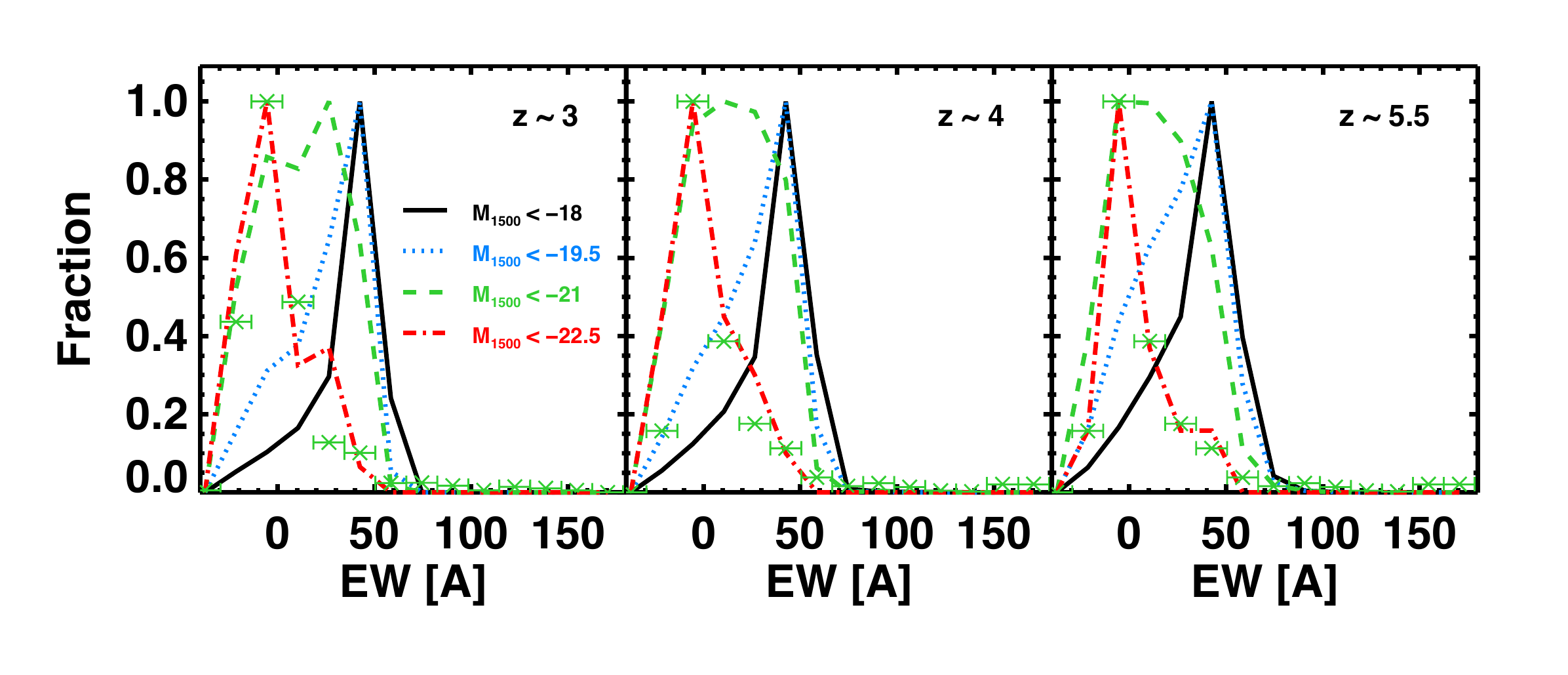}
\vskip-6ex
\caption{\lya equivalent widths distributions of UV-selected galaxies at z $=$ 3 (left), z $=$ 4 (center), and z $=$ 5.5 (right). In each panel, we show the distributions for different UV magnitude cuts, as labelled. The green crosses correspond to the data of \citet{cassata2014a} (rebinned) for galaxies selected with $M_{\rm 1500}$ $\lesssim$ -21 (see their Figure 1).}
\label{fig:muv_ew}
\end{figure*}

In order to clarify the disagreement between the model and the observed redshift evolution of $X_{\rm LAE}$ in LBGs, as well as the strong variation with the EW cuts, we now investigate the variation of the \lya equivalent width distributions with UV magnitude. Indeed, by construction, should the model perfectly match the observed EW distributions of UV-selected galaxies, the redshift evolution of the LAE fraction in LBGs would be recovered. These are shown in Figure \ref{fig:muv_ew} for z $=$ 3, z $=$ 4, and z $=$ 5.5 as a function of $M_{\rm 1500}$, and compared to the VUDS data of \citet{cassata2014a} (green crosses). 

First, we note that the model distributions shift to low EW values as brighter and brighter LBGs are considered. They span a range similar to the observed ones, from $\approx$ -40 \AA{} to 70 \AA. Unlike LAE surveys, which preferentially pick up low-continuum sources, UV-selected galaxies show very few high EWs ($\gtrsim$ 70 \AA). Interestingly, the tail of the distributions appears to be slightly shifted to larger EW values for fainter LBGs. This explains why the fraction of galaxies with EW $>$ 50 \AA{} is found to increase towards faint UV magnitude in Figure \ref{fig:frac_lae}, in agreement with the data of \citet{stark10}. However, the fraction of galaxies drops sharply at EW $\approx$ 50 \AA{} for all UV magnitude selections considered here, echoing the rapid variation of $X_{\rm LAE}$ for \lya equivalent width thresholds near 50 \AA, as discussed in Figure \ref{fig:frac_lae_z}.

\begin{center}
\begin{table*}
\begin{tabular}{cccccccc}
\hline
\hline
 & & z $=$ 3 & z $=$ 4 & z $=$ 5 & z $=$ 6 & z $=$ 6.6 & z $=$ 7.5 \\
\hline
 & log(M$_{\rm halo}^{\rm LAE}/\msun)$ & 10.7$^{\scriptstyle +0.8}_{\scriptstyle -0.3}$ & 10.6$^{\scriptstyle +0.6}_{\scriptstyle -0.3}$  & 10.4$^{\scriptstyle +0.4}_{\scriptstyle -0.2}$  & 10.2$^{\scriptstyle +0.4}_{\scriptstyle -0.2}$  & 10.2$^{\scriptstyle +0.4}_{\scriptstyle -0.2}$  & 10.1$^{\scriptstyle +0.4}_{\scriptstyle -0.2}$  \\
Faint LAEs/LBGs & log(M$_{\rm halo}^{\rm LBG}/\msun)$ & 10.7$^{\scriptstyle +0.8}_{\scriptstyle -0.3}$ & 10.5$^{\scriptstyle +0.6}_{\scriptstyle -0.3}$  & 10.4$^{\scriptstyle +0.4}_{\scriptstyle -0.2}$  & 10.2$^{\scriptstyle +0.4}_{\scriptstyle -0.2}$  & 10.1$^{\scriptstyle +0.4}_{\scriptstyle -0.2}$  & 10.1$^{\scriptstyle +0.4}_{\scriptstyle -0.2}$  \\
 & n$^{\rm LAE} \: /$ n$^{\rm LBG}$ & 0.96 & 0.91 & 0.91 & 0.95 & 0.98 & 1.00 \\
\hline
 & log(M$_{\rm halo}^{\rm LAE}/\msun)$ & 11.5$^{\scriptstyle +0.6}_{\scriptstyle -0.3}$ & 11.4$^{\scriptstyle +0.5}_{\scriptstyle -0.2}$  & 11.1$^{\scriptstyle +0.4}_{\scriptstyle -0.2}$  & 10.9$^{\scriptstyle +0.4}_{\scriptstyle -0.2}$  & 10.8$^{\scriptstyle +0.4}_{\scriptstyle -0.2}$  & 10.7$^{\scriptstyle +0.4}_{\scriptstyle -0.2}$  \\
Typical LAEs/LBGs & log(M$_{\rm halo}^{\rm LBG}/\msun)$ & 11.4$^{\scriptstyle +0.6}_{\scriptstyle -0.3}$ & 11.3$^{\scriptstyle +0.5}_{\scriptstyle -0.3}$  & 11.0$^{\scriptstyle +0.4}_{\scriptstyle -0.2}$  & 10.9$^{\scriptstyle +0.4}_{\scriptstyle -0.2}$  & 10.8$^{\scriptstyle +0.4}_{\scriptstyle -0.2}$  & 10.8$^{\scriptstyle +0.3}_{\scriptstyle -0.2}$  \\
 & n$^{\rm LAE} \: /$ n$^{\rm LBG}$ & 0.71 & 0.67 & 0.72 & 0.67 & 0.64 & 0.66  \\
\hline
 & log(M$_{\rm halo}^{\rm LAE}/\msun)$ & 12.2$^{\scriptstyle +0.3}_{\scriptstyle -0.2}$ & 12.1$^{\scriptstyle +0.3}_{\scriptstyle -0.2}$  & 11.8$^{\scriptstyle +0.3}_{\scriptstyle -0.2}$  & 11.7$^{\scriptstyle +0.3}_{\scriptstyle -0.2}$  & 11.5$^{\scriptstyle +0.1}_{\scriptstyle -0.1}$  & 11.4$^{\scriptstyle +0.2}_{\scriptstyle -0.2}$  \\
Bright LAEs/LBGs & log(M$_{\rm halo}^{\rm LBG}/\msun)$ & 12.1$^{\scriptstyle +0.4}_{\scriptstyle -0.3}$ & 12.0$^{\scriptstyle +0.4}_{\scriptstyle -0.2}$  & 11.7$^{\scriptstyle +0.4}_{\scriptstyle -0.2}$  & 11.6$^{\scriptstyle +0.3}_{\scriptstyle -0.2}$  & 11.5$^{\scriptstyle +0.2}_{\scriptstyle -0.1}$  & 11.4$^{\scriptstyle +0.4}_{\scriptstyle -0.2}$  \\
 & n$^{\rm LAE} \: /$ n$^{\rm LBG}$ & 0.24 & 0.21 & 0.31 & 0.28 & 0.22 & 0.33 \\
\hline
\end{tabular}
\caption{Properties of LAEs and LBGs at z $=$ 3, 3.7, 5, 6, 6.6 and 7.5 for three bins of \lya and UV luminosity. Here, LAEs are defined to have EW $>$ 0 \AA{} and they are classified as faint ($10^{41} \lt L_{\rm Ly\alpha} \lt 10^{42}$ erg s$^{-1}$), typical ($10^{42} \lt L_{\rm Ly\alpha} \lt 10^{43}$ erg s$^{-1}$), and bright ($10^{43} \lt L_{\rm Ly\alpha} \lt 10^{44}$ erg s$^{-1}$) according to their \lya luminosity. The faint, typical, and bright samples of LBGs are selected from their UV magnitude, $-15.8 > {\rm M_{1500}} > -18.3$, $-18.3 > {\rm M_{1500}} > -20.8$, and $-20.8 > {\rm M_{1500}} > -23.3$ respectively. M$_{\rm halo}^{\rm LAE}$ and M$_{\rm halo}^{\rm LBG}$ give the median host halo masses of LAEs and LBGs in each subsample. The subscripts and superscripts next to each value correspond to the 10th and 90th percentiles respectively. For each subsample (faint, typical and bright), the third row shows the ratio of the number of LAEs to the number of LBGs, n$^{\rm LAE} \: /$ n$^{\rm LBG}$. }
\label{tab1}
\end{table*}
\end{center}

Second, the VUDS EW distributions computed for galaxies with $M_{\rm 1500}$ $\lesssim$ -21 peak to lower EWs than the model distributions for the same UV magnitude range (green \modibis{dashed} curves), which is the reason why we find larger $X_{\rm LAE}$ than what is observed for EW $\geq$ 25 \AA{} (\modi{bottom} panels of Figure \ref{fig:frac_lae_z}). However, the EW distributions of \citet{cassata2014a} agree very well with the brightest LBGs in the model ($M_{\rm 1500}$ $<$ -22.5), as shown by the red \modibis{dot-dashed} curves.\\

Although our model catches well the trend between \lya equivalent widths and UV magnitude, the above study suggests that the exact scaling between these quantities is very sensitive to the cuts used to compute the \lya fractions in LBGs. In addition, some fine-tuning of the model is required to fully reproduce the observed UV/\lya cross-properties. In particular, the ratio of UV/\lya escape fraction is found to have the right scaling with relevant physical properties, like \hi column density, or SFR, in order to interpret the link between the observed properties of LAEs and LBGs, but a stronger differential UV/\lya dust extinction in UV-bright objects is needed to improve the quantitative agreement with the data.

\section{Host halos of LAEs/LBGs}
\label{sec:halos}
\vskip-1ex
In this section, we analyse the properties of the host halos of LAEs and LBGs as predicted by our model. While the study of the clustering of LAEs and LBGs will be addressed in a companion paper (Garel et al., in prep), here, we focus on the halo masses and halo occupation of LAEs and LBGs as a function of redshift to identify how these galaxies respectively trace the dark matter structures. We also briefly discuss how the dynamical range probed by LAEs depends on the EW selection. 

\subsection{Halo masses as a function of \lya/UV luminosities}
\label{subsec:mh_lae_lbg}

On the one hand, clustering analysis reveal that LAEs reside in rather low mass halos, with a median mass of $\sim 10^{11}$$\msun$ at z $\approx$ 3 \citep{gawiser2007a}. Overall, the measured biases of LAEs at z $\approx$ 3-7 correspond to dark matter halos in the range M$_{\rm h} = 10^{10-12} \msun$  \citep[][]{ouchi2010a}. On the other hand, \citet{hildebrandt2009a} find the typical halo mass of LBGs at z $=$ 3-5 to be $\gtrsim 10^{12}$$\msun$, namely about one order of magnitude more massive than those of LAEs \citep{hamana2004a,ouchi2004a,ouchi2005a,mclure2009}. Nevertheless, large uncertainties remain in the determination of the halo masses, notably for LAEs for which the selection and the detection limit can be quite different from one survey to another. Moreover, samples of narrow-band selected galaxies are likely to contain significant fractions of interlopers, which lead to an underestimation of the halo mass.  Correcting for such effect, \citet{kovac2007a} derive a similar bias for LAEs and LBGs at z $\approx$ 4. However, they report that LAEs are 2-16 times rarer than LBGs for a given halo mass. We propose to study these considerations with our model in the next paragraphs.

\begin{figure*}
\hskip-4ex
\begin{minipage}[]{0.45\textwidth}
\centering
\includegraphics[width=8.6cm,height=6.2cm]{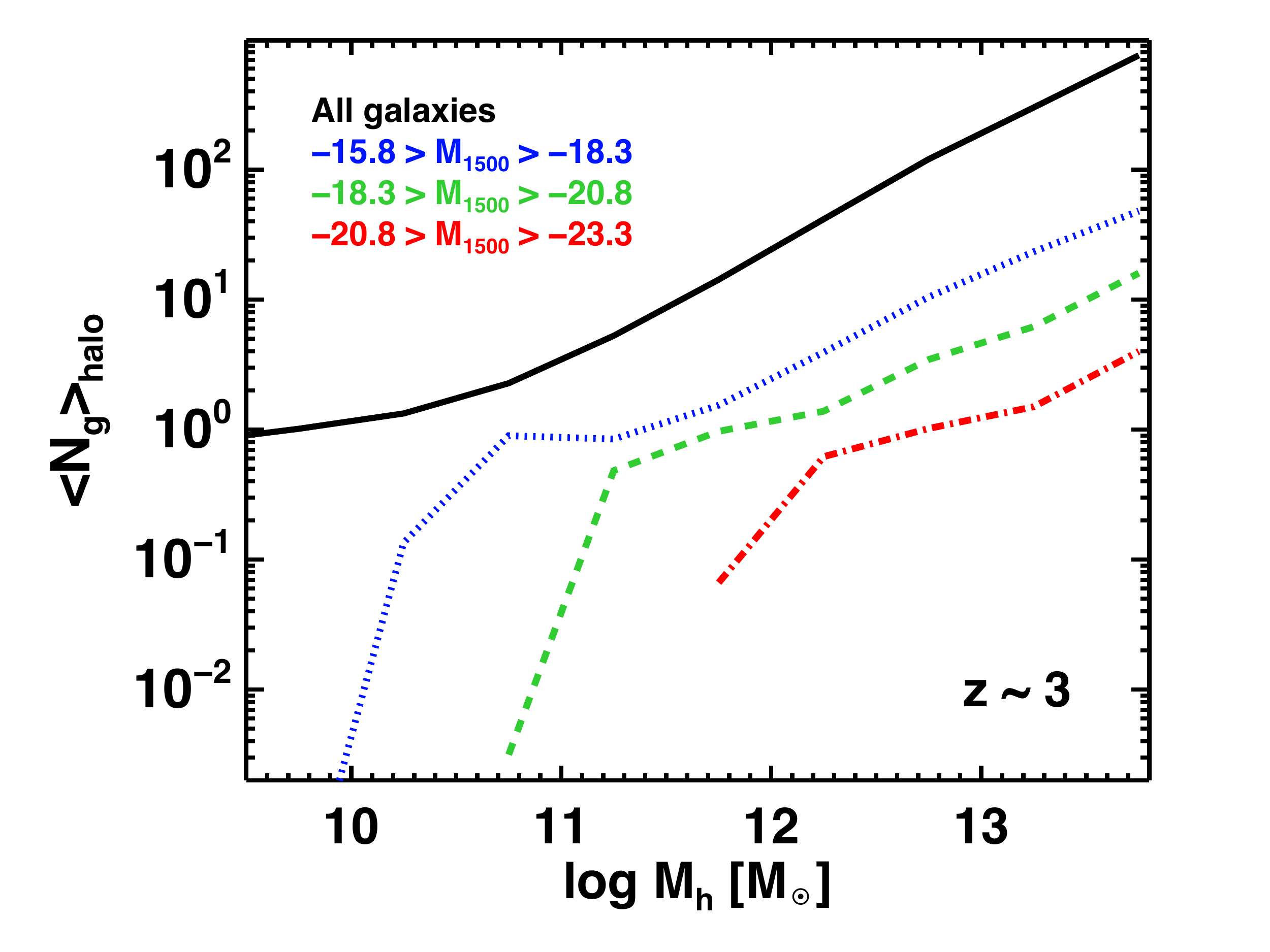}
\end{minipage}
\hskip10ex
\begin{minipage}[]{0.45\textwidth}
\centering
\includegraphics[width=8.6cm,height=6.2cm]{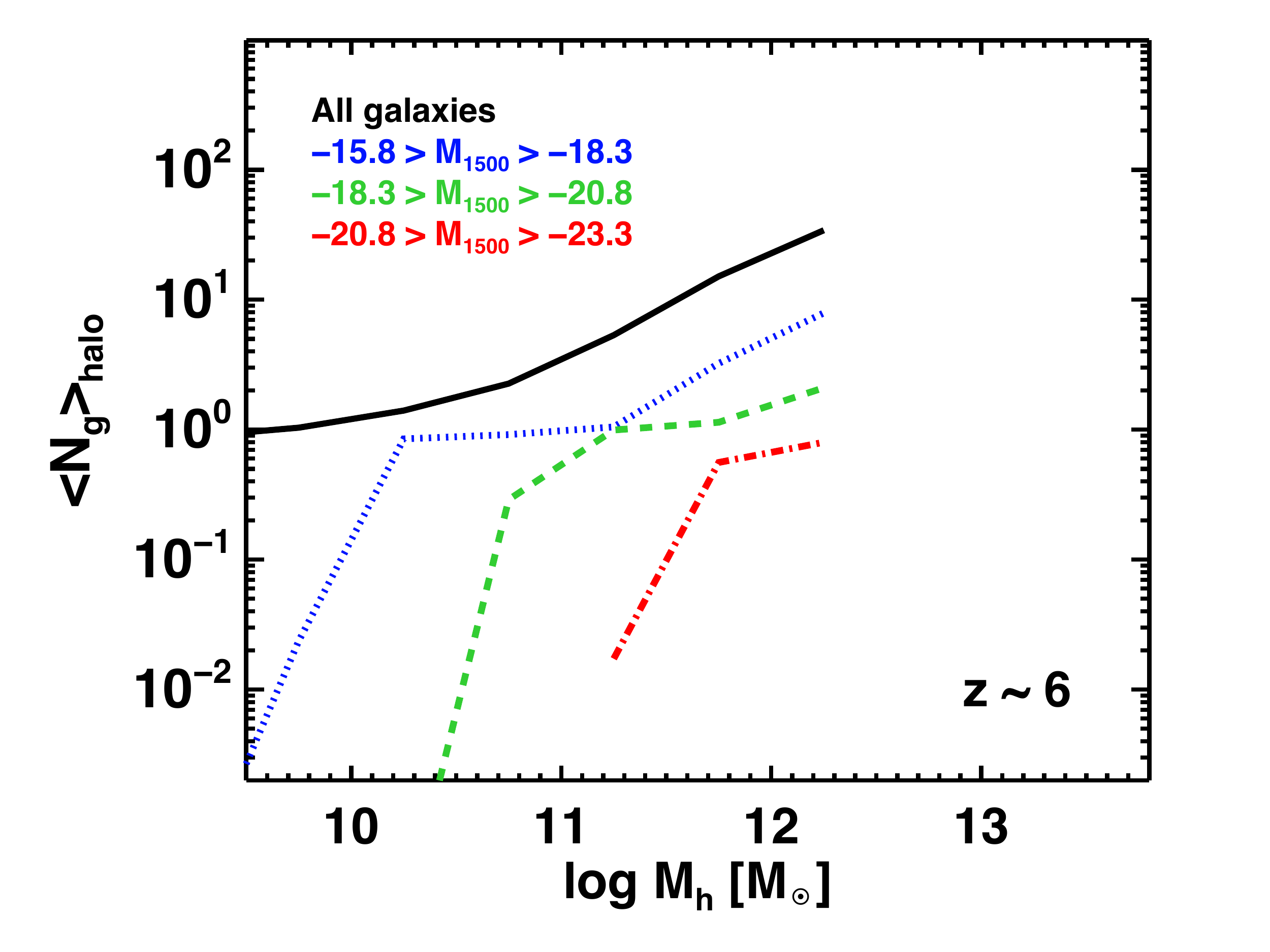}
\end{minipage}
\vskip-2ex
\caption{Halo occupation distribution of LBGs. The mean number of galaxies per dark matter halo, \modi{$\langle N_{\rm g} \rangle_{\raisebox{-1pt}{\scriptsize halo}}$}, as a function of halo mass is shown \modibis{by the black solid line} for z $=3$ (left panel) and z $=$ 6 (right panel). The coloured curves represent the mean halo occupation of faint ($-15.8 > {\rm M_{1500}} > -18.3$: \modibis{blue dotted line}), typical ($-18.3 > {\rm M_{1500}} > -20.8$: \modibis{green dashed line}), and bright ($ -20.8 > {\rm M_{1500}} > -23.3$: \modibis{red dot-dashed line}) LBGs.} 
\label{fig:focc_lbg}
\end{figure*}

\begin{figure*}
\hskip-4ex
\begin{minipage}[]{0.45\textwidth}
\centering
\includegraphics[width=8.6cm,height=6.2cm]{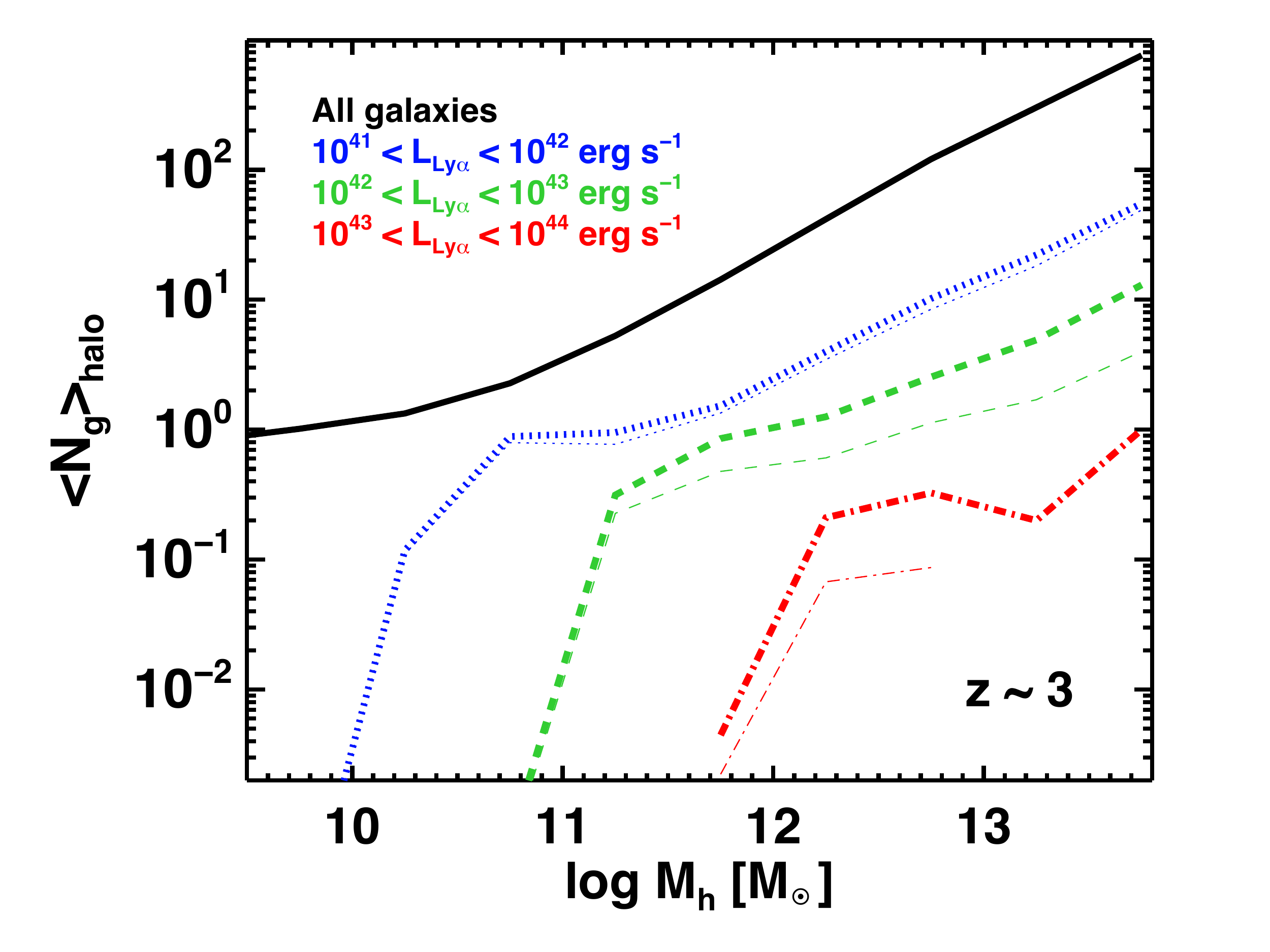}
\end{minipage}
\hskip10ex
\begin{minipage}[]{0.45\textwidth}
\centering
\includegraphics[width=8.6cm,height=6.2cm]{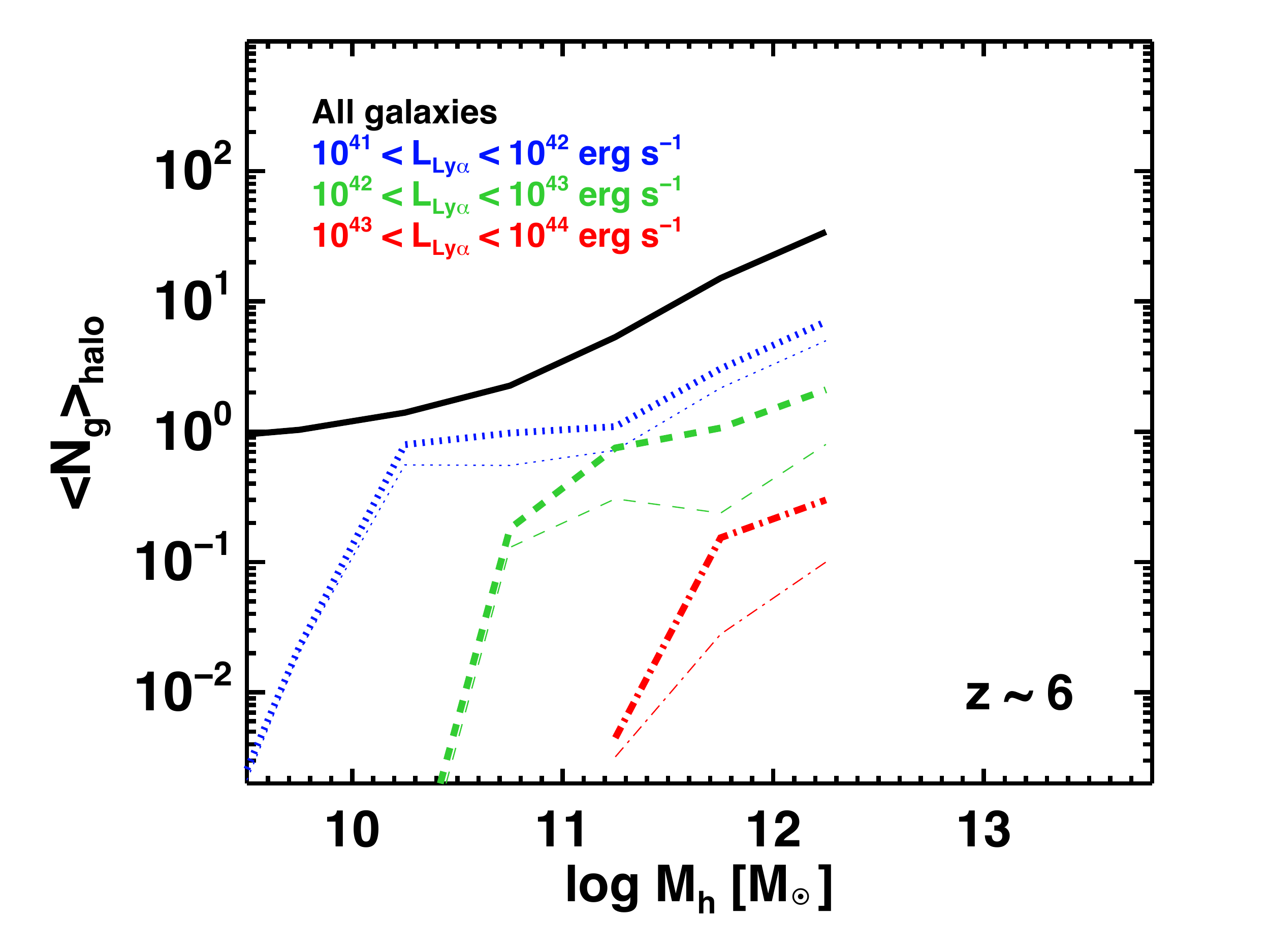}
\end{minipage}
\vskip-2ex
\caption{Halo occupation distribution of LAEs. The mean number of galaxies per dark matter halo, \modi{$\langle N_{\rm g} \rangle_{\raisebox{-1pt}{\scriptsize halo}}$}, as a function of halo mass is shown \modibis{by the black solid thick line} for z $=3$ (left panel) and z $=$ 6 (right panel). The coloured \modibis{thick} curves represent the mean halo occupation of faint ($10^{41} \lt L_{\rm Ly\alpha} \lt 10^{42}$ erg s$^{-1}$: \modibis{blue dotted line}), typical ($10^{42} \lt L_{\rm Ly\alpha} \lt 10^{43}$ erg s$^{-1}$: \modibis{green dashed line}), and bright ($10^{43} <$ $L_{\rm Ly\alpha} < 10^{44}$ erg s$^{-1}$: \modibis{red dot-dashed line}) LAEs, assuming no equivalent width selection. The \modibis{thin} curves are similar but for LAEs with EW $> 40$ \AA.} 
\label{fig:focc_lae}
\vskip-3ex
\end{figure*}

In Table \ref{tab1}, we present the median halo masses in three bins of \textit{observed} (\modi{i.e. dust-attenuated}) \lya luminosity and UV magnitude at redshifts z $=$ 3 to z $=$ 7.5 in our model. The \lya luminosity ranges are $10^{41} \lt L_{\rm Ly\alpha} \lt 10^{42}$ erg s$^{-1}$, $10^{42} \lt L_{\rm Ly\alpha} \lt 10^{43}$ erg s$^{-1}$, and $10^{43} \lt$ $L_{\rm Ly\alpha} \lt 10^{44}$ erg s$^{-1}$, and we term them faint, typical and bright LAEs. Similarly, the notation of faint, typical and bright LBGs will be used for galaxies with $-15.8 > {\rm M_{1500}} > -18.3$, $-18.3 > {\rm M_{1500}} > -20.8$, and $-20.8 > {\rm M_{1500}} > -23.3$. The limit values correspond to SFRs of $\sim 0.08, 0.8, 8$ and $80$ \msunyr{} according to the expected intrinsic scaling between $L_{\rm Ly\alpha}$, $M_{1500}$ and SFR (see Section \ref{subsec:expected_rel}).

\begin{figure*}
\hskip-4ex
\begin{minipage}[]{0.45\textwidth}
\centering
\includegraphics[width=8.6cm,height=6.2cm]{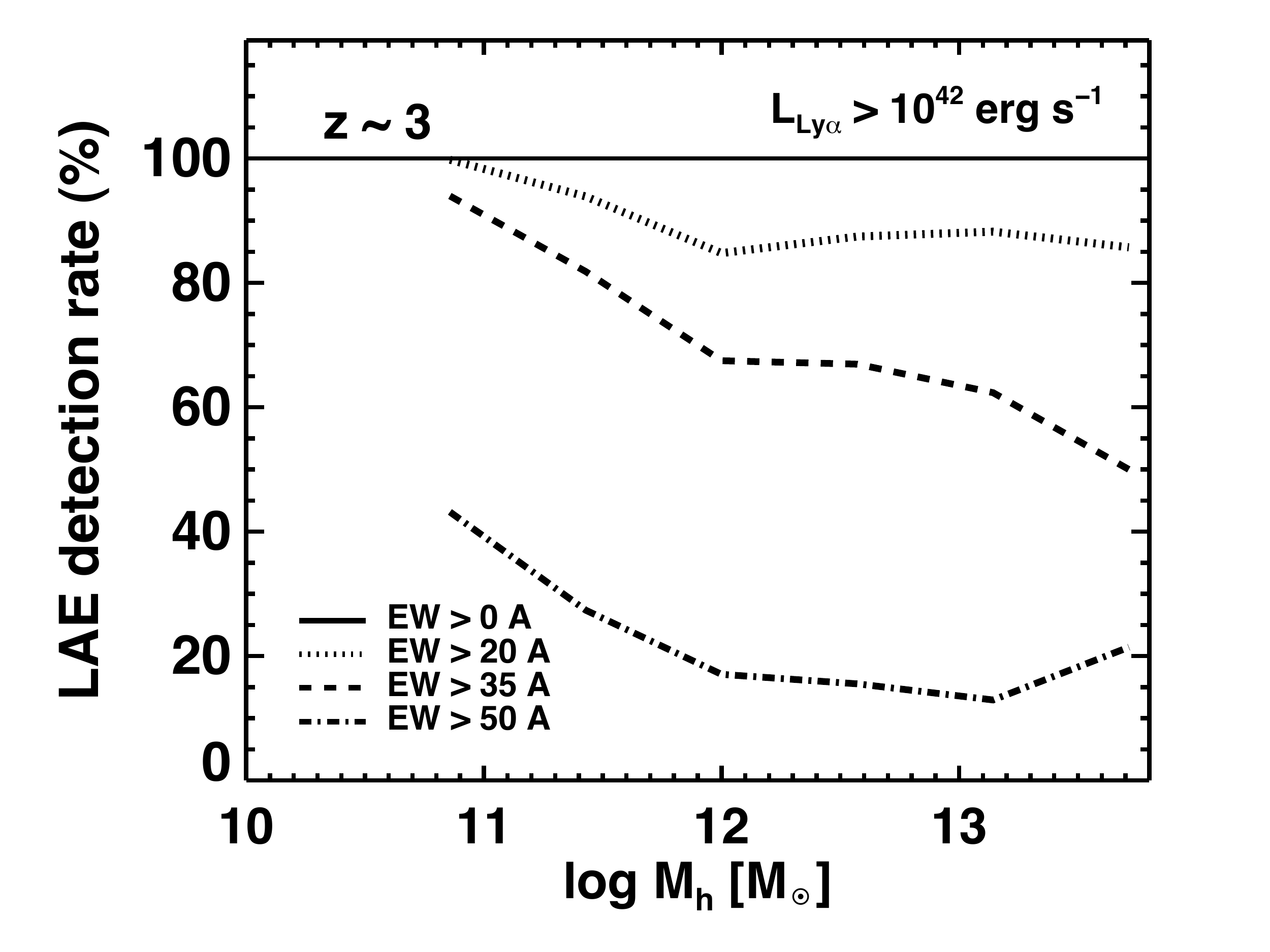}
\end{minipage}
\hskip4ex
\begin{minipage}[]{0.45\textwidth}
\centering
\includegraphics[width=8.6cm,height=6.2cm]{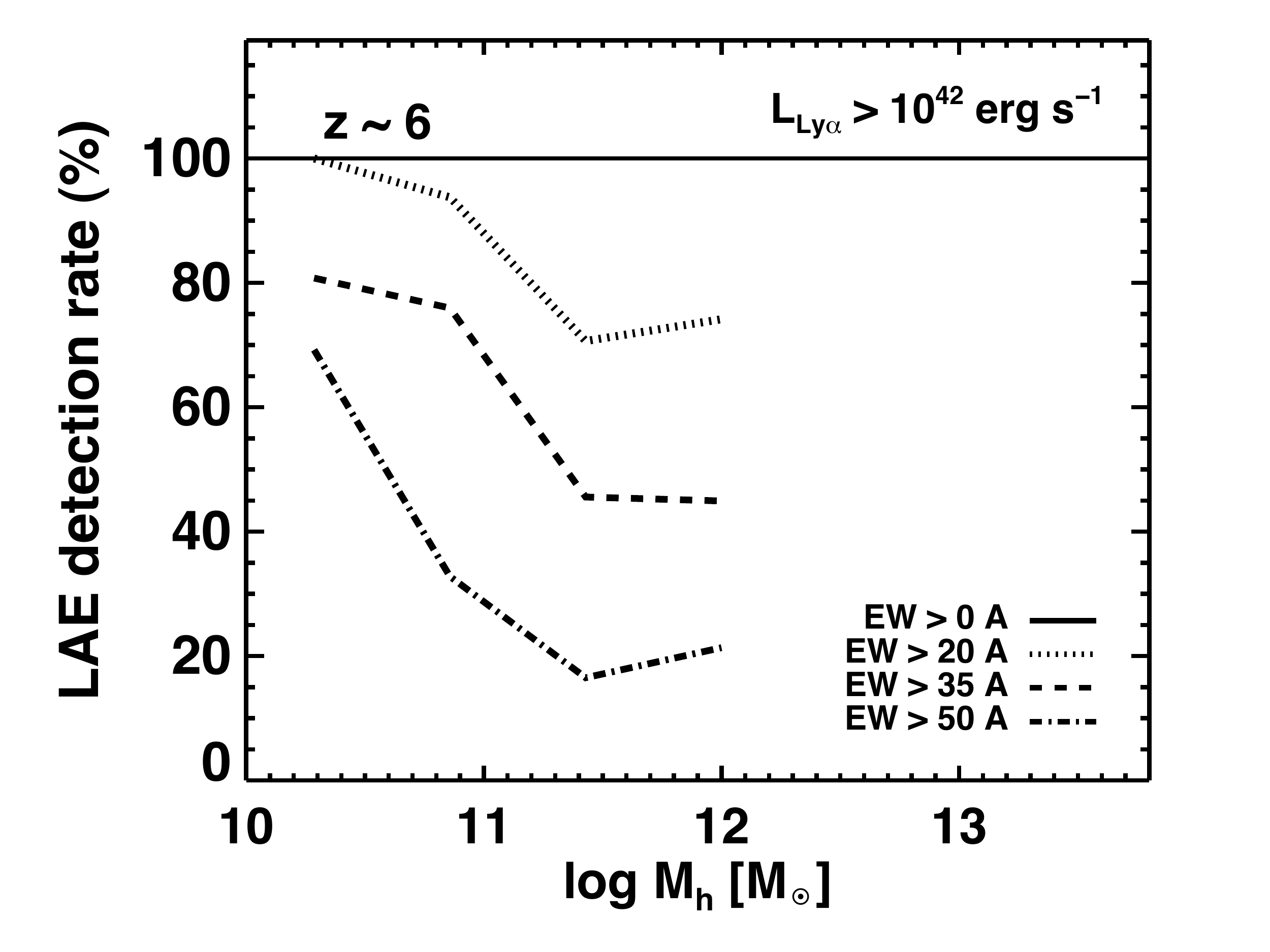}
\end{minipage}
\caption{Detectability of LAEs as function of host halo mass for various EW thresholds for z $\sim$ 3 (left panel) and z $\sim$ 6 (right panel). The 'LAE detection rate' is the ratio of the number of LAEs with EW $>$ 20, 35, and 50 \AA{} (dotted, dashed and dot-dashed curves respectively) to the number of galaxies with a \lya emission (i.e. EW $>$ 0 \AA). Here, we have only considered galaxies brighter than $L_{\rm Ly\alpha} > 10^{42}$ erg s$^{-1}$, which corresponds approximatively to the sensitivity limit of current photometric surveys of LAEs. Galaxies with such luminosities only inhabit halos that are massive enough, hence the apparent cut-off of the curves at low M$_{\rm h}$.} 
\label{fig:ew_mh}
\end{figure*}

First, from inspection of Table \ref{tab1}, we see that the host halo mass of LAEs and LBGs increases from faint to bright luminosity/magnitude at all redshifts. This is expected because the gas accretion rate is higher in more massive halos, so galaxies have higher SFR, hence higher intrinsic \lya and UV emission. Second, for the three luminosity/magnitude bins we defined, LAEs and LBGs inhabit very similar halos, in agreement with the results of \citet{kovac2007a}.
For instance, at z $=$ 3, the median halo mass is $5.10^{10}$$\msun$ for faint LAEs/LBGs, and $1-2.10^{12}$$\msun$ for bright LAEs/LBGs. This result is simply due to \lya and UV emission being both intrinsically powered by SFR, as discussed in Sec. \ref{subsec:expected_rel}.

Although we find very similar host halo masses for LAEs and LBGs at all luminosities considered here, their occupation rate of dark matter halos significantly evolves (see Table \ref{tab1}). The ratio of the number of LAEs to the number of LBGs, n$^{\rm LAE} \: /$ n$^{\rm LBG}$, is about one for faint objects because, at the faint end, dust extinction has little effect so nearly all LBGs are also LAEs, and their \textit{observed} UV magnitudes and \lya luminosities follows the expected relation given by Eq. \ref{eq:llya_muv_kennimf}. At the bright end, radiative transfer effects affect more strongly the \lya intensity than the UV so that only $20-30 \%$ of bright LBGs will have a bright \lya counterpart. \modi{In Table \ref{tab1}, we considered LAEs to have EW $>$ 0 \AA. If we assumed a higher EW threshold (e.g. 10-30 \AA), the ratio of the number of LAEs to LBGs would still be about one for faint sources but it would be less than $20-30 \%$ for bright galaxies, since UV bright sources have lower EWs than faint LBGs on average (as discussed in Figure \ref{fig:muv_ew}).}

Several reasons can be invoked to explain why LAEs are observationally measured to be located in less massive halos than LBGs: (i) the LAEs are selected in deeper surveys than LBGs, (ii) the LAE samples were highly contaminated, or (iii) the LAEs were selected in the tail of the EW distribution where \lya traces a very recent starburst, so their \lya luminosity is higher than expected for a given UV magnitude (see Eq. \ref{eq:llya_ew_exp}).

\subsection{Halo occupation of LAEs and LBGs}
In order to investigate how LBGs and LAEs respectively populate dark matter halos, we now turn our interest to the halo occupation of UV- and \lyat-selected galaxies. Figures \ref{fig:focc_lbg} and \ref{fig:focc_lae} plot the mean number of galaxies per halo, \modi{$\langle N_{\rm g} \rangle_{\raisebox{-1pt}{\scriptsize halo}}$}, as a function of halo mass, for z $\approx$ 3 (left panels) and z $\approx$ 6 (right panels). The black \modibis{solid} curves in all panels correspond to all galaxies. While our model predicts about one galaxy per halo at low M$_{\rm h}$, \modi{$\langle N_{\rm g} \rangle_{\raisebox{-1pt}{\scriptsize halo}}$} increases with halo mass and reaches about 1,000 (100) in the most massive structures at z $\approx$ 3 (z $\approx$ 6). 

Here, we again split galaxies according to their observed UV and \lya luminosities: the \modibis{thick dotted} blue, \modibis{dashed} green and \modibis{dot-dashed} red curves refer to faint, typical, and bright objects respectively. Although galaxies can form in halos with masses $\gtrsim 2.10^9$ \msun{} in our simulation, we can see that there is a minimum allowed host halo mass for a given UV or \lya luminosity threshold. For instance, almost no faint LBG ($-15.8 > {\rm M_{1500}} > -18.3$) inhabit halos less massive than $10^{10}$ \msun{} at z $=$ 3 (blue \modibis{dotted} curve in left panel of Figure \ref{fig:focc_lbg}). Moreover, the minimum halo mass increases towards brighter objects. Again, this is the direct consequence of the UV luminosity varying as the SFR, and the SFR being proportional to the gas accretion rate, and thus to the halo mass. 

Each halo mass bin is populated by one corresponding (central) galaxy of given luminosity and by satellites which belong to a fainter population: there is one central faint LBG per halo of \modi{log(M$_{\rm h}/\msun) \approx$ 10.5}, one central typical LBG per halo of \modi{log(M$_{\rm h}/\msun) \approx$ 11.5}, and one bright LBG per halo of \modi{log(M$_{\rm h}/\msun) \approx$ 12.5}. A similar behaviour is seen for LAEs (Figure \ref{fig:focc_lae}), but we note that the number of bright LAEs in massive halos is less than the number of bright LBGs. This echoes the results from Table \ref{tab1} where we found that bright LAEs were rarer than bright LBGs (see Section \ref{subsec:mh_lae_lbg}), and this is due to the stronger effect of dust attenuation for \lya than UV continuum which can turn intrinsically bright LAEs into much fainter objects (see Section \ref{subsec:expected_rel}). 

Still in Figure \ref{fig:focc_lae}, the \modibis{thin} curves correspond to strong emitters only (EW$ > 40$ \AA). For faint LAEs, this criterion does not remove many objects. However, \modi{$\langle N_{\rm g} \rangle_{\raisebox{-1pt}{\scriptsize halo}}$} starts decreasing more significantly for the typical and bright samples, especially for galaxies residing in massive halos. This indicates that massive halos are more likely to host weak emitters (low EW galaxies) than low-mass halos in our model. 

\subsection{The bias on LAE host halos introduced by EW cuts}
\label{subsec:detect_lae}

As noticed in Figure \ref{fig:focc_lae}, selecting LAEs above a given \lya EW especially removes galaxies in more massive halos. Here, we investigate this aspect more quantitatively. In Figure \ref{fig:ew_mh}, we plot the LAE detection rate as a function of host halo mass at z $\approx$ 3 (left panel) and z $\approx$ 6 (right panel), adopting various EW cuts. We consider only galaxies brighter than $10^{42}$ erg s$^{-1}$, and we define the LAE detection rate as the ratio of the number of LAEs with EW $>$ 20, 35, and 50 \AA{} (dotted, dashed and dot-dashed curves respectively) and the number of all LAEs (i.e. galaxies with \lya in emission, EW $>$ 0 \AA). 

We find that the LAE detection rate decreases at all halo masses as the EW threshold increases. Using a value of 20 \AA{} still allows to detect the vast majority of LAEs. However, many line emitters may be missed when larger values are adopted according to our model. Moreover, for a given EW cut, the drop of the LAE detection rate is more significant in massive halos than low-mass ones at both z $\approx$ 3 and z $\approx$ 6. As an example, for EW $=$ 35 \AA{} at z $\approx$ 3, it decreases from $\approx 90 \%$ at M$_{\rm h}=10^{11}$$\msun$ to 60 $\%$ at M$_{\rm h}=10^{13}$$\msun$. This behaviour has two main causes. 

First, central, intrinsically \lyat-bright, galaxies which are located in massive halos are often strongly affected by dust extinction which can significantly reduce, not only their \lya luminosity, but also their equivalent width. Second, the bulk of galaxies residing in massive halos have a lower intrinsic \lya equivalent width than sources located in low-mass halos in our model. They are often satellite galaxies, so they no longer accrete fresh gas from the IGM. Thus, they have less intense recent star formation, which leads to lower intrinsic ratio of ionising (i.e. \lyat) to UV-continuum photons. 

Therefore, the host halo population probed by narrow-band surveys using high EW selections \citep[e.g.][]{ouch08} may be biased towards low-mass halos. Nevertheless, we remind our model does not reproduce very well the EW distributions from narrow-band LAE surveys. As discussed in Section \ref{subsec:expected_rel}, we underpredict the number of high EW sources, i.e. EW $\gtrsim$ 100-150 \AA, so the curves for the LAE detection rates in Figure \ref{fig:ew_mh} should be shifted up by an amount which depends on the exact form of the EW distribution. On the other hand, these high-EW sources correspond to UV-faint galaxies according to observations so they will be located in low mass halos on average. Should they be added to our model, this could only have the effect of accentuating the trend between LAE detection rate and halo mass that we see in Figure \ref{fig:ew_mh}.

\section{Discussion - Summary}
\label{sec:summary}

In this paper, we have investigated the connection between Lyman-alpha Emitters and Lyman-Break galaxies from z $\approx$ 3 to z $\approx$ 7 with a semi-analytic model of galaxy formation. Observationally, LAEs are detected from their strong nebular \lya emission line whereas LBG surveys pick up strong near-UV stellar continuum. While both channels intrinsically trace the stellar formation in galaxies, it remains unclear how different are the galaxies probed by these two techniques.

We have used the GALICS hybrid model to describe the formation and evolution of galaxies, based on a new high resolution N-body dark matter simulation run with a set of cosmological parameters consistent with the WMAP-5 data release. The simulation box contains more than a billion particles in a representative volume of the Universe (100 h$^{-1}$ cMpc on a side) which allows us to study the statistical properties of currently observed LAEs and LBGs at high-redshift. We describe the radiative transfer of \lya and UV photons through dusty gas outflows using a simple expanding shell model. To account for the resonant scattering of \lya photons, our model is coupled to the library of \citet{schaerer2011a} which provides us with the \lya escape fraction and the emergent line profile based on 3D numerical Monte-Carlo radiation transfer simulations.

With this approach, we aimed at interpreting the large sets of observational constraints on LAEs and LBGs to better understand the link between these two galaxy populations. We first focused on the comparison of our model with \lya and UV observations of \lyat- and UV-selected galaxies. In a second part, we presented predictions on the host halo masses and halo occupation of LAEs and LBGs based on our model. 

We summarise the main results from our paper as follows.\\\\
(i) \modi{We can reproduce the luminosity functions of LAEs and LBGs between z $\approx$ 3 and z $\approx$ 7 reasonably well. At z $\gtrsim$ 6, the model may slightly underpredict the abundances of sources, as it only matches data with the lowest observed number densities. We note that the data, especially for the \lya LFs, are not always homogeneous due to different selections and degrees of contamination, so it is not straightforward to constrain models very accurately, in particular at the highest redshifts}.\\

(ii) Applying selection criteria similar to those used in the observations, the UV LFs of LAEs predicted by the model are in rather good agreement with the observed ones, except for LAE samples selected above high EW thresholds.
This highlights the fact that strong emitters (EW $\gtrsim$ 70\AA) are almost absent from our model, and we then investigate in more details the link between \lya and UV emission properties of LAEs, i.e. the \lya equivalent widths. \\
 
(iii) While various mechanisms, extensively discussed in the literature, can produce high EWs, we suggest that bursty star formation, rather than almost constant star formation, should bring our model in agreement with the observed EW distributions of LAEs. We show \modi{that} increasing the surface density SF threshold by an order of magnitude compared to the standard threshold can mimic stochastic star formation rate, needed to match the EW distributions. \\
 
(iv) We find that the fraction of strong line emitters (EW $>$ 50 \AA) in LBG samples increases towards faint UV magnitudes, as reported by \citet{stark10} at z $=$ 4-6. This trend is essentially due to resonant \lya photons being more affected by dust than UV continuum photons in bright LBGs because these objects have a higher \hi column density on average, echoing the results of \citet{verh08}.\\

(v) We study the redshift evolution of the fraction of strong (EW $>$ 55 \AA) and weak (EW $>$ 25 \AA) \lya line emitters, $X_{\rm LAE}$, within UV-selected samples. Overall, $X_{\rm LAE}$ seems to quickly evolve when varying UV magnitude limits and \lya equivalent width cuts. The predictions agree reasonably well with the data for strong emitters, but the low number statistics and the rapid variation of the shape/normalisation of the $X_{\rm LAE}$-redshift relation with UV magnitudes limits and \lya equivalent width cuts prevent us from drawing robust conclusions. In addition, as shown by the study of the model EW distributions as a function of UV magnitude, a stronger differential UV/\lya dust extinction in UV-bright objects is needed to improve the quantitative agreement with the data. \\

(vi)  We find that LAEs and LBGs in each sample are located in very similar halos, and that brighter sources inhabit more massive halos average. At z $=$ 3 for instance, faint LAEs and LBGs are hosted by $5 \times 10^{10}$ \msun{} halos, whereas bright objects reside in halos of $\gtrsim 10^{12}$ \msun.\\

(vii) The halo occupation rate of LAEs and LBGs is very similar, except for the bright sample, where LAEs appear to be about four times rarer than LBGs, in broad agreement with the observations of \citet{kovac2007a}. \\

(viii) More massive halos tend to host weaker \lya emitters on average in our model, which suggests that LAEs selected with high EW cuts will preferentially probe LAEs in low mass halos.\\

In this article, we have shown that our model can reproduce many observed statistical quantities at z $\approx$ 3-7, using a rather simple modelling of galaxies and \lya radiation transfer. Some additional ingredients may still be necessary to accommodate the tail of the EW distribution, possibly due to burstiness, and the exact scaling between UV luminosities and \lya equivalent width, by increasing the effect of dust in UV-bright objects. Nevertheless, within a single and coherent framework, we are able to interpret reasonably well the abundances of LAEs and LBGs, as well as the UV LFs of LAEs and the \lya fraction in UV-selected samples. 

The picture emerging from this study of the \lyat, UV, and halo properties of high redshift galaxies is consistent with the idea that LAEs are a subset of the LBG population. The apparent differences between these two populations would simply arise from the EW selection in LAE surveys, and the continuum detection limit used to select LBGs. In other terms, LAEs undetected in the UV should always be probed by deeper LBG surveys. At given SFR, LAEs and LBGs should have very similar UV magnitude, and the effect of dust on \lya photons will redistribute the most massive, intrinsically \lyat-bright, galaxies at fainter fluxes and lower equivalent widths, although these can still appear as LBGs \citep{shapley03}. As a consequence, LAEs that are observed are more likely to have lower dust extinction than LBGs. This picture seems consistent with the comparison study of the mid-IR properties of LAEs and LBGs \citep{yuma2010a}, the work of \citet{verh08} based on stellar population and \lya radiative transfer modelling, the hydrodynamical simulations of \citet{dayal2012}, and the results of \citet{cooke2009} and \citet{schaerer2011} who demonstrate that the \lya emission properties of UV-selected galaxies can be statistically inferred from broad-band photometry and SED modelling techniques \citep[see][for a review]{dunlop2013}. 

Nonetheless, LAEs remain quite powerful at probing faint (i.e. low-mass) high redshift star-forming galaxies, especially strong emitters which would require very deep observations to be detected in UV continuum surveys. In addition, the observational and theoretical study of the \lya line profile and morphology of LAEs hold great potential in probing the distribution, the content, and the kinematics of the gas in the interstellar medium, but also at larger scale in the circumgalactic and intergalactic media \citep*[see e.g.][]{barnes2014}. Increased numerical power along with larger data samples extending to fainter luminosities and higher redshift, expected from ongoing and forthcoming surveys (e.g. MUSE, HETDEX, KCWI or Hyper Suprime-Cam), will undoubtedly help refine our understanding of \lya emitters, and galaxy formation and evolution at high redshift in general.\\

\section*{Acknowledgements}
We thank Daniel Schaerer for his helpful comments on an earlier version of the manuscript. TG is the recipient of an Australian Research Council SuperScience Fellowship. This work was granted access to the HPC resources of CINES under the allocation 2012-c2012046642 made by GENCI (Grand Equipement National de Calcul Intensif). LMD acknowledges support from the Lyon Institute of Origins under grant ANR-10-LABX-66. MH acknowledges the support of the Swedish Research Council, Vetenskapsr{\aa}det and the Swedish National Space Board (SNSB).

\bibliographystyle{mn2e}
\bibliography{biblio_lae-lbg}


\label{lastpage}

\end{document}